\documentclass[twocolumn]{aastex62}
\usepackage{booktabs}
\usepackage{amsmath}
\usepackage{array}
\usepackage{float}
\usepackage{natbib}
\usepackage{booktabs}
\usepackage{amsmath}
\usepackage{array}
\usepackage{wasysym}
\newcommand{\msun}{M$_\odot$}

\newcommand{\msy}{M$_\odot$~yr$^{-1}$}
\newcommand{\mmsun}{\rm M_\odot}

\newcommand{\mbh}{M$_\bullet$}
\newcommand{\mmbh}{\rm M_\bullet}

\newcommand{\MSig}{$M-\sigma$}
\newcommand{\MKs}{$M-L_{K_{s},sph}$}

\newcommand{\Htwo}{H$_{2}$}
\newcommand{\HI}{\ion{H}{1}}
\newcommand{\HII}{\ion{H}{2}}

\newcommand{\Fe}{[\ion{Fe}{2}]}

\newcommand{\SII}{[\ion{S}{2}]}

\newcommand{\OIII}{[\ion{O}{3}]}
\newcommand{\NII}{[\ion{N}{2}]}
\newcommand{\SiVI}{[\ion{Si}{6}]}
\newcommand{\CaVIII}{[\ion{Ca}{8}]}

\newcommand{\SIX}{[\ion{S}{9}]}

\newcommand{\SiI}{\ion{Si}{1}}
\newcommand{\CaII}{\ion{Ca}{2}}

\newcommand{\Ha}{H\,$\alpha${}}
\newcommand{\Hb}{H\,$\beta${}}

\newcommand{\pab}{Pa\,$\beta${}}

\newcommand{\brg}{Br\,$\gamma${}}

\newcommand{\ecs}{erg~cm\pwr{-2}~s\pwr{-1}}
\newcommand{\es}{erg~s\pwr{-1}}

\newcommand{\kms}{km~s\pwr{-1}}

\newcommand{\hk}{$H-K$}

\newcommand{\pwr}[1]{$^{#1}$}

\makeatletter
\newcommand*{\rom}[1]{\expandafter\@slowromancap\romannumeral #1@}
\makeatother
\makeatletter
\def\blfootnote{\xdef\@thefnmark{}\@footnotetext}
\makeatother

\newcolumntype{R}[1]{>{\RaggedLeft\arraybackslash}p{#1}}
\newcolumntype{C}[1]{>{\centering\arraybackslash}p{#1}}

\shorttitle{NGC~5728 Ionization Cones : II - Kinematics}
\shortauthors{Durr\'{e} \& Mould}
\begin{document}

\title{The AGN Ionization Cones of NGC~5728 : II - Kinematics}
\author[0000-0002-2126-3905]{Mark Durr\'{e}}
\author[0000-0003-3820-1740]{Jeremy Mould}
\affil{Centre for Astrophysics and Supercomputing, Swinburne University of Technology, P.O. Box 218, Hawthorn, Victoria 3122, Australia}
\email{mdurre@swin.edu.au}
\defcitealias{Durre2018a}{Paper~I}
\begin{abstract}
	We explore the gas morphology and excitation mechanisms of the ionization cones of the Type II Seyfert galaxy NGC~5728. Kinematics derived from near-IR and optical data from the SINFONI and MUSE IFUs on the VLT reveal AGN-driven outflows powered by a super-massive black hole (SMBH) of mass $3.4 \times 10^{7}$ \msun, bolometric luminosity of $1.46 \times 10^{44}$~\es, Eddington ratio $3.3 \times 10^{-2}$ and an accretion rate of $2.7~\times 10^{-2}$~\msy. The symmetric bicone outflows show rapid acceleration to $\pm250$ \kms{} at $\sim250$ pc, then decelerating to $\sim130$ \kms{} at 500 pc from the AGN, with an estimated mass outflow rate of 38 \msy; the mass ratio of outflows to accretion is 1415. The kinetic power is $1.5 \times 10^{42}$ \es{}, 1\% of the bolometric luminosity. Over the AGN active lifetime of $\sim$10\pwr{7} years, $1.6 \times 10^{8}~\mmsun$ of gas can become gravitationally unbound from the galaxy, a large proportion of the gas mass available for star formation in the nuclear region. The bicone internal opening angle (50\degr.2) and the inclination to the LOS (47\degr.6) were determined from \Fe{} line profiles; the outflow axis is nearly parallel to the plane of the galaxy. This geometry supports the Unified Model of AGN, as these angles preclude seeing the accretion disk, which is obscured by the dusty torus.
\end{abstract}
\section{Introduction}
This is the second of two papers studying the narrow-line region (NLR) of the Seyfert 2 galaxy NGC~5728 using data acquired from the  SINFONI and MUSE Integral Field Unit (IFU) spectroscopic instruments on ESO's VLT telescope. This galaxy has a complex nuclear structure, with a spectacular ionization cones emanating from the hidden active galactic nucleus (AGN), an inner star forming ring and spiraling central dust lanes \citep{Gorkom1982,Schommer1988,Wilson1993, Mediavilla1995,Capetti1996,Rodriguez-Ardila2004,Rodriguez-Ardila2005,Dopita2015a}. \cite{Urry1995}, in the seminal review paper on AGN unification, cited NGC~5728 as the paradigm of a Type 2 AGN with ionization cones; these are predicted by the Unified Model of AGN, as radiation from the accretion disk is collimated by the surrounding dusty torus to impinge on the ISM, exciting the gas by photo-ionization. This object is a prime target for investigation, particularly using observations in the near infrared, allowing penetration of obscuring dust. 

In our previous paper \citep[hereafter `Paper I']{Durre2018a}, we explored the near infrared (NIR) flux distributions and line ratios of \Fe, \Htwo, \SiVI{} and hydrogen recombination emission lines, plus the optical \Ha, \Hb, \OIII, \SII{} and \NII{} emission lines, to characterize gas excitation mechanisms, gas masses in the nuclear region and extinction around the AGN. In this paper, we will use these data to examine the kinematics of the  cones, determining mass outflow rates and power and relate these to the AGN bolometric luminosity and the amount of gas that can be expelled during the AGN activity cycle. We will also determine the stellar kinematics and relate them to the gas kinematics. The super-massive black hole (SMBH) mass will be determined by several methods and the Eddington ratio deduced. 

Our conclusions from \citetalias{Durre2018a} were as follows:

\begin{itemize}
	\item Dust lanes and spiraling filaments feed the nucleus, as revealed by the \textit{HST} structure maps.
	\item A star-forming ring with a stellar age of 7.4 -- 8.4 Myr outlines a nuclear bulge and bar.
	\item A one-sided radio jet, impacting on the inter-stellar medium (ISM) at about 200 pc from the nucleus, is combined with radio emission from the supernova (SN) remnants in the SF ring.
	\item Ionized gas in a bi-cone, traced by by  emission lines of \HI{}, \Fe{} and \SiVI, extends off the edges of the observed field. MUSE \OIII{} data shows that the full extent of the outflows is over 2.5 kpc from the nucleus. The cones are somewhat asymmetric and show evidence of changes in flow rates, plus impact on the ISM, combined with obscuration.
	\item The AGN and BLR is hidden by a dust bar of size $64 \times 28$ pc, with up to 19 magnitudes of visual extinction, with an estimated dust temperature at the nuclear position of $\sim870$~K. Extinction maps derived from both hydrogen recombination and \Fe{} emission lines show similar structures, with some indication that \Fe-derived extinction does not probe the full depth of the ionization cones. There is  evidence for dust being sublimated in the ionization cones, reducing the extinction. This is supported by extinction measures derived from MUSE \Ha/\Hb{} emission, which show much decreased absorption in the ionization cones.
	\item An extended coronal-line region is traced by \SiVI, out to $\sim300$ pc from the nucleus. This is excited by direct photo-ionization from the AGN, plus shocks from the high-velocity gas. Higher-ionization species (\CaVIII{} and \SIX{}) are closely confined around the AGN.
	\item NIR line-ratio active galaxy diagnostics show mostly AGN-type activity, with small regions of LINER and TO modes at the peripheries of the cones and where the dusty torus obscures the direct photo-ionization from the BH. MUSE optical diagnostics (BPT diagrams) show a clean star-formation/AGN mixing sequence.
	\item The \Htwo{} gas is  spatially independent of the cones, concentrated in an equatorial disk in the star-forming ring, but also showing entrainment along the sides of the bicone. The warm \Htwo{} has a mass of 960 \msun, with an estimated $6\times10^8~\mmsun$ of cold \Htwo{} in the field of view. Line-ratio diagnostics indicate that the gas is excited by thermal processes (shocks and radiative heating of gas masses) to temperatures in the range 1400-- 2100~K, with an increased fluorescent excitation towards the SF ring.
	\item With 100 pc of the nucleus, the ionized gas mass (derived from the \brg{} emission) is $8 \times 10^5~\mmsun$ and the cold gas mass (from extinction calculations) is $4.9 \times 10^6~\mmsun$; however this could be underestimated due to dust sublimation in the cones.
\end{itemize}

\section{Observations, Data Reduction and Calibration}
\label{sec:ngc5728Observations}
The data acquisition, reduction and calibration procedure was described in detail in \citetalias{Durre2018a}. 3D data cubes were produced from the SINFONI and MUSE observations, as laid out in Table \ref{tbl:ngc5728datacubes}. Subsequent cube manipulation and computation is done using the the cube data viewer and analysis application \texttt{QFitsView} \citep{Ott2012}, which incorporates the \texttt{DPUSER} language. In this paper, we use the standard cosmology of H$_0 = 73$ \kms Mpc\pwr{-1}, $\Omega_{Matter}=0.27$ and $\Omega_{Vacuum}=0.73$. All images are oriented so that North is up, East is left, as indicated in Figure \ref{fig:ngc5728stellarkinematics1}.
\begin{table}[!htbp]
	\centering
	\caption{3D Data Cubes from IFU observations}
	\label{tbl:ngc5728datacubes}
	\begin{tabular}{lccc}
		\toprule
		Instrument &        Filter        &    FOV (\arcsec)     &    Plate Scale (\arcsec)    \\
		           & $\lambda$ range (nm) &    FOV (pc)         & R ($\Delta\lambda/\lambda$) \\ \midrule
		SINFONI~\tablenotemark{a}    &          J           &         8          &    0.25\tablenotemark{*}    \\
		           &      1095--1402      &        1600        &            2400             \\
		SINFONI~\tablenotemark{b}    &         H+K          &        3.8         &    0.1\tablenotemark{*}     \\
		           &      1452--2457      &        760         &            1560             \\
		MUSE~\tablenotemark{c}       &          V           &        68.2        &             0.2             \\
		           &       475--935       &       13640        &            1800             \\ \bottomrule
		%		WiFeS   &          B           &   $25 \times 38$   &              1              \\
		%          &       350--570       & $5000 \times 7600$ &            3000             \\
		%		WiFeS   &          R           &   $25 \times 38$   &              1              \\
		%          &       540--702       & $5000 \times 7600$ &            7000             \\ 
	\end{tabular}
{\footnotesize\tablenotetext{*}{SINFONI plate scale is re-binned to half size in the final data cube.}
\tablenotetext{a}{ESO program ID/PI 093.B-0461(A)/Mould}
\tablenotetext{b}{ESO program ID/PI 093.B-0057(B)/Davies}
\tablenotetext{c}{ESO program ID/PI 097.B-0640(A)/Gadotti}}
\end{table}
\section{Results}
\subsection{Deriving Kinematic Parameters}
We use the \texttt{QFitsView} \textit{velmap} function; this fits a Gaussian curve for every spaxel at the estimated central wavelength and full width half maximum (FWHM). This generates maps of the best fit continuum level (C), height (H), central wavelength ($\lambda$) and FWHM. These are readily converted to the dispersion ($\sigma$), the flux (F) and the emission equivalent width (EW).

Map values are accepted or rejected based on ranges on each of the derived parameters produced by the procedure, based on visual inspection of the fluxes and kinematic structures i.e. the derived values of continuum, line flux, LOS velocity or dispersion have to be within certain values. Single pixels may be rejected because of large excursions from their neighbors; the source of anomalous values is usually noise spikes in the data or low flux values, causing poor model fits. Rejected pixels are then interpolated over from neighbors or masked out, as appropriate.

\subsection{Stellar Kinematics}
\label{sec:ngc5728StellarKinematics}
\begin{figure*}[!htbp]
	\centering
	\includegraphics[width=0.7\linewidth]{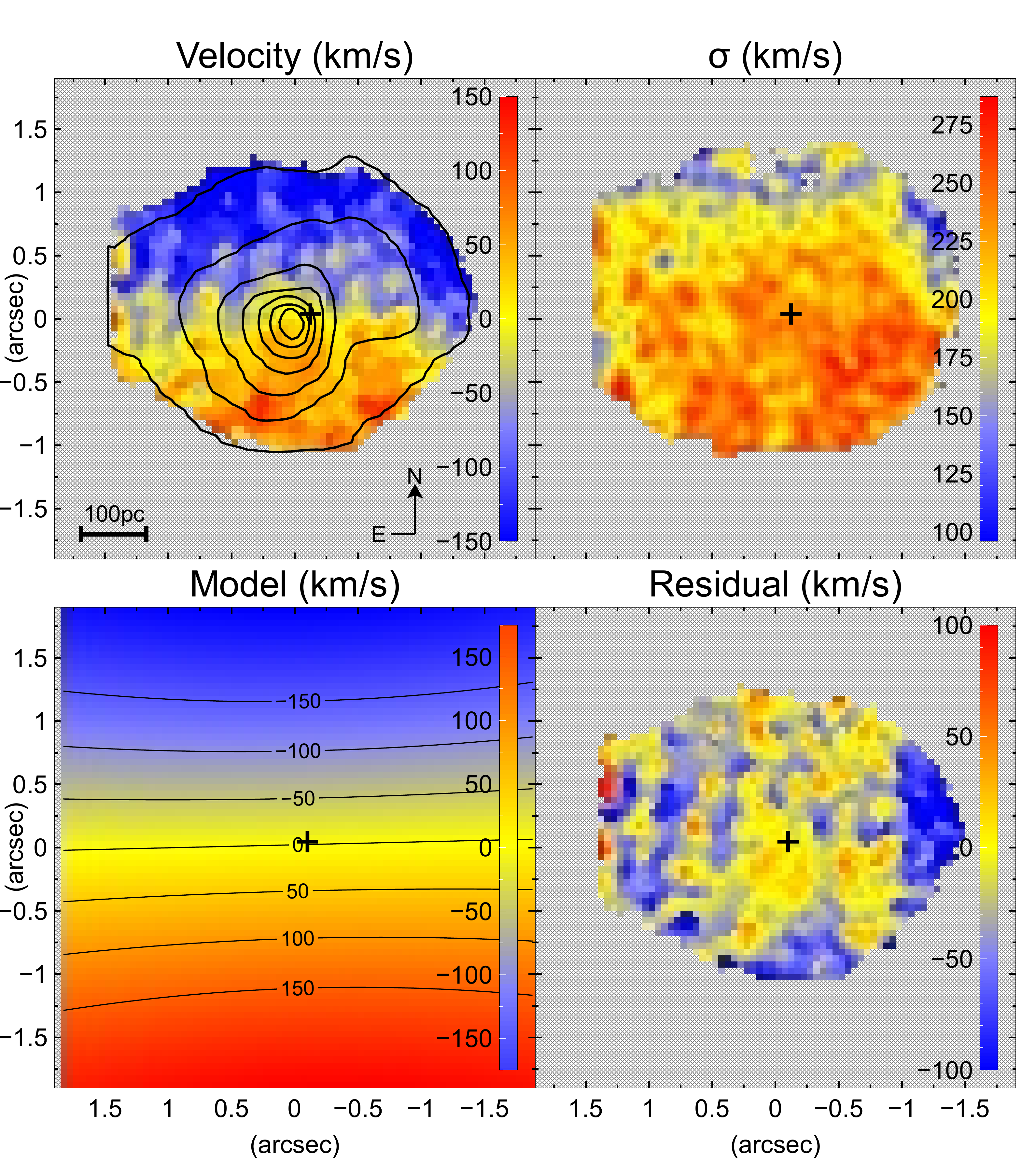}
	\caption{Top row : Stellar LOS velocity and dispersion from the \SiI{} absorption line (1589 nm). Contours are the H-band normalized flux, from 30 to 90\% of maximum, in steps of 10\%. Velocity and dispersion values in \kms. Bottom row : Plummer model fit and residual. This shows the ordered, tumbling rotation of the circumnuclear stellar structure, mis-aligned to the main galactic plane rotation axis by $\sim60\degr$.}
	\label{fig:ngc5728stellarkinematics1}
\end{figure*}

The stellar kinematics help to disentangle the complex nuclear structure and also give an estimate for the SMBH mass. They are derived from the \SiI{} line in the \textit{H}-band at 1589.2 nm using the \textit{velmap} procedure from \texttt{QFitsView}. This procedure is designed for emission lines; the spectrum was inverted to turn an absorption line into a peak. The resulting (cleaned) velocity and dispersion map is presented in Fig. \ref{fig:ngc5728stellarkinematics1}. This shows a simple rotation around the east-west axis; the average dispersion in the central 0\arcsec.5 radius is $230\pm26$ \kms{} and the systemic velocity is $2984\pm55 $\kms.  Combined with the wavelength calibration and the barycentric velocity corrections (derived above), our systemic velocity is 2937 \kms{}, compared to the \HI{} 21 cm velocity of 2786 \kms{} \citep{Roth1991} and the \Ha-derived velocity of 2804 \kms{} \citep{Catinella2005}. The difference can be ascribed to the \HI{} measuring the velocity of the whole galaxy and the \Ha{} velocity including gas with a non-zero peculiar velocity.

The stellar velocity field derived from the \SiI{} line is modeled using the Plummer potential,  \citep{Plummer1911} \citep[see e.g.][]{Riffel2011}.
\begin{equation}
V_{LOS} = V_{Sys} + \dfrac{\sqrt{\dfrac{R^{2}GM}{(R^{2}+A^{2})^{3/2}}}\left[\sin(i)\cos(\Psi-\Psi_{0})\right]}{\left[\cos^{2}(\Psi-\Psi_{0})+\dfrac{\sin^{2}(\Psi-\Psi_{0})}{\cos^{2}(i)}^{3/2}\right]} \label{eqn:odraPlummerProfile}
\end{equation}
where $ V_{LOS} $ and $ V_{Sys} $ are the observed line of sight and systemic velocities, \textit{R} is the projected distance from the nucleus, $\Psi$ is the position angle, $\Psi_0$ is position angle of the line of nodes, \textit{G} is the gravitational constant, \textit{A} is the scale length, \textit{i} is the disk inclination (\textit{i}=0 for face-on) and \textit{M} is the enclosed mass. Including the location of the kinematic center, this equation has 6 free parameters. The kinematic center and line of nodes can be well-constrained from the observations. Standard  techniques (e.g. Levenberg–-Marquardt least-squares fitting algorithm) are used to solve for the parameters. 

The model was fitted with the generalized reduced gradient algorithm (`GRG Nonlinear') implemented in the \texttt{MS-Excel} add-on \textit{Solver}. We centered the kinematics on the AGN location. The model and residual results are also shown in Fig. \ref{fig:ngc5728stellarkinematics1}. The axis PA is 1\degr.4 (i.e. the rotational equator is nearly edge-on to the LOS), with negligible inclination (-0\degr.39), i.e. the ring is essentially face-on, and a scale length of $\sim395$ pc (i.e. off the edge of the field). A slice through the measured velocity field along the axis produces a velocity gradient of 205 \kms{} over 140 pc, which implies an enclosed mass of $\sim3.5 \times 10^{8}~\mmsun$, assuming simple Keplerian rotation. 

The kinematics of the inner ring and bar match (to first order) those of the molecular hydrogen (see Section \ref{sec:ngc5728h2kinematics}); however, if this ring of star formation is a disk structure, then it appears \textit{not} to be rotating about its normal axis, which would be almost along the LOS. Instead the Plummer model shows that the rotation axis is almost east-west and the ring is `tumbling', mis-aligned to the main galactic plane rotation axis by $\sim60\degr$.

\begin{figure*}[!htbp]
	\centering
	\includegraphics[width=0.7\linewidth]{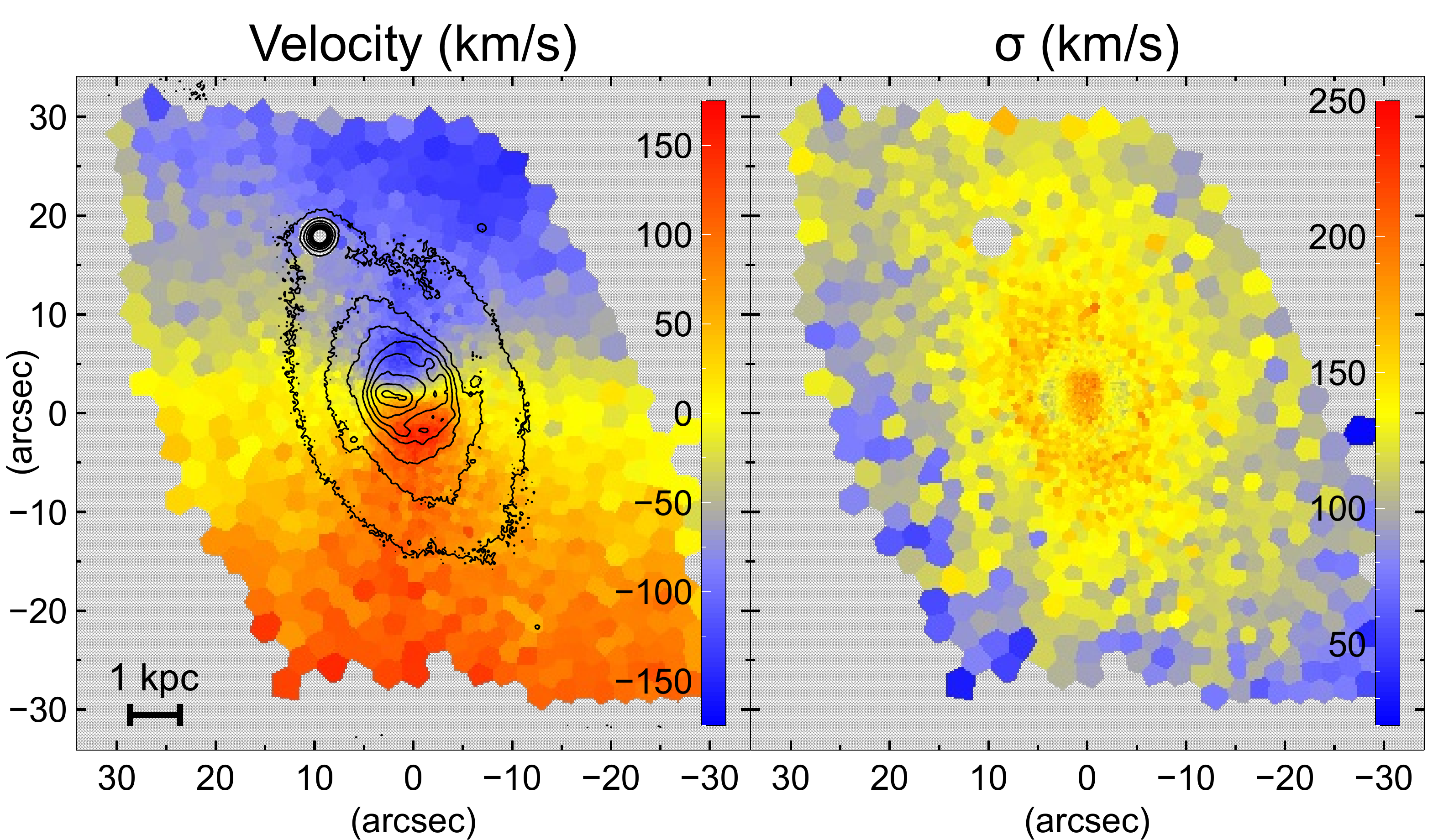}
	\caption{Top row : Stellar LOS velocity and dispersion from the \CaII{} absorption line (from MUSE data). Contours are the \textit{V}-band surface brightness magnitude, in steps of 0.5 mag from 17.5 to 21 mag deg\pwr{-2}. Velocity and dispersion values in \kms. Values have been smoothed in the inner $28\arcsec \times 28\arcsec$. Note that the kinematics and photometry are aligned, and the stellar dispersion is suppressed in the SF ring. The feature in the NE quadrant in the photometry is a foreground star.}
	\label{fig:ngc5728stellarkinematics2}
\end{figure*}

A large-scale plot of the stellar kinematics was obtained using the MUSE data cube, using the 8544~\AA{} \CaII{} triplet absorption feature (using the \textit{velmap} procedure in \texttt{QFitsView}). The derived velocity and dispersion is shown in Fig. \ref{fig:ngc5728stellarkinematics2}. The `Weighted Voronoi Tessellation' (WVT) \citep{Cappellari2003} was used to increase the S/N for pixels with low flux; we use the \textit{voronoi} procedure in \texttt{QfitsView}. This aggregates spatial pixels in a region to achieve a common S/N, in this case set to 60. This needs both signal and noise maps, which were, respectively, the white-light image and from the square-root of the average variance statistics from the MUSE data. Within the central 1\arcsec, the velocity dispersion is $\sigma_{*} = 178.7 \pm 10.1$ \kms, again compatible with the values derived above for the CO band-heads.

The rotational axis is skewed from the outer regions to the inner, but the change is not discontinuous; this is different from the long-slit results from \cite{Prada1999}, who found that the core is counter-rotating with respect to the main galaxy. The dispersion values show a suppression at the SF ring and then an inwards rise; the inner reduction was also seen by \cite{Emsellem2001} using the ISAAC spectrograph of the VLT. The skewed velocity field and central dispersion drop are characteristic of inner bars that are rotating faster than the primary bar;  at the inner Lindblad resonance (ILR) of the primary bar (which is presumed to be the location of the SF ring), the gas accumulates (driven there by the gravity torque of the bar), and then forms stars. The dispersion of the young stars reflects the gas dispersion, and is lower than that of the old stars. The inner rise is then due the greater prevalence of old bulge stars, plus the presence of the SMBH. (F. Combes, B. Ciambur, pers. comm.).

\subsection{Gaseous Kinematics}
\subsubsection{Line-of-sight Velocity and Dispersion}
\label{sec:ngc5728losvd}
The kinematics of gas around the nucleus diagnose outflow, inflow and rotational structures in the nuclear region; combined with the emission-line flux, we can compute the energetics of this gas. Different species have different kinematics structures, e.g. outflows are prominent in hydrogen recombination and other ionized emission lines, whereas molecular hydrogen tends to be in a rotating gas disk around the center.

\begin{figure*}
	\centering
	\includegraphics[width=.7\linewidth]{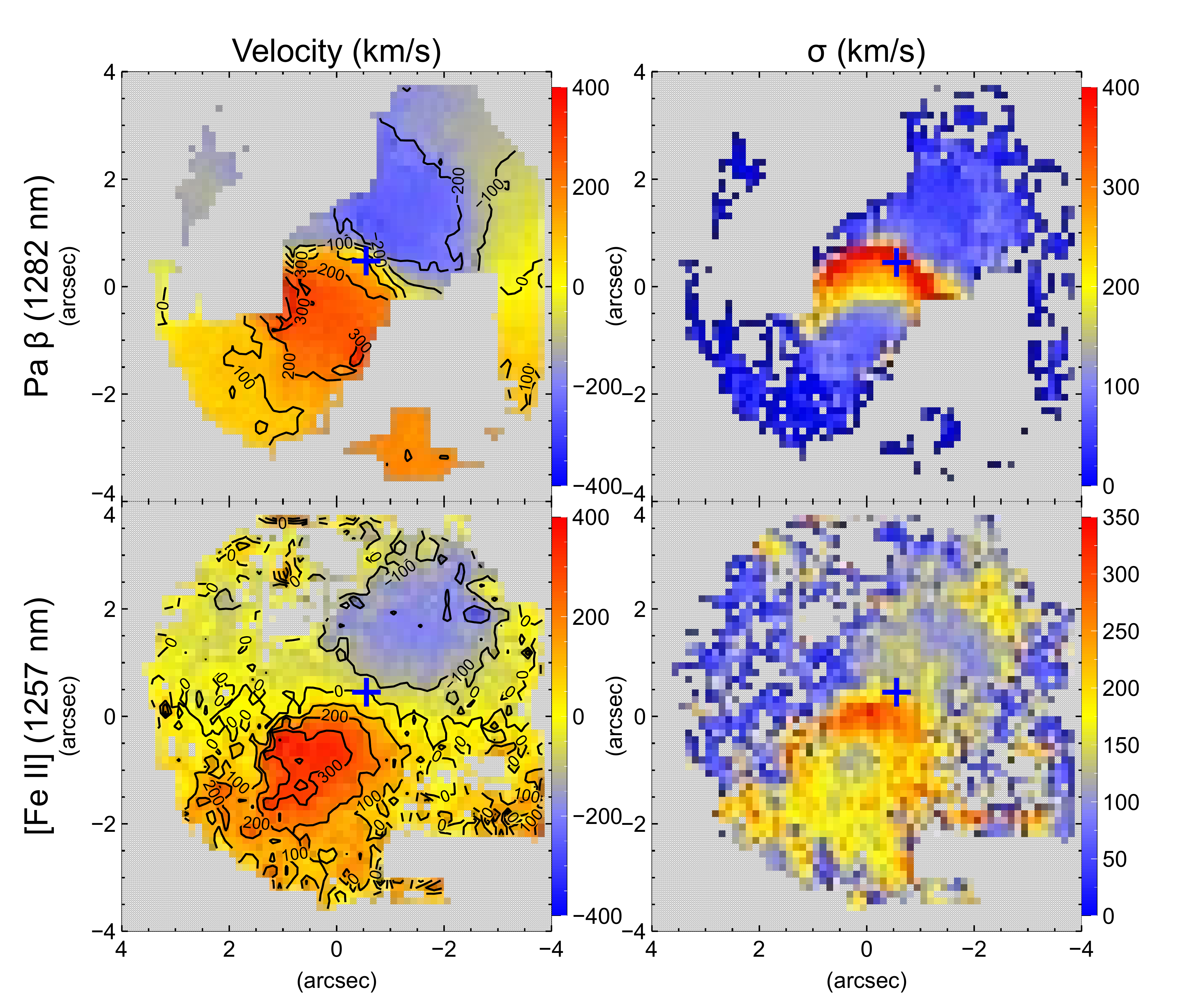}
	\caption{Kinematics for \pab{} and \Fe{} (1257 nm). Flux units 10\pwr{-16} \ecs, EW units in nm. Both species show the strong bi-polar flow in velocity, with the nuclear dispersion being much higher for hydrogen recombination than \Fe{}. The zero velocity point is set at the AGN location. White pixels are where the emission line was too weak to derive the kinematics.}
	\label{fig:ngc5728gaskinematics3}
\end{figure*}

\begin{figure*}
	\centering
	\includegraphics[width=.7\linewidth]{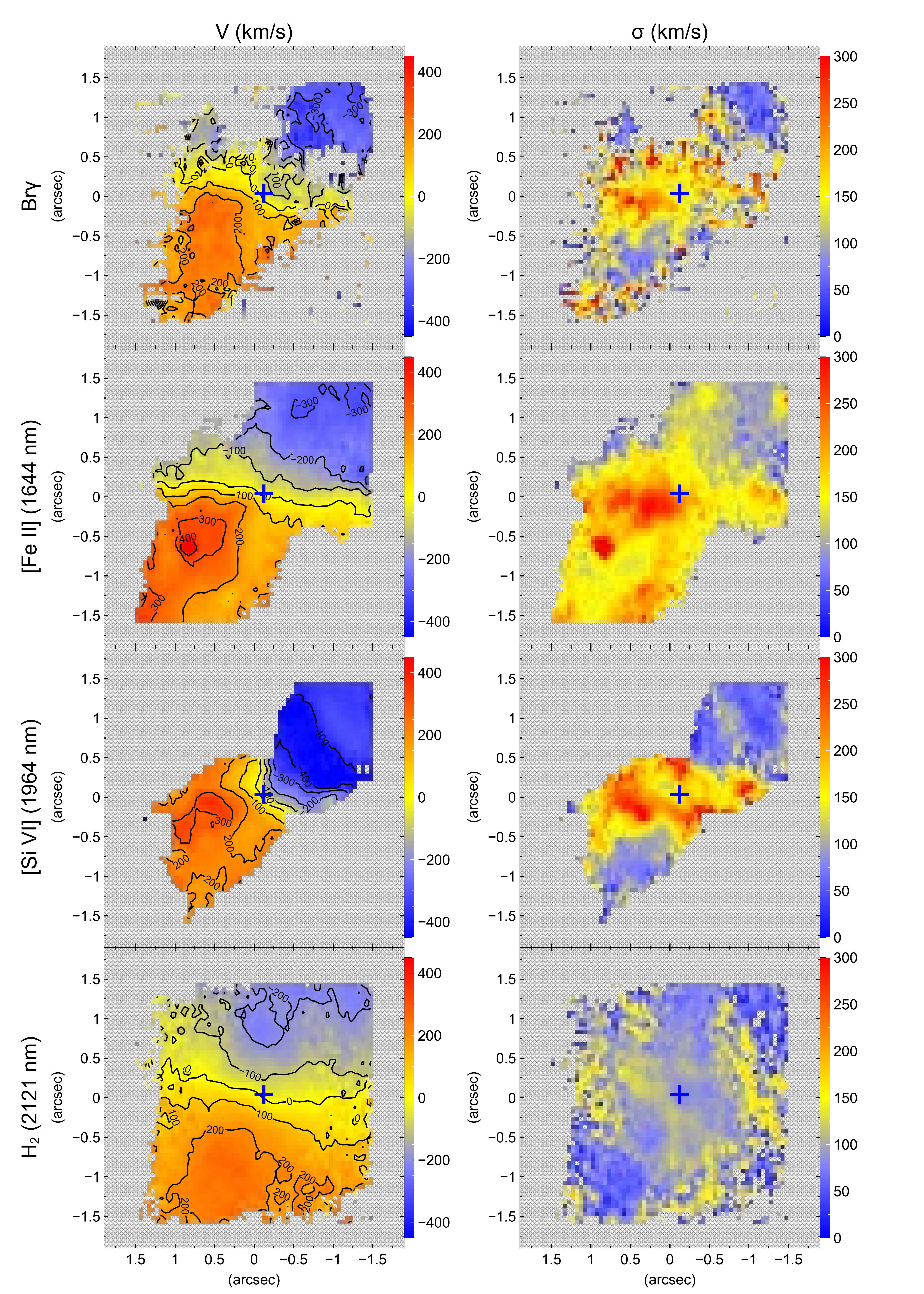}
	\caption{As for Fig. \ref{fig:ngc5728gaskinematics3} for \brg, \Htwo{} (2121 nm), \Fe{} (1644 nm) and \SiVI.}
	\label{fig:ngc5728gaskinematics4}
\end{figure*}

In Figs. \ref{fig:ngc5728gaskinematics3} and  \ref{fig:ngc5728gaskinematics4} we present the line-of-sight velocity and dispersion for the main emission lines in the \textit{J} and \textit{H+K} band. We use the kinematics of these lines to delineate outflows, rotations and (possible) feeding flows. We set the velocity zero-point at the location of the AGN; see section 3.5 of \citetalias{Durre2018a}.

From the \pab{} kinematics, the velocity field reaches a maximum of about -250 \kms{} at 1\arcsec.1 NW ($\sim220$ pc projected distance) and 270 \kms{} at 1\arcsec.4 ($\sim280$ pc) SE of the nucleus, then decelerates to -150/110 \kms{} at double the respective distances from the nucleus. This profile suggests gas accelerated away from the AGN by radiation pressure, then decelerating as it interacts with the ISM and the radiation pressure drops off with increasing distance from the source. High absolute velocity values in the northern and southern parts of the field are due to the star-forming ring rotation, aligned with the stellar kinematics along a PA of $\sim10\degr$ (see Section \ref{sec:ngc5728StellarKinematics}). The \Fe{} velocity field shows a similar pattern to the \pab{} field, however the peak velocity is further from the AGN.

%Because of the telluric feature, the \Fe{} velocity field, while it displays a similar spatial extent, is biased towards higher recession velocity. A manual check of the velocity fit along the outflows, where the offending spectral pixels are masked out, shows that the velocity range is in line with the \pab{}, i.e. $\pm200$ \kms.
	
The \brg, \Fe{} 1644 nm and \SiVI{} velocity and dispersion maps show very similar structure, as is also the case for the flux and EW. Again, the \Htwo{} kinematics do not align  with the other species (see Section \ref{sec:ngc5728h2kinematics}). For \Fe{} 1257 nm, the higher dispersion is more extended; around the SE cone, the higher values outline the cone, supporting an emission mechanism of a shock boundary. The \brg, \Fe{} 1644 nm and \SiVI{} dispersion plots show the same structure at higher spatial resolution.

\begin{figure}[!h]
\centering
\includegraphics[width=1\linewidth]{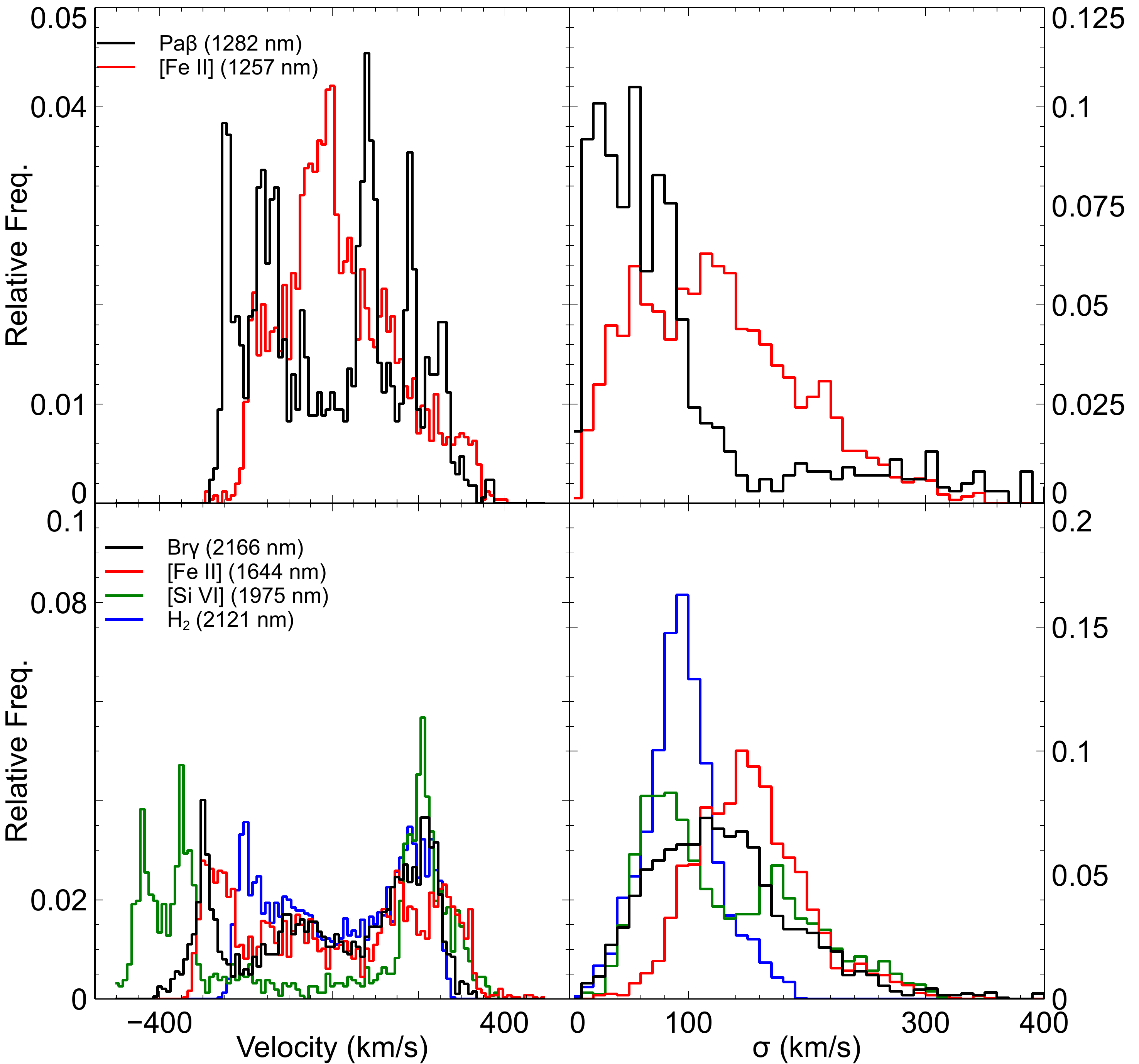}
\caption{Top panels: Velocity (left) and dispersion (right) distribution histograms for \pab{} and \Fe{} (1257 nm). The higher \Fe{} dispersion is from line-broadening by LOS components of the outflow velocity and turbulent entrainment. Bottom panels: Velocity (left) and dispersion (right) distribution histograms for \brg, \Htwo{} (2121 nm), \Fe{} (1644 nm) and \SiVI. The \Htwo{} is kinematically colder than the other species.}
\label{fig:ngc5728gaskinematics5}
\end{figure}

We present histograms of the velocity and dispersion values in Figs. \ref{fig:ngc5728gaskinematics5}. The \HII{} and \SiVI{} velocity histograms show the characteristic double peak of outflows or rotation, which is also shown in the \Fe{} 1644 nm line. The negative \SiVI{}  velocities are somewhat higher than the \brg{} velocity; for the receding velocities, \Fe{} has higher values than \brg{} and \SiVI{}, with a modal value of $\sim200$ \kms; there is a small group of pixels with a velocity of $\sim$-300 -- -400 \kms. The \Fe{} 1257 nm line shows a stronger concentration around zero velocity; this is because the \textit{J}-band map takes in a larger area than the \textit{H+K}-band map, where there is a large partially-ionized volume of gas which is not in the outflow and therefore has a lower velocity.

The \Fe{} dispersion distribution is more extended than \brg{}, as seen both in the maps and histograms. As it has a lower ionization potential than \HI, it is only found in partially ionized media. In this case, it is on the edge of the cones, which will broaden the line from the LOS components of the  velocity, and by turbulent entrainment at the cone boundaries. This is clearly seen in the map of the \Fe{} 1257 nm dispersion, showing a ring of higher dispersion around the SE outflow. The \SiVI{} dispersion values seem to be somewhat bi-modal, with high values more or less coincident with the high values for \Fe{} and lower values elsewhere; this may reflect the production mechanisms, i.e. photo-ionization where directly illuminated by the accretion disk, with low dispersion, and shocks (high dispersion) on the outflow boundaries.

The high dispersion values observed for \pab{} perpendicular to the cones across the nucleus can be attributed to the beam-smearing, causing the LOS to intersect both the approaching and receding outflows and broadening the line artificially; the Seyfert classification for this object was altered from Sy 1.9 (i.e. broad \Ha{} emission lines present) to Sy 2 (no broad lines present) \citep{Shimizu2018}.

\begin{figure}[!h]
\centering
\includegraphics[width=1\linewidth]{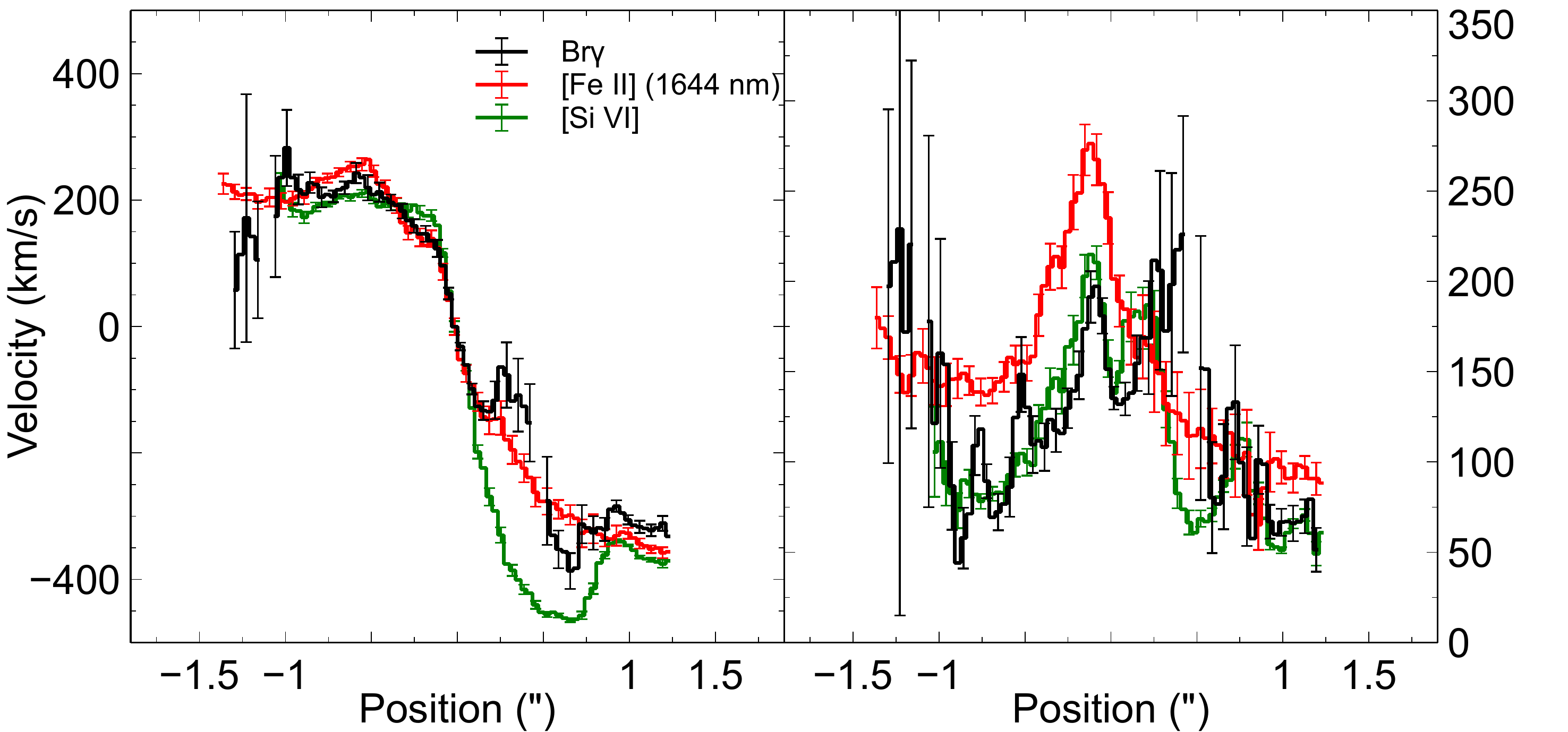}
\caption{LOS velocity and dispersion along the bicone centerline, centered on the AGN location, showing the steep gradient near the AGN and the asymptotic flattening further out.}
\label{fig:ngc5728gaskinematics7}
\end{figure}

We plot the velocity and dispersion for the \textit{H+K}-band lines in a slice along the bicone centerline in Fig. \ref{fig:ngc5728gaskinematics7}. The velocity profile is virtually identical for all species, except for a region in the NW outflow (approaching) with a higher velocity for \SiVI.

\begin{figure*}
	\centering
	\includegraphics[width=.7\linewidth]{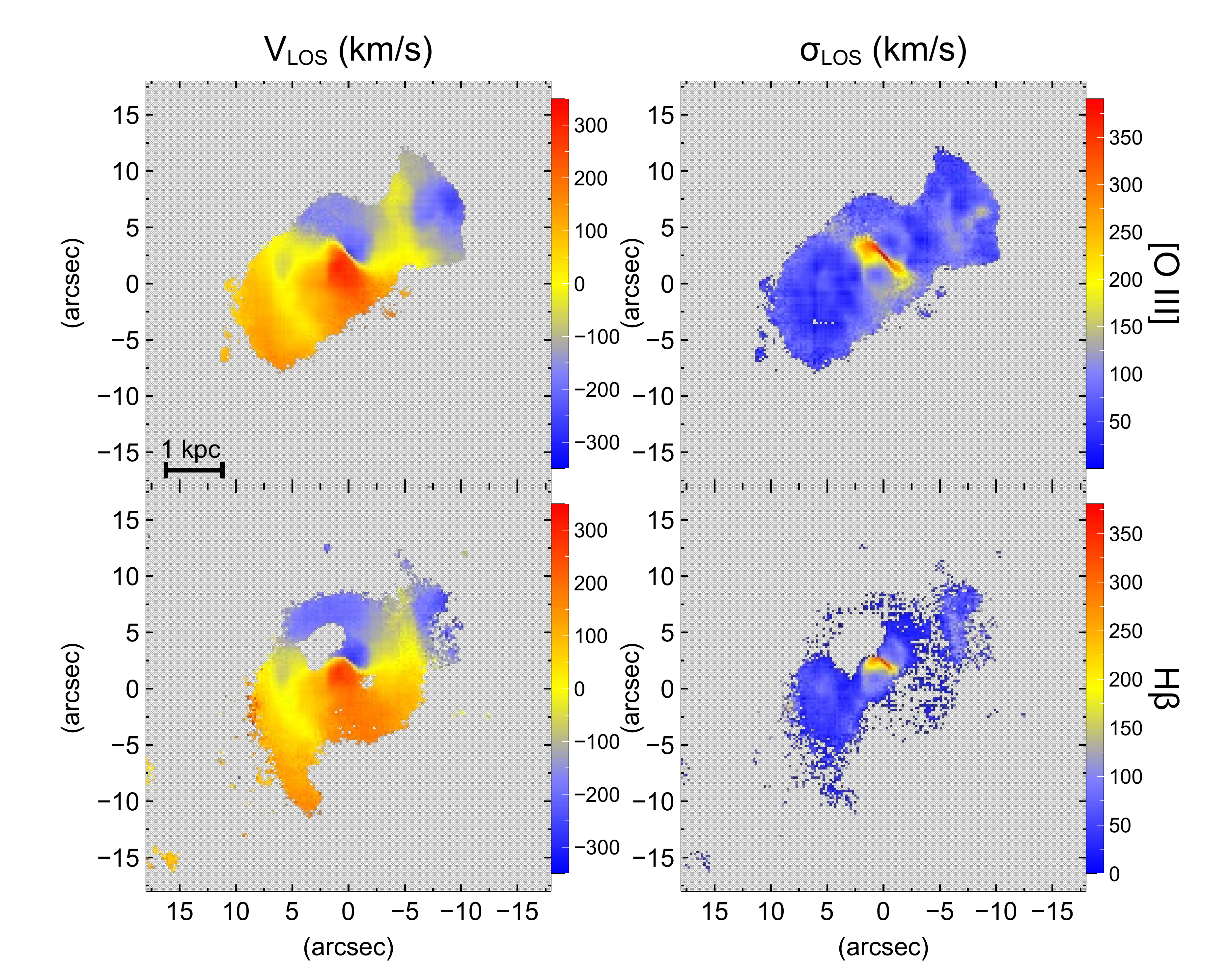}
	\caption{\OIII{} and \Hb{} kinematics from MUSE. Hydrogen recombination emission delineates both the outflow and the SF ring, whereas \OIII{} is only excited in the outflow.}
	\label{fig:ngc5728musekinematics3}
\end{figure*}

For comparison, the \OIII{} and \Hb{} flux and kinematic maps from the MUSE archival data (seeing-limited at $\sim0\arcsec.7$) are also presented here. These were produced in the same manner as the SINFONI kinematic maps. Figs. \ref{fig:ngc5728musekinematics3} shows the inner $9 \times 9$ kpc. The \OIII{} delineates the bicone, whereas the \Hb{} also includes the SF ring; the distortion in the LOS velocity field for \Hb{} from the combination of the SF ring and the outflows is clearly seen. Kinematic values are the same as the SINFONI observations; however the high values of the LOS dispersion aligned NE/SW across the nucleus are due to the overlapped outflow edges measured with limited (0\arcsec.2) plate scale resolution.

\subsubsection{Molecular Hydrogen Kinematics}
\label{sec:ngc5728h2kinematics}
Molecular hydrogen (\Htwo) is the direct source material for star formation; its location and kinematics help diagnose SF processes and AGN/SF linkages.  Most \Htwo{} structures in Seyfert galaxies are in rotating disks \citep{Diniz2015}. As we illustrated in Fig. \ref{fig:ngc5728gaskinematics4}, the \Htwo{} velocity structure is aligned with a PA of $\sim350\degr$ while the bicone has a PA of $\sim315\degr$; the ordered \textit{PV} diagram also suggests simple rotation (see Section \ref{sec:ngc5728pvdiagram}). Interestingly, this orientation is not in the plane either of the whole galaxy, or the inner structure (either disk or bar), unlike other examples \citep[see e.g. NGC~4151,][]{Storchi-Bergmann2010}. 

The \Htwo{} LOS velocities and dispersions are kinematically colder than for \Fe, \brg{} and \SiVI{} (Fig. \ref{fig:ngc5728gaskinematics5}), reaching a maximum velocity of $\sim\pm$250 \kms{} vs. $\sim\pm$300--400 \kms{} for the ionic species, and a median (maximum) dispersion of $\sim$95 \kms{} (200 \kms) vs. $\sim$130 \kms{} (350 \kms) for \brg. These trace different astrophysics; a cold molecular gas ring with kinematically associated star-formation vs. AGN-driven outflows.

\begin{figure*}[!htpb]
\centering
	\includegraphics[width=1\linewidth]{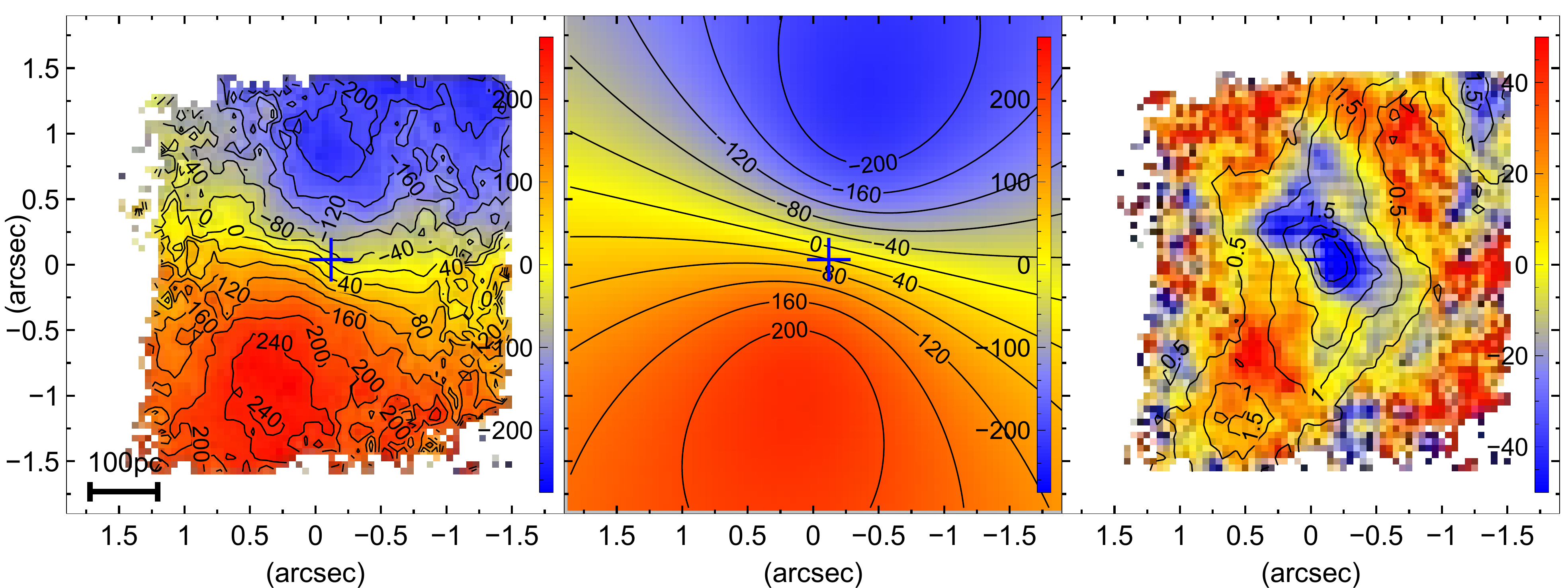}
	\caption{Plummer model for \Htwo{} velocity field with unconstrained parameters. Left panel: \Htwo{} velocity field. Middle panel: Best fit Plummer model. Right panel: Residual velocity with \Htwo{} equivalent width contours overplotted (in nm). All velocities shown in \kms.}
	\label{fig:ngc5728h2plummermodel1}
	\includegraphics[width=.7\linewidth]{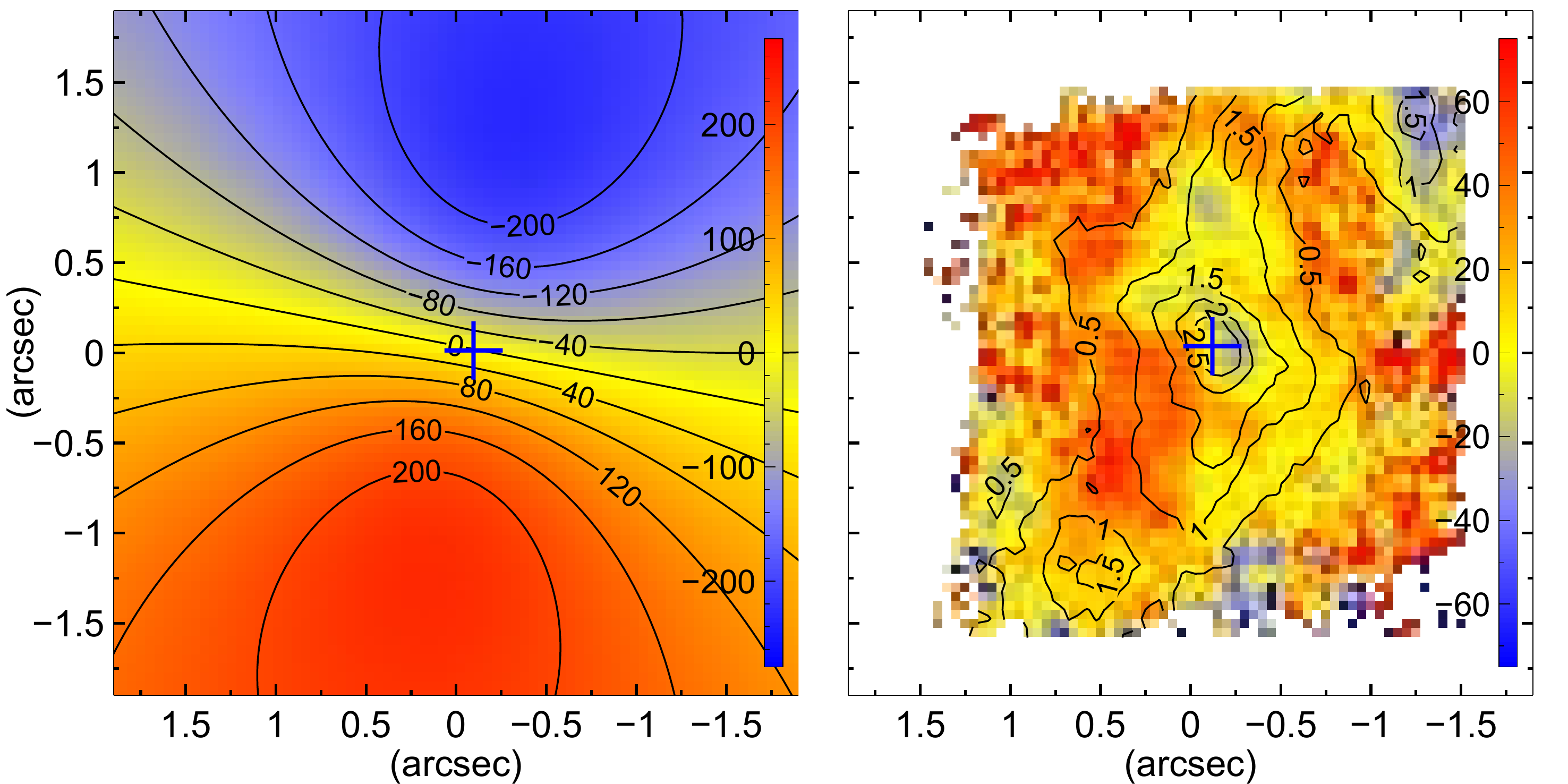}
	\caption{Plummer model for \Htwo{} velocity field with kinematic center and line of nodes constrained. Left  panel: Best fit Plummer model. Right panel: Residual velocity with \Htwo{} equivalent width contours overplotted (in nm). All velocities shown in \kms. These show ordered rotation broadly aligned with the stellar kinematics, plus the component of entrainment along the bicone edges.}
	\label{fig:ngc5728h2plummermodel2}
\end{figure*}

We fitted a Plummer potential model to the \Htwo{} velocity field, using the method described in Section \ref{sec:ngc5728StellarKinematics}. Initially, the five parameters (kinematic center [$X_0, Y_0$], line of nodes, inclination and scale length) were left reasonably unconstrained. The fitted alignment was 347\degr{}, with the kinematic center located 0\arcsec.09 from the AGN position (equivalent to 17 pc projected distance). Fig. \ref{fig:ngc5728h2plummermodel1} shows the result. The model rotation axis is aligned at PA = 77--257\degr{} (polar alignment 13--193\degr), with a very low inclination, i.e. almost edge-on. The scale length is approximately 0\arcsec.9, which corresponds to 180 pc. 

We also fitted the model where we constrained the kinematic center to the position of the AGN (assuming that the SMBH is the center of rotation). The polar alignment was also constrained to align with the main NS ridge of the equivalent width plot (the line of nodes constrained to between 260--265\degr). The results are shown in Fig. \ref{fig:ngc5728h2plummermodel2}. The scale length and inclination are very similar to the unconstrained fit. Both residuals plots show a region to the east to SE of the nucleus of the gas disk which shows negative residual velocities; these are hypothesized to be gas entrained on the edges of the bicone. This is seen in the pronounced clockwise `twist' of both the flux and velocity contours from the center. This is supported by the CO(1-0) radio emission morphology (which traces the extended molecular hydrogen), as reported by \cite{Combes2002} from observations using the IRAM-30 m telescope.

\subsubsection{Channel Maps}
\label{sec:ngc5728channelmaps}
\begin{figure}[!htbp]
	\centering
	\includegraphics[width=1\linewidth]{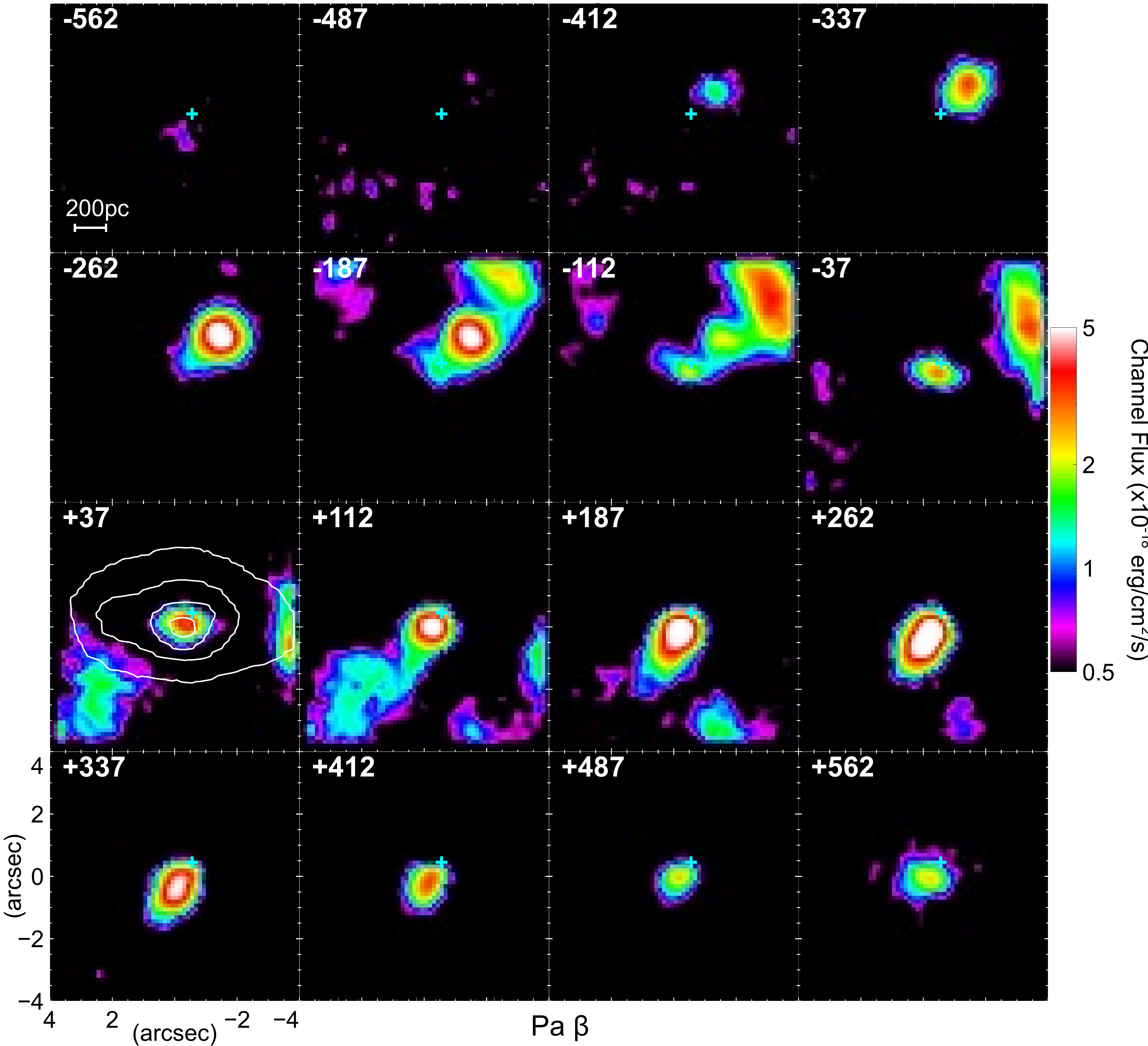}
	\caption{Channel map for \pab. Each channel has a width of 75 \kms, with the channel central velocity shown in the top left corner of each plot. The cyan cross is the position of the AGN. Color values are fluxes in units of 10\pwr{-18} \ecs, plotted with log scaling; all channels have the same maximum and minimum values. Positive velocities are receding, negative are approaching. The white contour on the +37 \kms{} channel is the \textit{J}-band continuum flux, at levels of 30, 50, 70 and 90\% of maximum. Note the 200 pc scale on the -562 \kms{} channel. The bicone structure is prominent in the NW (approaching) and SE (receding), plus the SF ring component at the peripheries. The outflows, aligned NW-SE are visible, as well as the SF ring emission from -187 to +187 \kms.}
	\label{fig:ngc5728channelmapsj2}
	\centering
	\includegraphics[width=1\linewidth]{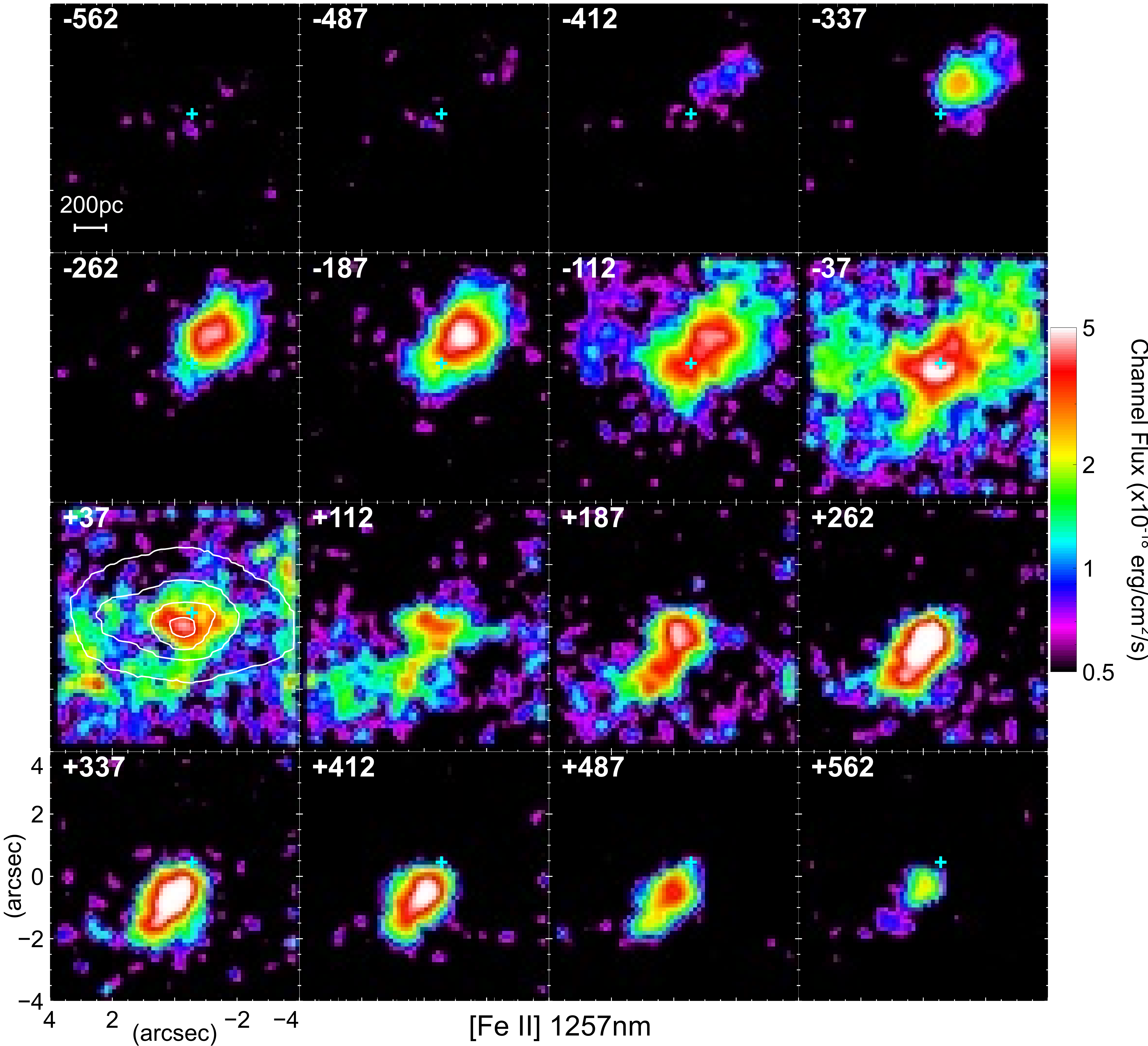}
	\caption{As for Fig. \ref{fig:ngc5728channelmapsj2}, for \Fe{} 1257 nm. \\}
	\label{fig:ngc5728channelmapsj1}
\end{figure}

\begin{figure}[!htbp]
	\centering
	\includegraphics[width=1\linewidth]{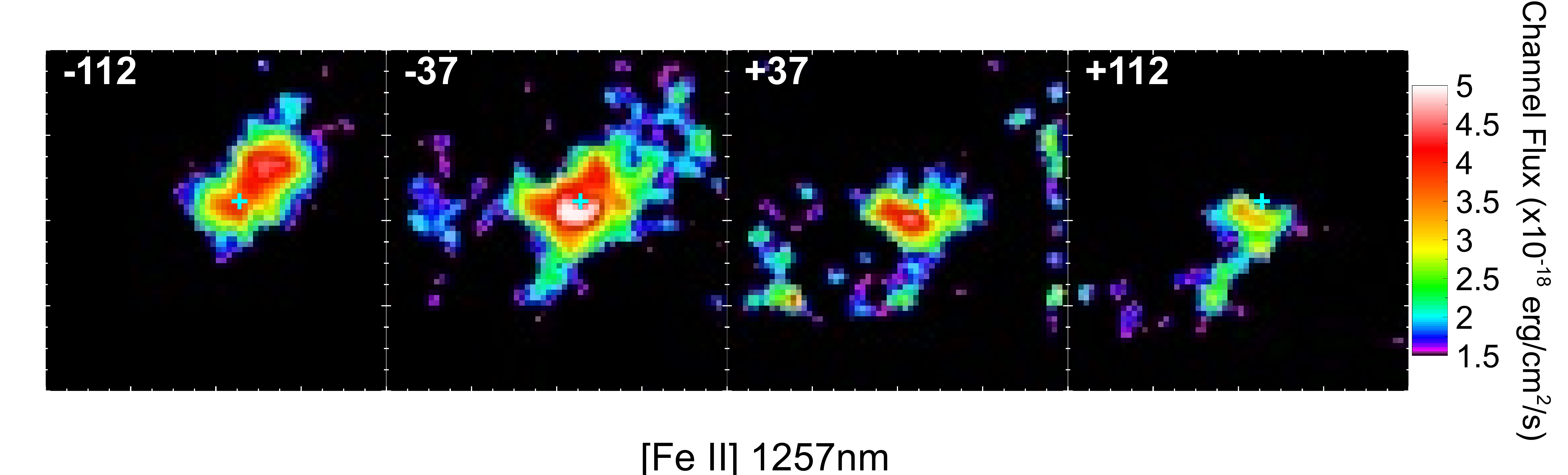}
	\caption{\Fe{} channel maps for low velocity with reduced scale range to enhance structure visibility. The `X'-shaped bicone boundaries are visible in the -37 \kms{} channel.}
	\label{fig:ngc5728channelmapsj4}
\end{figure}

\begin{figure}
	\includegraphics[width=1\linewidth]{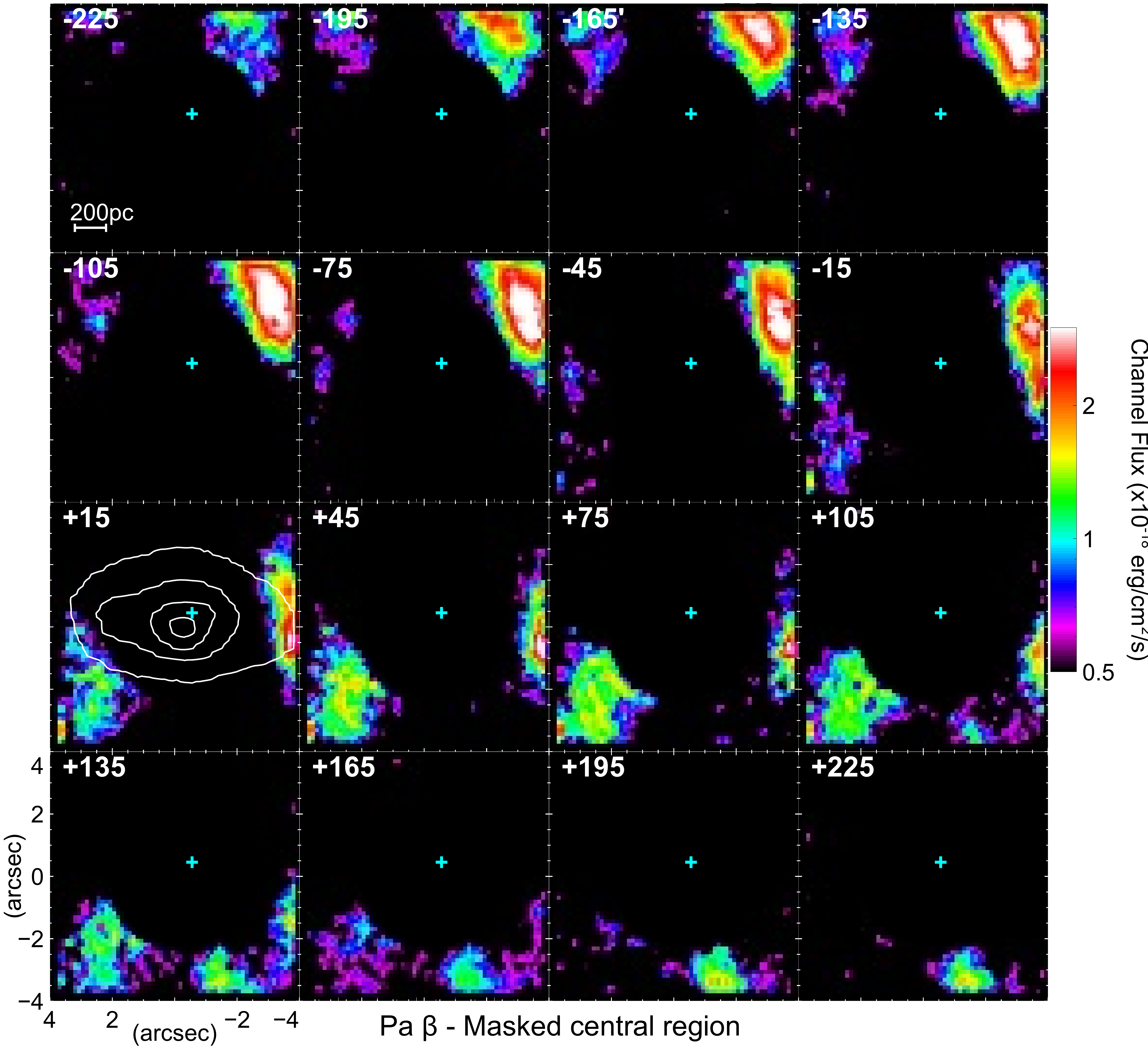}
	\caption{As for Fig. \ref{fig:ngc5728channelmapsj2}, for \pab, with the central 2\arcsec.5 around the AGN masked out. This enhances the SF ring features. The channel map velocity intervals are now 30 \kms.}
	\label{fig:ngc5728channelmapsj3}
	\centering
	\includegraphics[width=1\linewidth]{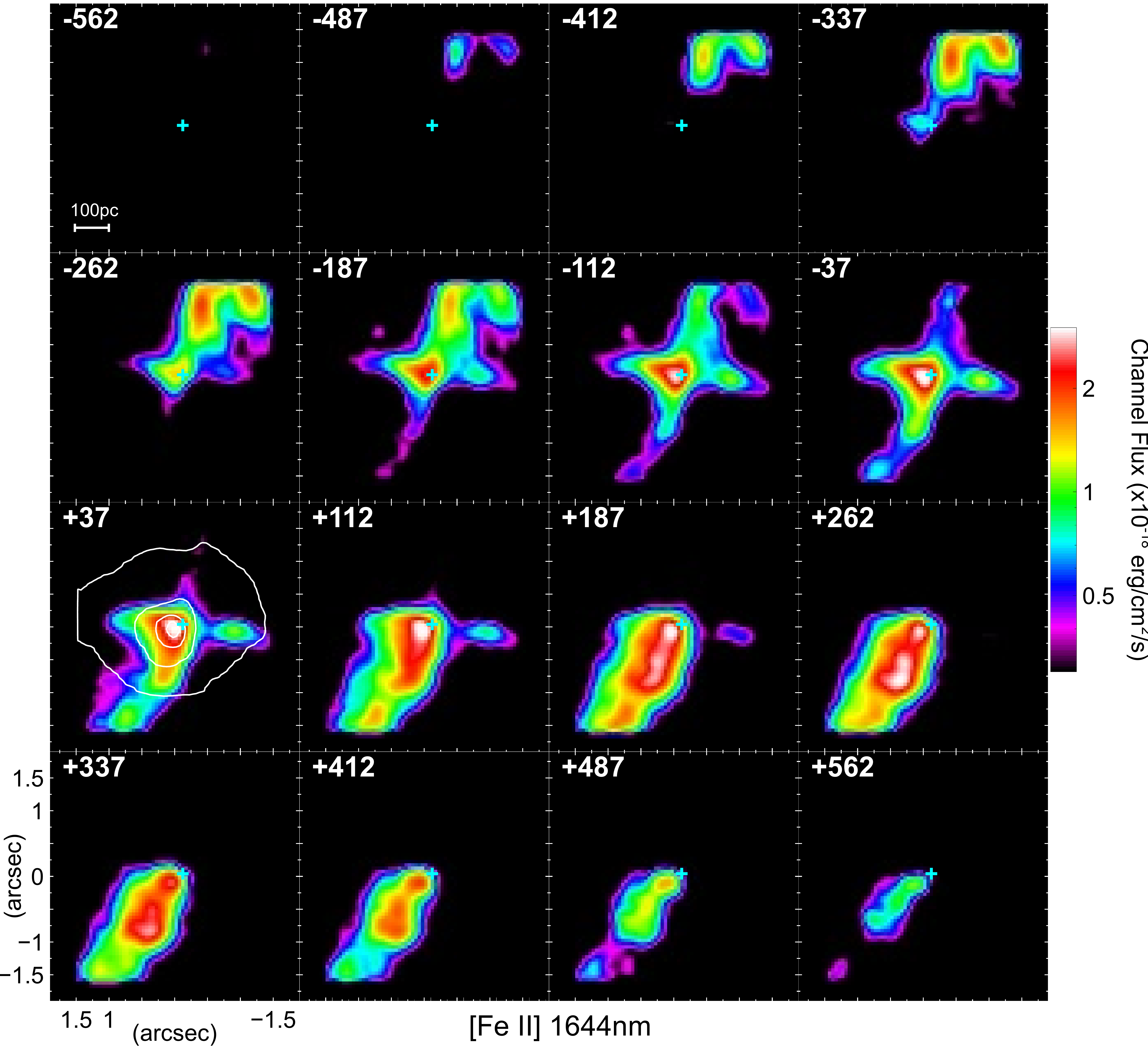}
	\caption{Channel map for \Fe{} 1644 nm. The white contour on the +37 \kms{} channel is the \textit{H}-band continuum flux, at levels of 30, 50, 70 and 90\% of maximum. Note the 100 pc scale on the -562 \kms{} channel. The cross-shaped feature outlining the outflows is especially prominent in the -262 to +37 \kms{} channels. The \Fe{} emission is from partially ionized and shocked gas, which are the boundaries of the outflows where the accretion disk radiation is weaker and the shear turbulence between the bicone and the ISM is greatest.}
	\label{fig:ngc5728channelmapshk1}
\end{figure}

\begin{figure}
	\centering
	\includegraphics[width=1\linewidth]{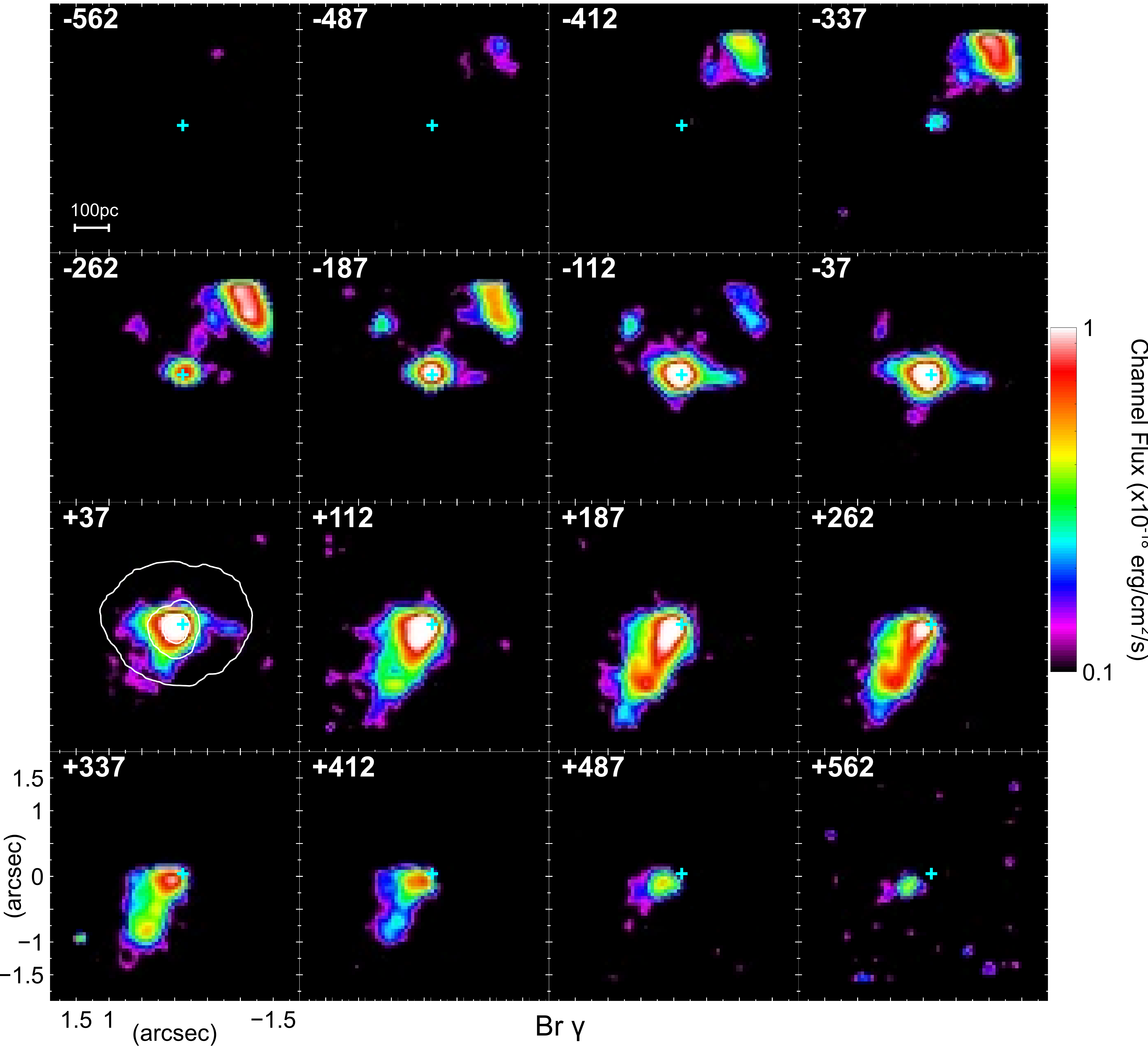}
	\caption{As for Fig. \ref{fig:ngc5728channelmapshk1}, for \brg. The white contour on the +37 \kms{} channel is now the \textit{K}-band continuum flux, at levels of 30, 50, 70 and 90\% of maximum. This is confined to the bicone interior.}
	\label{fig:ngc5728channelmapshk2}
	\centering
	\includegraphics[width=1\linewidth]{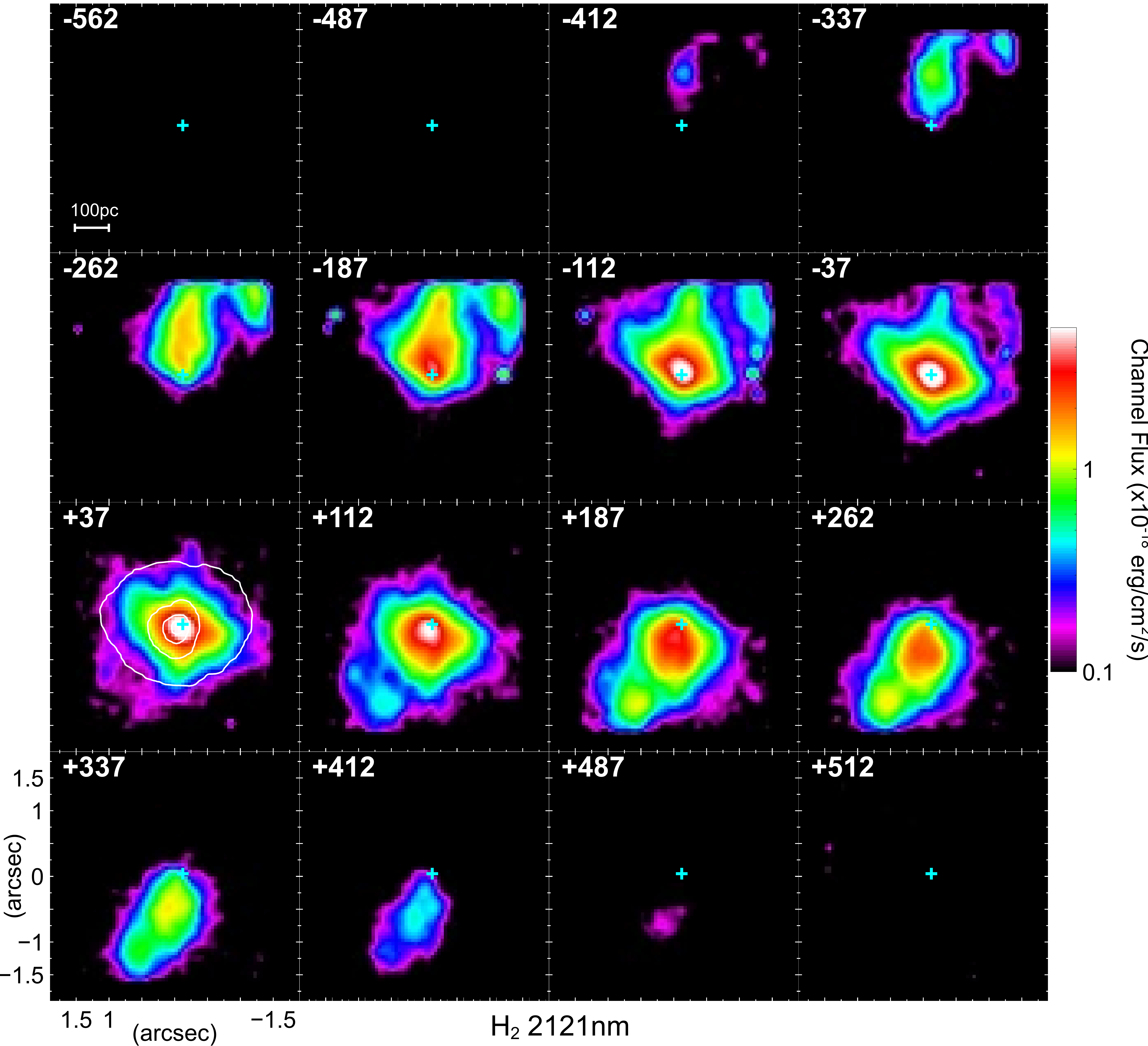}
	\caption{As for Fig. \ref{fig:ngc5728channelmapshk2}, for \Htwo{} 2121 nm. The kinematics are oriented more NS, especially in the negative velocity channels, with some entrainment effect by the outflows.}
	\label{fig:ngc5728channelmapshk3}
\end{figure}

\begin{figure}[!htbp]
	\centering
	\includegraphics[width=1\linewidth]{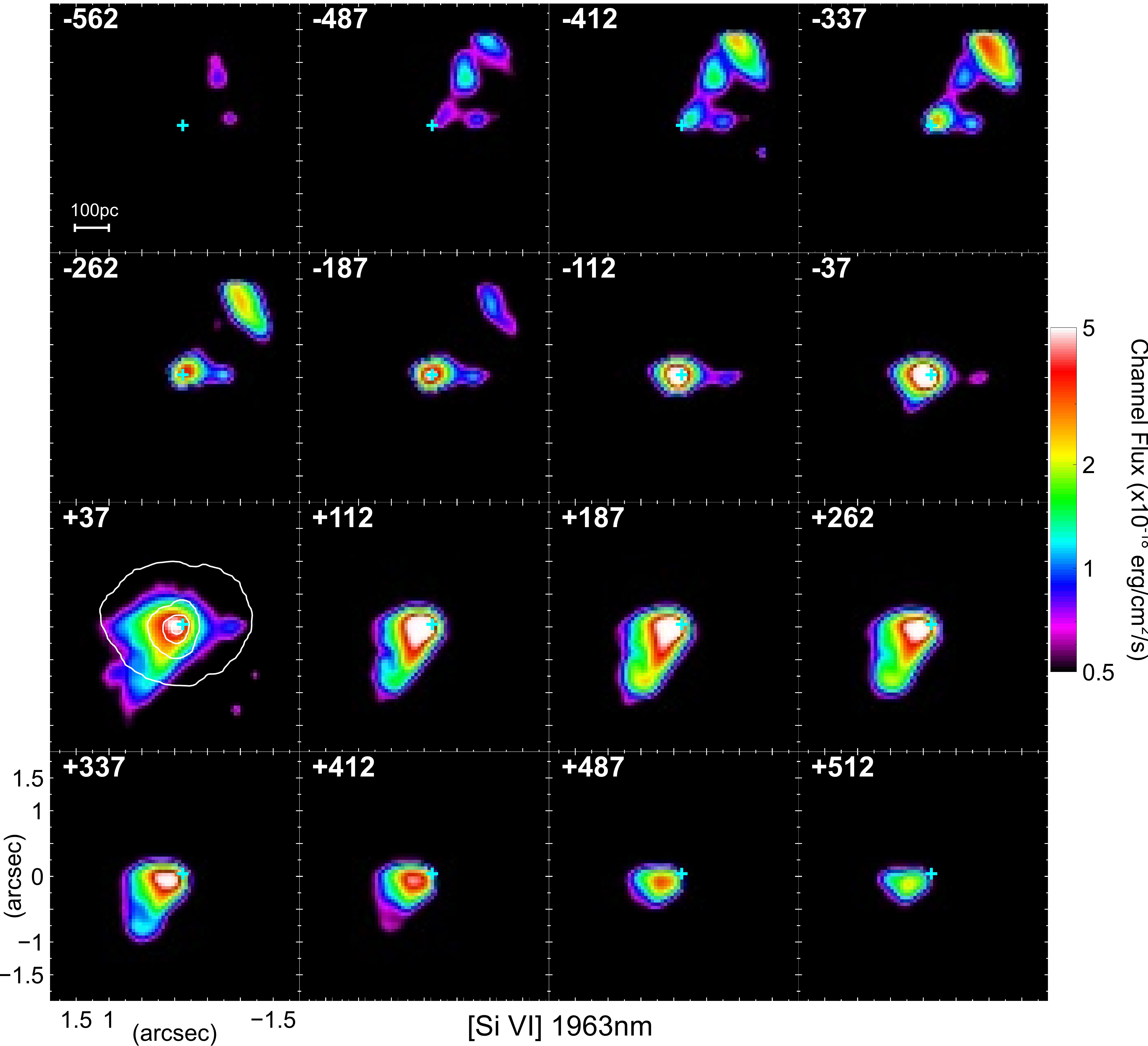}
	\caption{As for Fig. \ref{fig:ngc5728channelmapshk2}, for \SiVI{} 1964 nm, strictly confined to the bicone interior closer to the AGN than for \brg.}
	\label{fig:ngc5728channelmapshk4}
\end{figure}

We further analyzed the gas kinematics using channel maps and position-velocity diagrams. These are similar, and are both used to reveal kinematic structures that are not visible in the LOS velocity and dispersion plots. The former slice the velocity axis across the LOS, while the latter slice it parallel to the LOS. Figs. \ref{fig:ngc5728channelmapsj2} to \ref{fig:ngc5728channelmapshk4} show the channel maps for the spectral lines \pab, \Fe{} 1257 nm (both from the \textit{J} data cube), \Fe{} 1644 nm,  \brg, \Htwo{} 2121 nm and \SiVI{} 1964 nm (from the \textit{H+K} data cube). The maps were constructed by subtracting the continuum height, derived from the velocity map function \textit{velmap}, from the data cube. The spectral pixels are velocity binned and smoothed to reduce noise.

The \textit{J}-band channel maps are at a larger scale, encompassing more of the star-forming ring, as well as the inner bicone, whereas the \textit{H+K}-band maps show greater detail in the bicone. To separate the bicone and SF ring kinematics, we masked out the central 2\arcsec.5 around the AGN location where the bulk of the bicone emission is located (Fig. \ref{fig:ngc5728channelmapsj3} - the channels are now 30 \kms). There is still some impact from the cones (e.g. at the +45 and -135 \kms{} channels), but the maximum recession velocity is now in the SW quadrant, moving to the NE over the 450 \kms{} range, rather than the SE--NW direction of the bicone. This velocity range and orientation are compatible with stellar kinematics derived from the \SiI{} absorption line (see Section \ref{sec:ngc5728StellarKinematics} for a discussion of the anomalous kinematics). The star-forming ring has a greater velocity than the cones, because its axis of rotation is perpendicular to our LOS, therefore we measure the full rotation velocity. By contrast, the cones are inclined to our LOS (see Section \ref{sec:NGC5728OutflowKinematics}) and we only measure the component velocity. 
	
	%The \textit{J}-band \Fe{} channel maps (Fig. \ref{fig:ngc5728channelmapsj1}) are affected by the telluric absorption feature that could not be fully corrected; this shows as a reduced flux in the +112 \kms{} channel. 
The low velocity channels (-112 -- +112 \kms) show the emission across most of the field; this is due to the common minimum value for the scaling for all channels, to enhance the high-velocity channels. These channels have been replotted in Fig \ref{fig:ngc5728channelmapsj4} to suppress this noisy background; the -37 \kms{} show the `X'-shaped outline of the bicones. %The +37 \kms{} channel has low flux levels, due to the imperfectly corrected telluric feature.  
	

\begin{figure}[!htbp]
\centering
\includegraphics[width=1\linewidth]{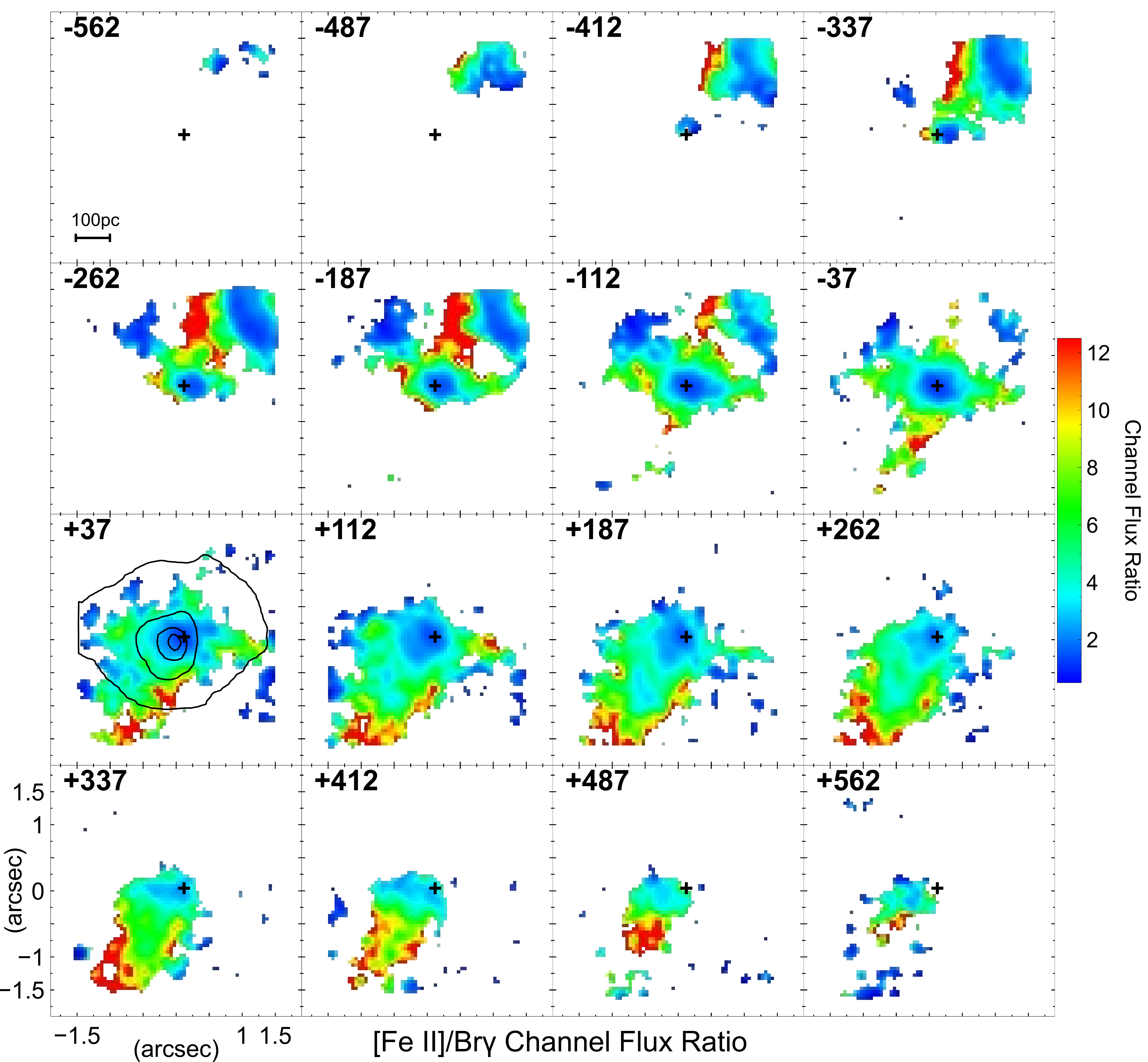}
\caption{Flux ratio for \Fe{} 1644 nm/\brg{} for each velocity channel. The color-bar shows the flux ratio, with  enhanced values (green and blue) at the peripheries of the cones in the higher velocities, showing enhanced \Fe{} emission in partially ionized and shocked gas.}
\label{fig:ngc5728channelmapsratiohk1}
\centering
\end{figure}

\begin{figure}[!htbp]
\centering
	\includegraphics[width=1\linewidth]{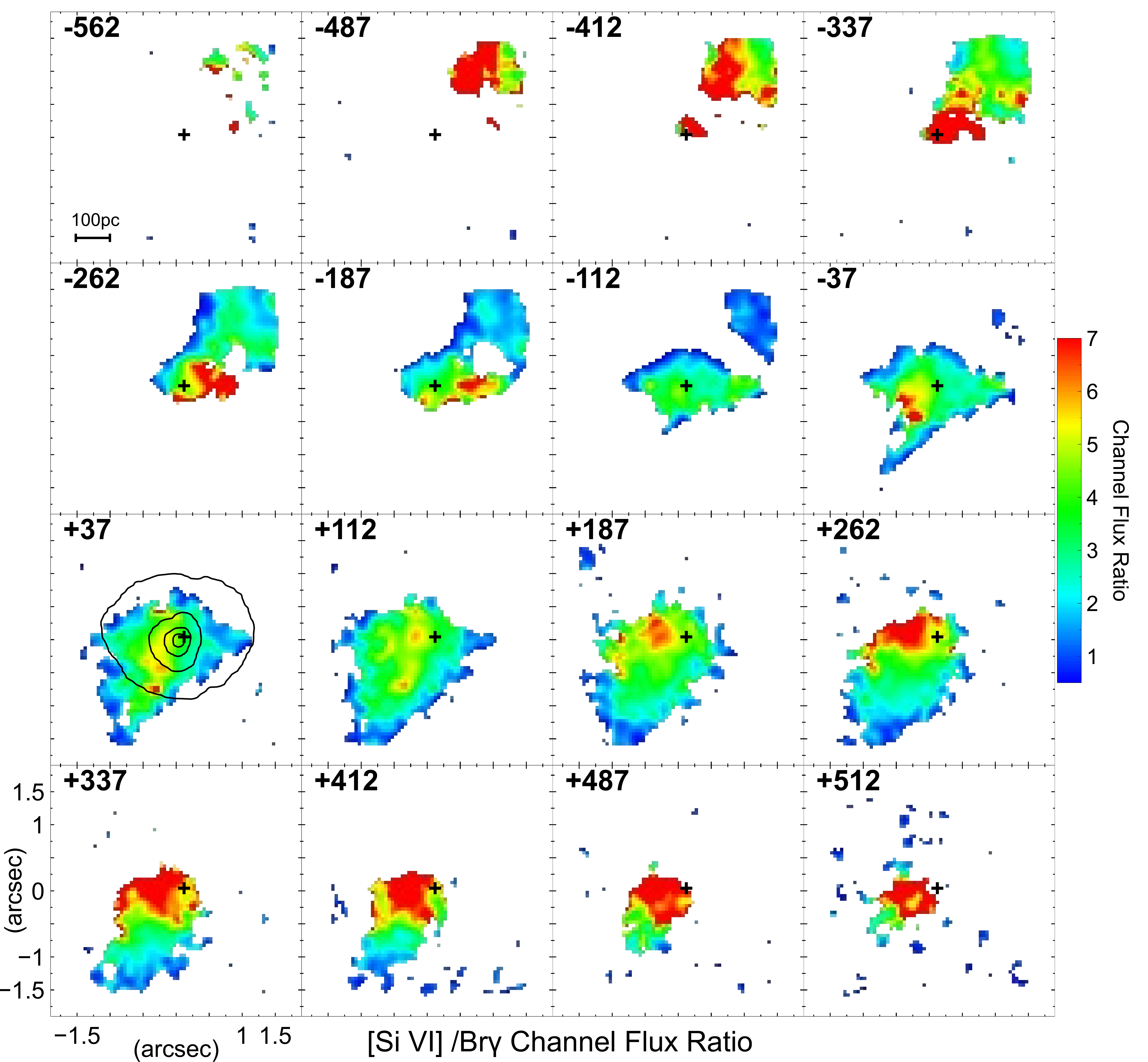}
	\caption{Flux ratio for \SiVI/\brg{} for each velocity channel, showing the more central concentration for the \SiVI{} from direct photo-ionization from the AGN.}
	\label{fig:ngc5728channelmapsratiohk2}
	%\end{figure}
%\begin{figure}[!htbp]
\centering
\includegraphics[width=1\linewidth]{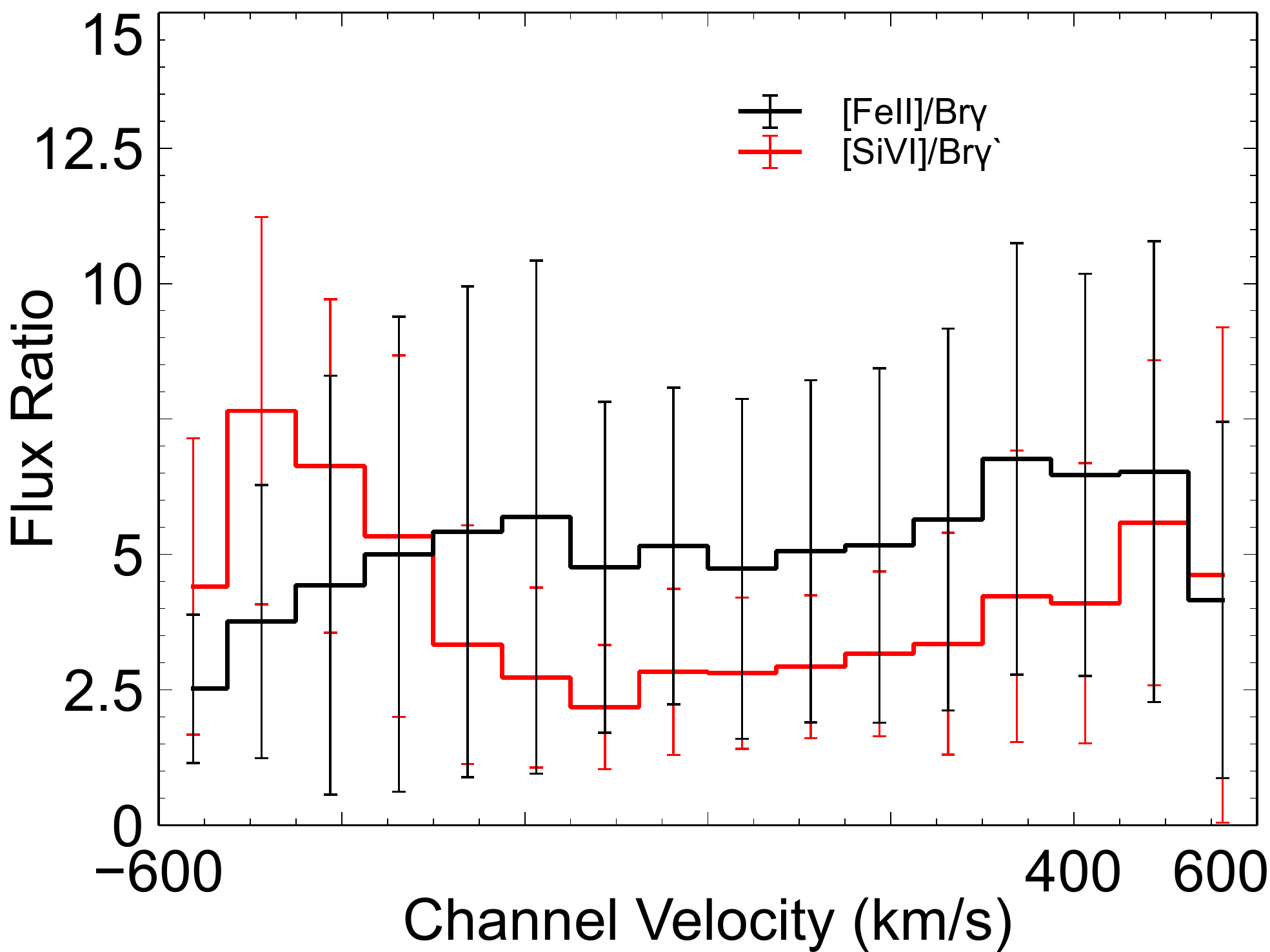}
\caption{Average flux ratios for each velocity channel, with standard deviation as the uncertainty, showing no significant differences across velocities. This contradicts the model of \cite{Murayama1998} of a clumpy matter-bound CLR.}
\label{fig:ngc5728channelmapsratiohk3}
\end{figure}

Assuming that the \Fe{} and \SiVI{} gas is co-moving with the \brg{}, we can plot the channel flux ratio for those species. These are presented in Figs. \ref{fig:ngc5728channelmapsratiohk1} and \ref{fig:ngc5728channelmapsratiohk2}. The \Fe/\brg{} channel ratio clearly shows that in the SE cone, the \Fe{} `outlines' the \brg{} emission, i.e. the flux ratios are higher to the SE, whereas the \SiVI/\brg{} channel ratios are higher towards the nucleus. A check whether there is a difference between the outflows is shown in Fig. \ref{fig:ngc5728channelmapsratiohk3}, which plots the average flux ratio for each velocity channel (with the standard deviation as uncertainties). There is a trend in the flux ratios towards the receding (positive) velocities for \Fe/\brg, and an opposite trend for \SiVI/\brg, but with large uncertainties.

\subsubsection{Position-Velocity Diagrams}
\label{sec:ngc5728pvdiagram}
\begin{figure}[!htbp]
	\centering
	\includegraphics[width=1\linewidth]{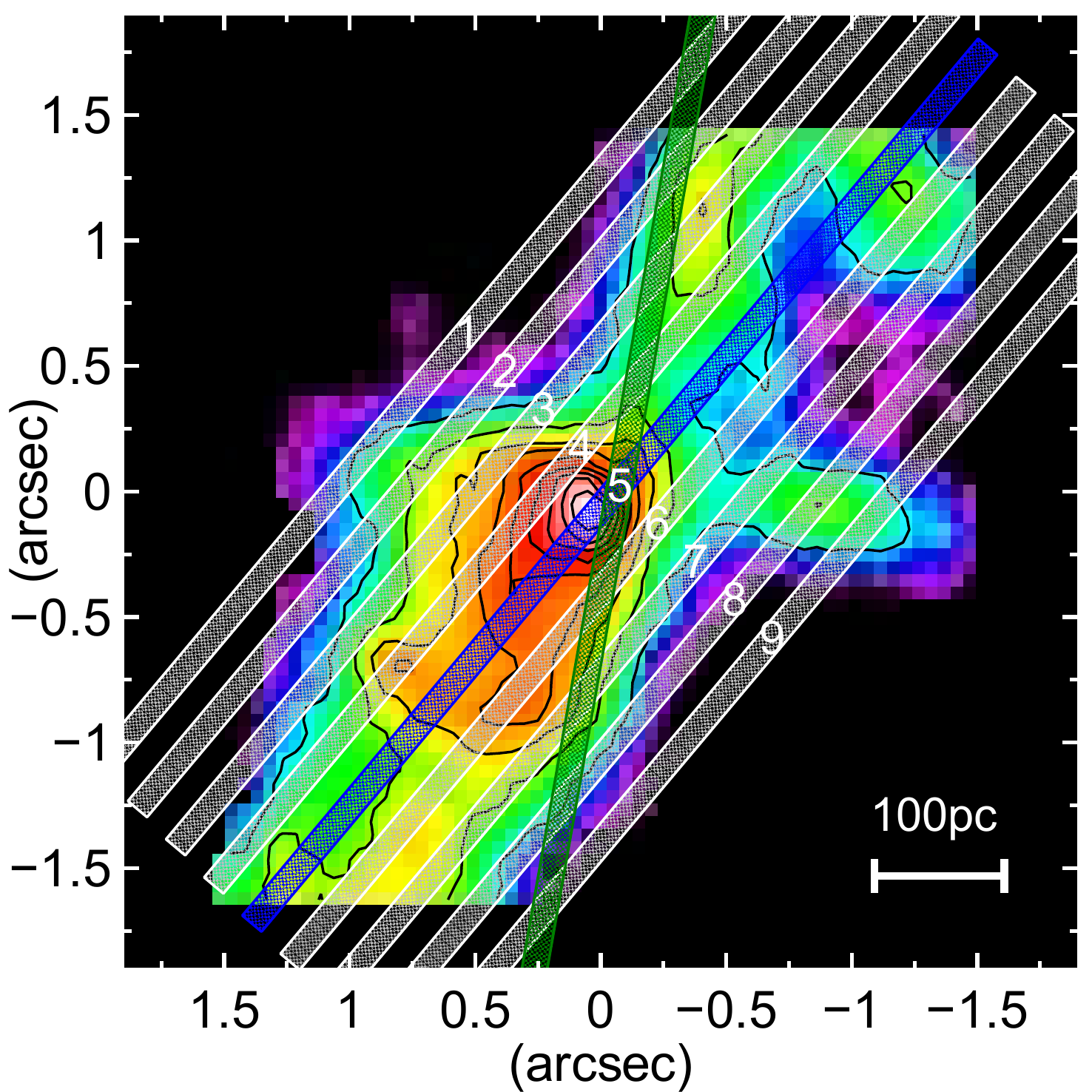}
	\caption{Longslit alignment, superimposed on \Fe{} flux map; the blue box is the slit for \brg, \SiVI{} and \Fe{} 1644 nm, the green for \Htwo. The white boxes are the off-nucleus pseudo-slits through the cones, with the numbers corresponding to the plots in Figs. \ref{fig:ngc5728pvdiagram2}, \ref{fig:ngc5728pvdiagram3} and \ref{fig:ngc5728pvdiagram4}.}
	\label{fig:ngc5728pvlayout}
\end{figure}

Similarly to the channel maps, the position-velocity (\textit{PV}) diagram can help to identify features in the cones, like velocity and dispersion anomalies, that are not apparent in the standard kinematic maps. The \textit{PV} diagram is generated by a `pseudo-longslit' (the \textit{longslit} function in \texttt{QFitsView}) along the cone  or the plane of rotation of the molecular gas. For \brg, \SiVI{} and \Fe{}, the longslit was laid along a PA = 140--320\degr{} (NW--SE) centered on the AGN and aligned with the velocity extrema; for \Htwo{} the line was along a PA = 170--350\degr{}, along the flux ridge line (presumably the plane of rotation). All slits are 2 pixels wide. The alignment of the longslits is shown in Fig. \ref{fig:ngc5728pvlayout}; slit 5 (blue and green) the one aligned with the AGN.
	
\begin{figure}[!htbp]
	\centering
	\includegraphics[width=1\linewidth]{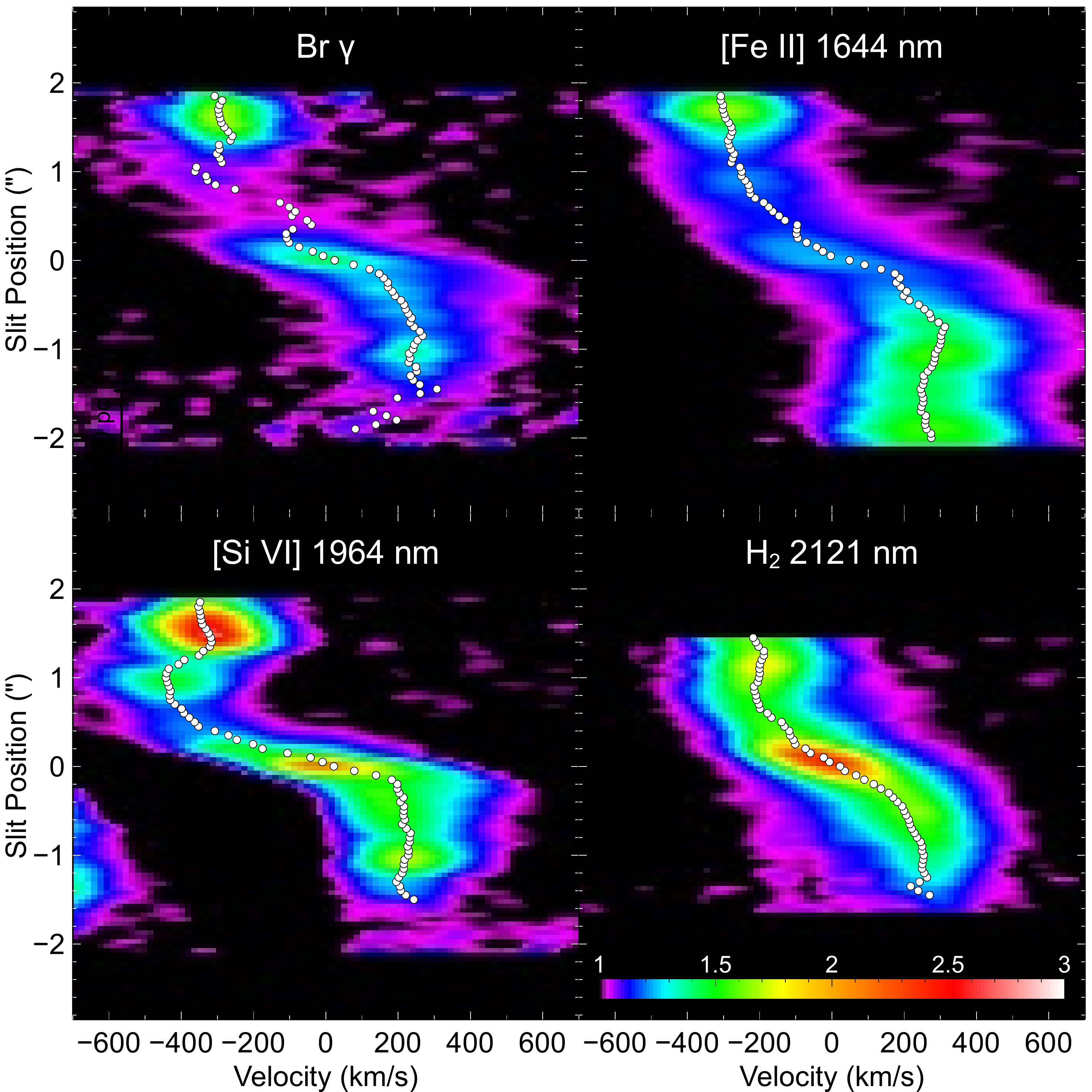}
	\caption{\textit{PV} diagrams for \brg, \Fe{}, \SiVI{} and \Htwo, as described in the text. The slit position values are negative in the SE/S direction to positive in the NW/N position. The flux maps have different intensity for contrast enhancement. The LOS velocity derived from the \textit{velmap} procedure is overlaid as white points. The \brg{} diagram shows significant discontinuities in the negative velocity outflow vs. the other species; this is also noted for the \OIII{} emission, as may be due to turbulent entrainment or flickering AGN activity}
	\label{fig:ngc5728pvdiagram1}
	%\end{figure}
	%\begin{figure}[!htbp]`
	\centering
	\includegraphics[width=1\linewidth]{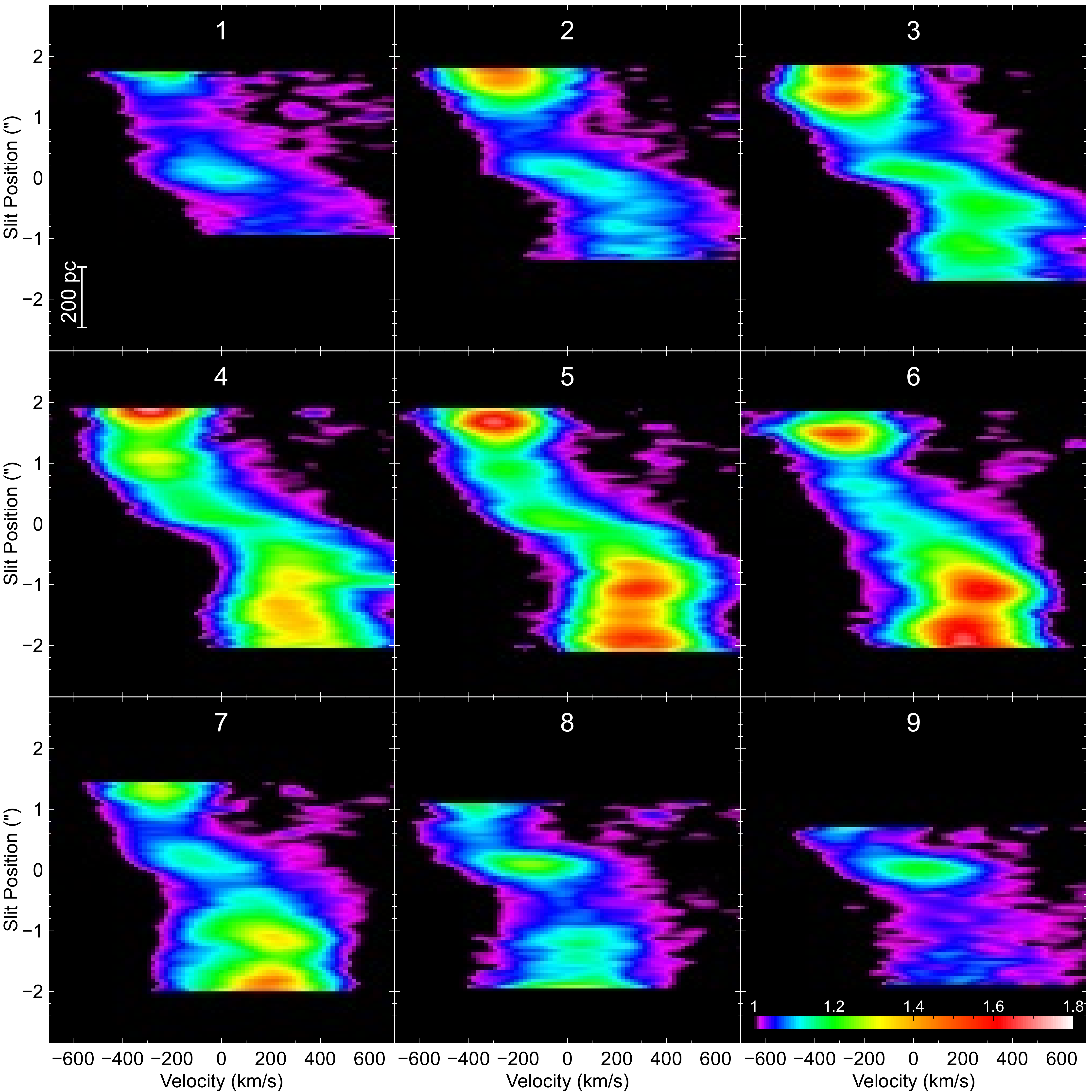}
	\caption{Off-nucleus \textit{PV} diagrams for \Fe{} through the outflow. The plots are numbered corresponding to the slits in Fig. \ref{fig:ngc5728pvdiagram2}. The flux maps all have the same intensity scale (color-bar in panel 9). }
	\label{fig:ngc5728pvdiagram2}
\end{figure}

Fig. \ref{fig:ngc5728pvdiagram1} shows the resulting plots for the central slit; the wavelength (on the X-axis) is converted to a velocity difference from the central pixel, and the Y-axis is the distance along the longslit in arcsec. To enhance the emission line features, the flux values have been divided by the median value along the slit position, otherwise the continuum variations would obscure the details; this is similar to the equivalent width of the emission line. The LOS velocity derived from the \textit{velmap} procedure is overlaid on the relative flux; these delineate the centerline of the PV diagram at each position.

The \textit{PV} diagrams show a velocity gradient at the nucleus of $\sim$330 (\brg), 475 (\SiVI) and 440 (\Fe) \kms{} per 100 pc, with \Htwo{} having a lower gradient of 175 \kms{} per 100 pc. The \Htwo{} velocity then levels off to a constant value of approximately $\pm$ 200 \kms, presumably as it co-rotates with the stars. In the positive (receding) velocity direction, the \brg, \Fe{} and \SiVI{} all plateau or slightly decline at roughly the same velocity. However, in the negative (approaching) velocity direction, the \brg{} does not continue smoothly, presumably because of low flux values introducing uncertainties; the \Fe{} velocity continues rising and the \SiVI{} shows a decline from a maximum of $\sim$430 \kms{} to $\sim$320 \kms. It is noted that all three species plateau and the flux brightens at the edge of the field at this velocity; it is hypothesized that the source of this emission is the radio jet interacting with the ISM co-rotating with the inner disk, moving towards us at this velocity, rather than the direct outflow.

\begin{figure}[!htbp]
	\centering
	\includegraphics[width=1\linewidth]{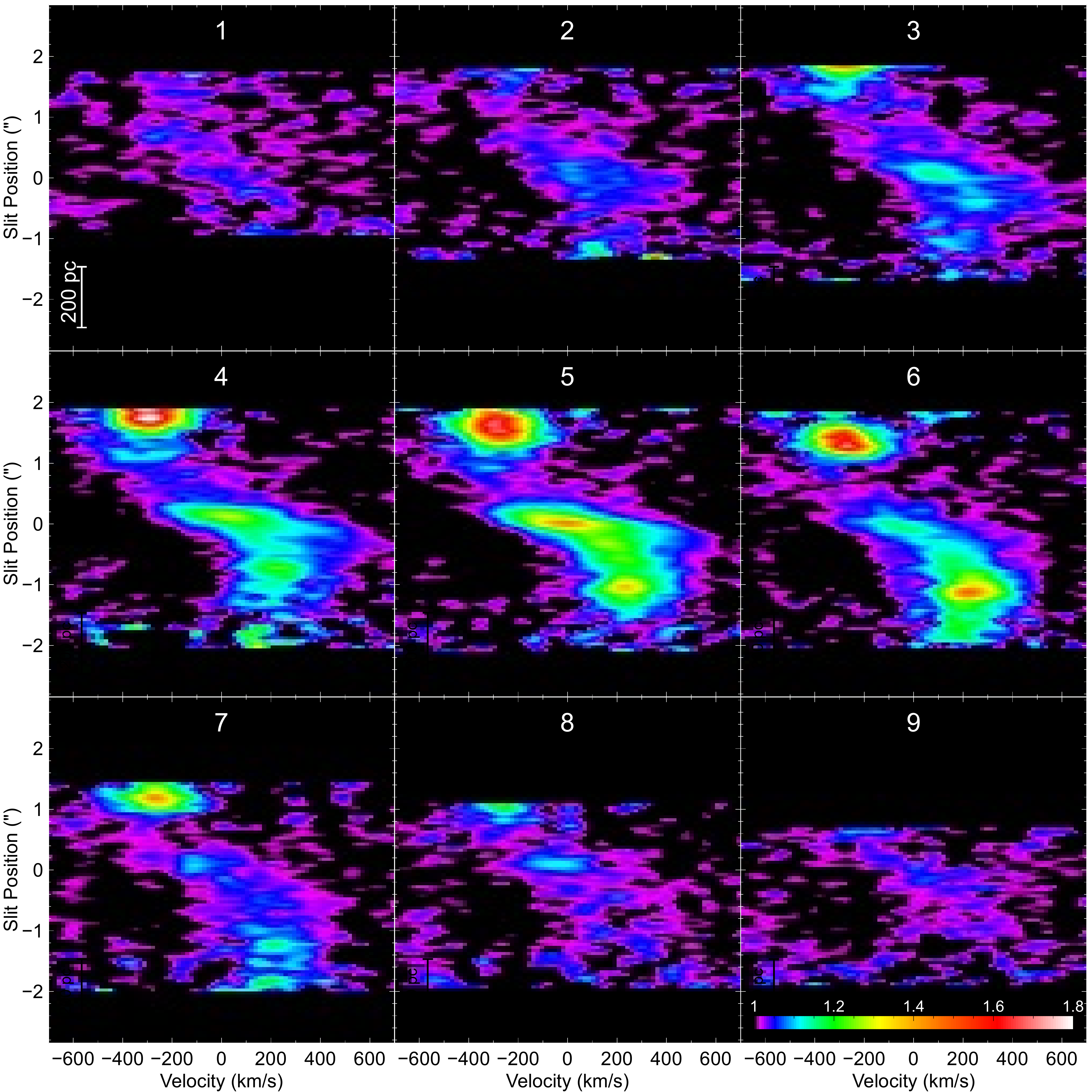}
	\caption{Off-nucleus \textit{PV} diagrams for \brg{} through the cones. The plots are numbered corresponding to the slits in Fig. \ref{fig:ngc5728pvdiagram2}. The flux maps all have the same intensity scale (color-bar in panel 9). There are significant discontinuities, especially in the negative velocity flows; the main `hot-spot' is the impact location on the star-forming ring.}
\label{fig:ngc5728pvdiagram3}
\end{figure}

\begin{figure}[!htbp]
\centering
\includegraphics[width=1\linewidth]{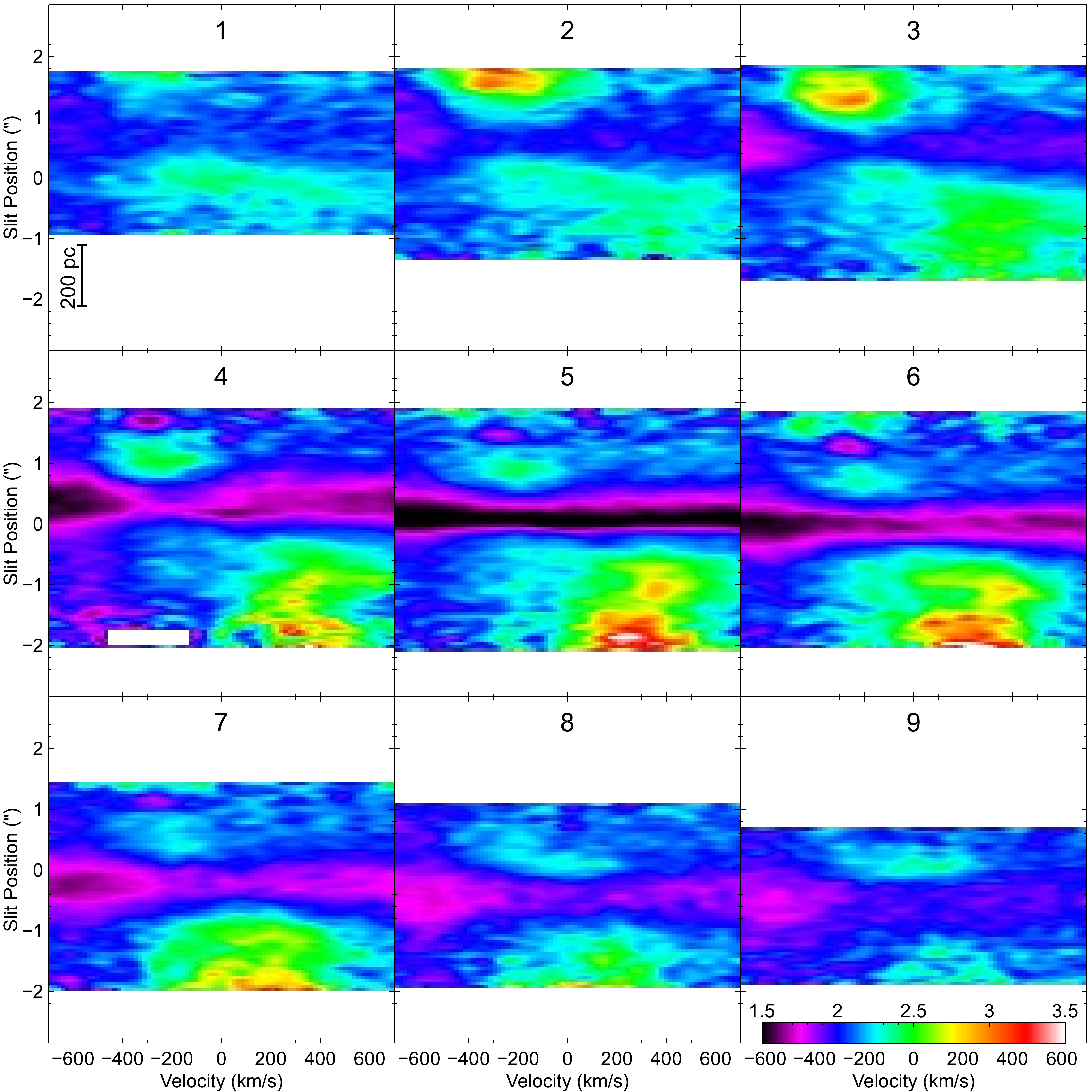}
\caption{Flux ratio \Fe/\brg{} \textit{PV} diagrams. The flux ratios all have the same range, given by the color-bar in panel 9. Higher values are where \Fe{} flux is relatively higher than \brg; this shows the reduced importance close to the nucleus and increased ratio further along the cones.}
\label{fig:ngc5728pvdiagram4}
\end{figure}

Figs. \ref{fig:ngc5728pvdiagram2} and  \ref{fig:ngc5728pvdiagram3} shows the \textit{PV} diagrams for \Fe{} and \brg{} at off-axis slits, aligned with the main axis at distances of 3 pixels in X and Y, i.e.  0\arcsec.2 = 40 pc apart. For both lines, panel 5 is identical to the corresponding diagrams in Fig. \ref{fig:ngc5728pvdiagram1}. For \Fe, the off-axis plots look similar to the on-axis diagram. The \brg{} diagrams show a consistent patch of high flux in panels 4, 5 and 6; this is radio jet-ISM impact location. Figs. \ref{fig:ngc5728pvdiagram4} shows the PV diagram of the flux ratios \Fe/\brg; in this case the values have \textit{not} been divided by the median. These diagrams show that \Fe{} has a relatively lower flux near the nucleus and becomes more important further along the outflows as the ionizing flux drops; e.g. panels 4, 5 and 6 have low ratios ($\sim1.5$) around slit position 0\arcsec{} and high values ($\sim3$) at -1\arcsec.75.
	
\section{Discussion}
\subsection{The Super-massive Black Hole}
\label{sec:ngc5728smbh}
The black hole mass can be derived using various scaling relationships from observed galaxy parameters. The best-known, and most well-established, uses the stellar velocity dispersion (the \MSig{} relationship). \cite{Graham2013} gives several scaling values, depending on the galaxy morphology. As it is not clear the exact morphology of the complicated nuclear structure, we will use the regression values for all galaxies in the data set, (with the dispersion $\sigma$ measured in \kms):
\begin{multline}
\log(M_{BH}) =(8.15\pm0.05) \\+ (6.07\pm0.31)~\log\left( \dfrac{\sigma}{200}\right) \pm 0.41 \label{eqn:odraSMBH1}
\end{multline}
We obtained the stellar dispersion for a central 1\arcsec{} aperture using several method, and compared them with literature values; these are given in Table \ref{tbl:ngc5728stellardisp}. As can be seen from Fig. \ref{fig:ngc5728stellarkinematics1}, the exact choice of aperture can also affect the measured dispersion.

\cite{Lin2018}, using the same SINFONI data set, derived the central stellar velocity dispersion of $164\pm4$ \kms{} from the CO band-heads. Our solution used the NIFS Gemini spectral library of near-infrared late-type stellar templates (Version 2) \citep{Winge2009}, which has 20 stars observed at 2.13\AA, re-binned to 1\AA. We used the penalized pixel-fitting (pPXF) method of \cite{Cappellari2004}; we found the dispersion to be the same as the \cite{Lin2018} result, within errors. 
	
We also derived the dispersion using the \CaII{} triplet templates from the Miles SSP models, using the Padova isochrones and the Kroupa Universal initial mass function, for a range of metallicities\footnote{\url{http://www.iac.es/proyecto/miles/}}; this is a suite of 350 template spectra. The pPXF method was again employed. As well, we compared this with the \textit{velmap} derived dispersion on the \CaII{} 8544.4\AA{} absorption line.

\begin{table*}[]
\begin{center}	
	\caption{Stellar dispersion derived by different methods.}
	\label{tbl:ngc5728stellardisp}
	\begin{tabular}{llccc}
		\toprule
		Spectral line      & Source             & Method & $\sigma$ & $\Delta\sigma$ \\ \midrule
		\CaII{} triplet    & This work          & T      & 156      & 21             \\
		CO band-heads      & This work          & T      & 154      & 26             \\
		\CaII{} 8544.4 \AA & This work          & V      & 179      & 10             \\
		\SiI{} 1588.8 nm   & This work          & V      & 221      & 29             \\
		                   &                    &        &          &                \\
		CO band-heads      & \cite{Lin2018}     & T      & 164      & 4              \\
		Stellar kinematics & \cite{Wagner1988}  & X      & 210      & 15             \\
		Stellar kinematics & \cite{McElroy1995} & X      & 209      & \dots{}        \\ \bottomrule
		\multicolumn{5}{l}{Method (Col. 3): T - template fitting using pPXF - see text for description of template sets.}\\
		 \multicolumn{5}{l}{V - simple Gaussian fit using the \textit{velmap} procedure from \texttt{QFitsView}.}\\
		 \multicolumn{5}{l}{X - cross-correlation technique of \cite{Tonry1979}.}\\
		 \multicolumn{5}{l}{Dispersion values for this work measured in 1\arcsec{} radius aperture.}\\
	\end{tabular}
\end{center}
\end{table*}
Fitting a single absorption line possibly overestimates the dispersion; the dispersion derived from the cross-correlation technique might also have aperture effects, especially considering the large velocity gradients over the central region. We will use the average of the three pPXF template fitting methods, which gives $158\pm34$~\kms{}. For comparison, we will also use the value for the \CaII{} single-line fit.

The exact choice of regression parameters used in the \MSig{} relationship for different galaxy morphologies, is of minor importance, as the errors in the dispersion measurement are dominant in the final calculation of the SMBH mass.

The \textit{K}-band bulge luminosity relationship (\MKs) is of the form:
\begin{multline}
log(M_{BH})=(9.05\pm0.12) \\+ (-0.44\pm0.08)~\log \left( M + 25\right) \pm0.44
\end{multline}
using the parameters from \cite{Graham2013}. The \MKs{} value was derived from the 2MASS $K_S$ image by fitting a 2D Gaussian to the central bulge and measuring the flux in counts, then converting to an absolute magnitude using the photometric zero-point and the distance modulus.

\begin{figure}[!htbp]
	\centering
	\includegraphics[width=1\linewidth]{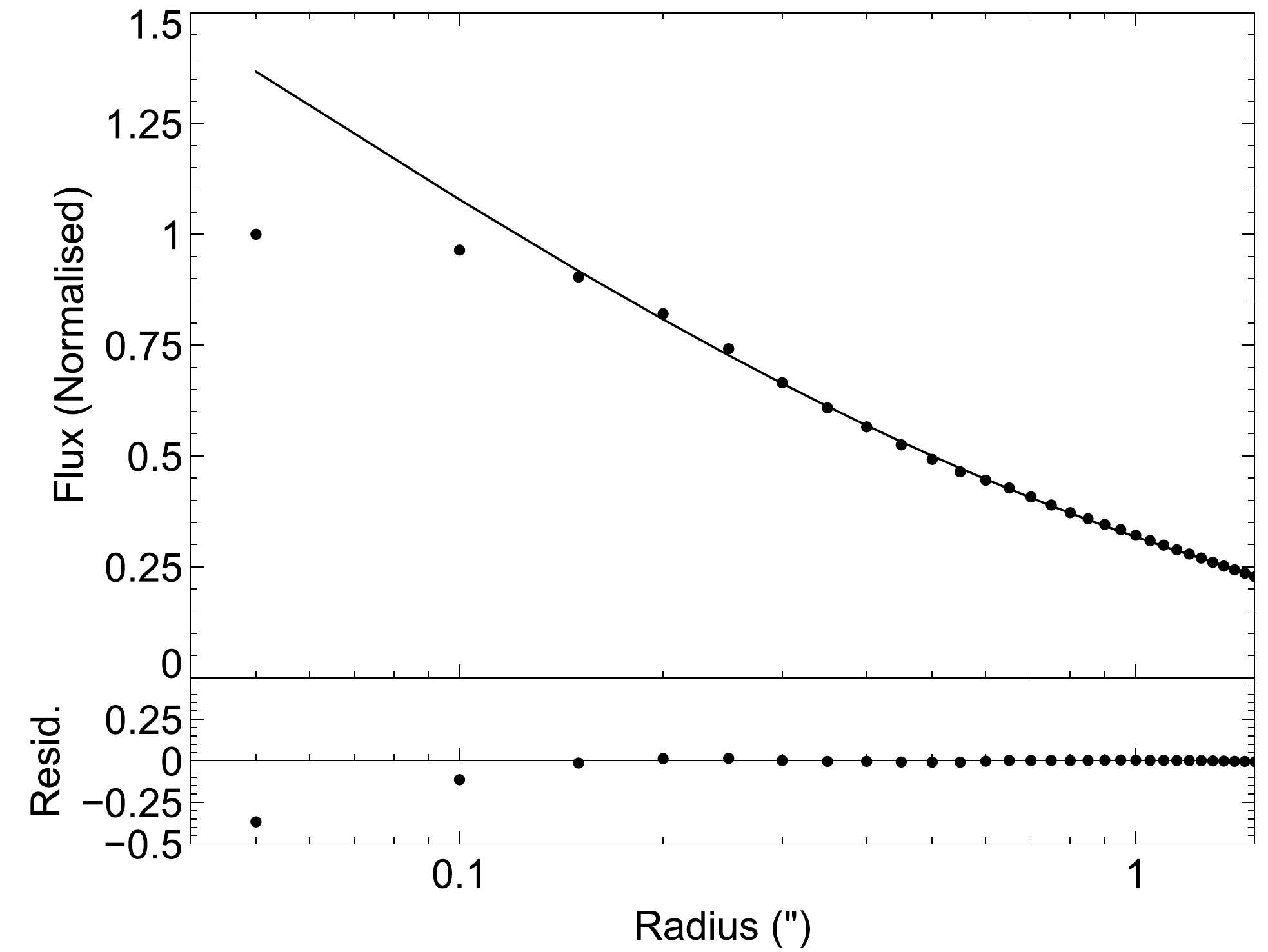}
	\caption{Top panel : radial profile data (points) with the S\'{e}rsic fit (index = 3.54). Bottom panel : residual (fit--data). Radius in logarithmic scale, flux is normalized. This implies a SMBH mass of $\sim3.4 \times 10^{8}~\mmsun$.}  
	\label{fig:ngc5728sersicfit}
\end{figure}

The S\'{e}rsic index ($n$) is fitted to the light profile, and has the scaling relationship \citep{Savorgnan2013}:
\begin{multline}
\log(M_{BH})=(8.37\pm0.3) \\+ (2.23\pm1.5)~\log \left(\dfrac{n}{3}\right) \pm 0.27
\end{multline}
The index was fitted to the \textit{K} continuum flux radial profile from the SINFONI data cube (see Fig. \ref{fig:ngc5728sersicfit}). The radial profile shows signs of a turn-over in the inner few pixels, indicative of a `core-S\'{e}rsic' profile  \cite[see][]{Graham2013,Savorgnan2013}; the S\'{e}rsic index fit excludes the inner two points; these points are also below the spatial resolution. We also compared this with the core-S\'{e}rsic fit for the 2MASS $K_S$ image.

\begin{figure}[!h]
	\centering
	\includegraphics[width=1\linewidth]{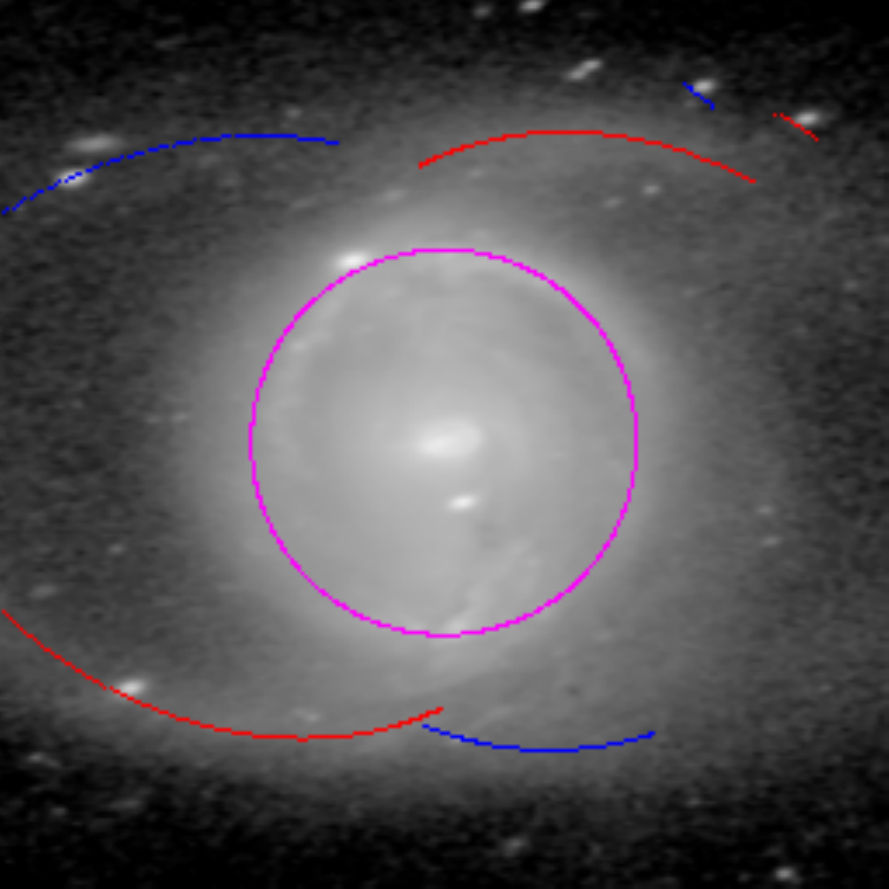}
	\caption{NGC~5728 de-projected image, showing the fitted logarithmic spiral arcs (red and blue) and the `Outer Ring' (magenta circle); the pitch angle relationship gives a SMBH mass of $\sim3.4 \times 10^{8}~\mmsun$}
	\label{fig:ngc5728spiralarcs}
\end{figure}

The spiral arm pitch angle is related to the SMBH mass in the form \citep{Davis2017a}:
\begin{multline}
\log(M_{BH}) = (7.01\pm0.07) \\- (0.171\pm0.017)~[|\phi|-15\degr]\pm0.30 \label{eqn:odraSMBH7}
\end{multline}
where $\phi$ is the spiral pitch angle in degrees. The spiral arm pitch angle was measured from the faint arms that come from the end of the bar, not the `Outer Ring' structure. I thank Dr. Benjamin Davis for providing the data, using the software \texttt{SPIRALITY} \citep{Shields2015a,Shields2015}, \texttt{SpArcFiRe} \citep{Davis2014} and \texttt{2DFFT} \citep{Davis2012}. The de-projected galaxy image, with the fitted logarithmic spiral arcs, is shown in Fig. \ref{fig:ngc5728spiralarcs}; the de-projection angle is 43\degr. 

Table \ref{tbl:ngc5728SMBHMass} gives the black hole mass measurements determined by the four different methods. The values show a scatter of about 1 dex. The \MKs{} value has uncertainties due to obscuration and a lack of a well-defined `classical' bulge and the presence of a nuclear bar; obscuration may also affect the derived S\'{e}rsic index.  We will use the value derived by the template fitting, $\mmbh~=~3.4(+13.6,-2.7) \times 10^{7}~\mmsun$. The mass derived from the 2MASS $K_S$ image S\'{e}rsic index is compatible with this value.

\begin{table*}[!htbp]
\begin{center}
		\caption{SMBH Mass derived from scaling relationships.}
	\label{tbl:ngc5728SMBHMass}
	\begin{tabular}{@{}lrrr@{}}
		\toprule
		Method                 &               Value &     log(\mbh) & $ \mmbh (\times10^{8}~\mmsun) $ \\ \midrule
		\MSig\tablenotemark{a} &     $158\pm24$ \kms & $7.53\pm0.70$ &             $0.34(+1.36,-0.27)$ \\
		\MSig\tablenotemark{b} &     $179\pm10$ \kms & $7.86\pm0.43$ &             $0.72(+1.26,-0.46)$ \\
		\MKs                   & $-23.25\pm0.01$ Mag & $8.13\pm0.49$ &              $1.36(+4.2,-0.45)$ \\
		$M-n$\tablenotemark{c} &                3.54 & $8.53\pm0.42$ &              $3.39(+8.87,-1.3)$ \\
		$M-n$\tablenotemark{d} &                1.56 & $7.74\pm0.59$ &              $0.55(+2.1,-0.14)$ \\
		$M-\phi$               &    $7.0\pm0.3\degr$ & $8.37\pm0.37$ &             $2.34(+3.15,-1.34)$ \\ \bottomrule
	\end{tabular}
\tablenotetext{a}{Template fitting using pPXF.}
\tablenotetext{b}{\textit{velmap} fit to \CaII{} 8544 \AA.}
\tablenotetext{c}{Fit to SINFONI \textit{K} image.}
\tablenotetext{d}{Fit to 2MASS $K_S$ image.}
\end{center}
\end{table*}

The `sphere of influence' of a SMBH (i.e. the volume of space over which the gravity of the SMBH dominates those of the stars) is given by the formula:
\begin{align}
R_{I} & =  \dfrac{G \textrm{\mbh}}{\sigma^{2}}\\
R_{I}(pc) & =  4.305 \times 10^{5}~\dfrac{\textrm{\mbh~[10$^{8}$~\msun]}}{\sigma^{2}~\textrm{[\kms]}}
\end{align}

The bolometric luminosity derived from the X-ray flux \citep{Winter2012}, the Eddington luminosity (i.e. the maximum luminosity possible from a SMBH of a given mass), plus the Eddington ratio and the accretion rate, are given by:
\begin{align}
\log(L_{Bol}) &= 1.12~\log(L_{X}~[14-195~\rm{keV}]) - 4.23\\
L_{Edd} &= 1.3 \times 10^{38}~\textrm{\mbh/\msun}~\rm{erg~s^{-1}}\\
R_{Edd}&= L_{Bol} /L_{Edd}\\
\dot{M}_{acc} &= \dfrac{L_{Bol}}{c^2 \eta}\\
&=1.84\times10^{-46}~L_{Bol}~[\rm{erg~s^{-1}}] \label{eqn:Macc}
\end{align}
where $ L_{X} $ is the X-ray luminosity, $L_{Bol}$ is the bolometric luminosity, \mbh{} is the BH mass and $\dot{M}_{acc}$ is the accretion rate. The rate is given in \msy{} using the constant in Equation \ref{eqn:Macc} and a standard efficiency rate $\eta=0.1$. 

Using the 14--195 keV X-ray luminosity from \cite{Davies2015} and SMBH mass calculated above, this gives:
\begin{align}
L_{X}(14-195) & = 1.62 \times 10^{42}~\textrm{erg s\pwr{-1}}  \nonumber          &  \\
L_{Bol}       & = 1.46 \times 10^{44}~\textrm{erg s\pwr{-1}} \nonumber           &  \\
L_{Edd}       & = 4.42~(+17.7,-3.04) \times 10^{45}~\textrm{erg s\pwr{-1}}\nonumber &  \\
R_{Edd}       & = 3.3~(+7.3,-2.6) \times 10^{-2} \nonumber                       &  \\
\dot{M}_{acc} & = 2.7~\times 10^{-2}~\textrm{\msun~yr\pwr{-1}}\nonumber          &  \\
R_I           & = 5.9~\textrm{pc}\nonumber                                        &
\end{align}

Eddington ratios have a large range; the trend, as outlined by \cite{Ho2008}, is for higher luminosity AGNs to have higher ratios. \cite{Ricci2014a}, in their discussion of emission line properties in 10 early-type galactic nuclei, noted that $R_{Edd} < 10^{-3}$ for all the galaxies that had AGN; however this sample were all LINERs. \cite{Storchi-Bergmann2010} found the ratio for NGC~4151 was 0.012 and \cite{Fischer2015} found 0.12 for Mrk~509 (both Seyfert 1.5); these have a higher bolometric luminosities than NGC~5728. \cite{Vasudevan2010} gives an Eddington ratio for NGC~5728 of 0.028, the same as the value found here. Their range of values of the ratio for the complete sample of 63 \textit{Swift}/BAT X-ray AGNs are $6\times10^{-3}\leq R_{Edd}\leq 7.3\times10^{-2}$. \cite{Ho2008} gives a median $R_{Edd}$ for LLAGN of $1.1\times10^{-3}$ for Seyfert 1-type galaxies, and $5.9\times10^{-6}$ for Seyfert 2s; the bolometric luminosity for this galaxy ($1.46 \times 10^{44}$ \es) places it rather above the range for LLAGN.

In summary, NGC~5728 has a moderate-luminosity Seyfert 2 AGN, with Eddington ratio and accretion rate within the range of values found in the literature, powered by SMBH of $3.4\times10^7~\mmsun$.
\subsection{Outflow Modeling and Kinematics}
Are the kinematics seen in the ionization cones dominated by outflows or rotations? Following \cite{Muller-Sanchez2011}, the main arguments for outflows are:
\begin{itemize}
	\item The velocity gradients across the field are too high to be explained by a physically reasonable gravitational potential. The velocity field shows a ${\Delta}V~\simeq~700$ \kms{} over $\sim$200~pc, which implies an enclosed mass of $\sim2.2~\times~10^{9}~\mmsun$ \citep[see e.g.][modelling NGC~1068]{Das2007}
	\item The NLR and stellar velocity fields are aligned for rotations - this is not the case with this object.
	\item The approaching (brighter) radio jet is aligned with blue-shifts, as shown by the channel maps presented above.
	\item \cite{Muller-Sanchez2011} found an anti-correlation between the outflow opening angles and maximum velocities in their fits of 6 (out of 7) Seyfert galaxies, i.e. the narrower the opening angle, the faster the velocities. The values derived in this work are consistent with their data (see below).
	\item As shown in section \ref{sec:ngc5728bicone}, the bicone does not intersect the plane of the inner nuclear ring but does intersect the plane of the main disk. At $>$ 200 pc from the AGN, the \OIII{} velocity map from the MUSE data shows a re-orientation, taking on the velocity field of the main disk plane, as seen in the stellar \CaII{} velocity map (Fig. \ref{fig:ngc5728stellarkinematics2}). In other words, there are two velocity fields; an outflow close to the AGN and a rotation further out.
	\item The velocity profile shows signatures of radial acceleration, followed by deceleration. The gas velocity does \textit{not} asymptotically tend to zero (as would be expected for rotation), but shows `coasting', i.e. reasonably constant velocity of $\pm100$ \kms{} after $\sim300$ pc; this is consistent with the modeling of NGC~1068 and NGC~4151 \citep{Crenshaw2000,Crenshaw2000a}, as well as \cite{Muller-Sanchez2011} (NGC~1068, NGC~3783, and NGC~7469).
\end{itemize}

\subsubsection{Velocity Models}
There are various physical models of the gas kinematics in outflows from AGN. \cite{Crenshaw2000} modeled the outflow NGC~1068 as a hollow bicone, with emitting material evacuated along the axis - the cone walls were clearly visible in the STIS G430L spectrum as a bifurcation. A similar model was applied to NGC~4151 \citep{Das2005}. \cite{Bae2016} extended this to generalized bicone models, simulating emitted line profiles. They found that velocity dispersion increases as the intrinsic velocity or the bicone inclination increases, while apparent velocity (i.e., velocity shifts with respect to systemic velocity) increases as the amount of dust extinction increases, since dust obscuration makes a stronger asymmetric velocity distribution along the LOS.

\cite{Das2005} also found that the outflow velocities could be modeled heuristically by a linear or square-root law, i.e. the the velocity increases out from the origin to a turn-over point and then decreases back to systemic, in the form $v = k r$ or $v = k \sqrt{r}$ for the acceleration and $v = v_{max} - k' r$ or $v = v_{max} - k' \sqrt{r}$ for the deceleration, where $k$ and $k'$ are constants, $v_{max}$ is the turn-over velocity and $r$ is the radial distance. In their study of AGN outflow systems, \cite{Muller-Sanchez2011} could model their galaxies with a linear acceleration law (except for Circinus), and 3 of them (NGC~1068, NGC~3783, and NGC~7469) with a linear deceleration law. \cite{Liu2015} modeled IRAS~F04250–-5718 and Mrk~509 as filled cones with constant velocity, in which case the maximum velocity dispersion would be expected along the bicone axis. 

\cite{Fischer2013} modeled the inclination of 53 Seyfert galaxies from NLR kinematics using a full hollow bicone from \OIII{} imaging and STIS spectroscopy from \textit{HST}. NGC~5728 kinematics were classified as `Ambiguous', which means that it has anti-symmetric velocities, but could not classified as having a bicone. They explained this as either the two cones are not identical or that the filling factor is not 1 within the hollow cone and zero outside. The PV diagrams presented above show that the outflowing gas does fill the cones. 

\begin{figure}
	\centering
	\includegraphics[width=1\linewidth]{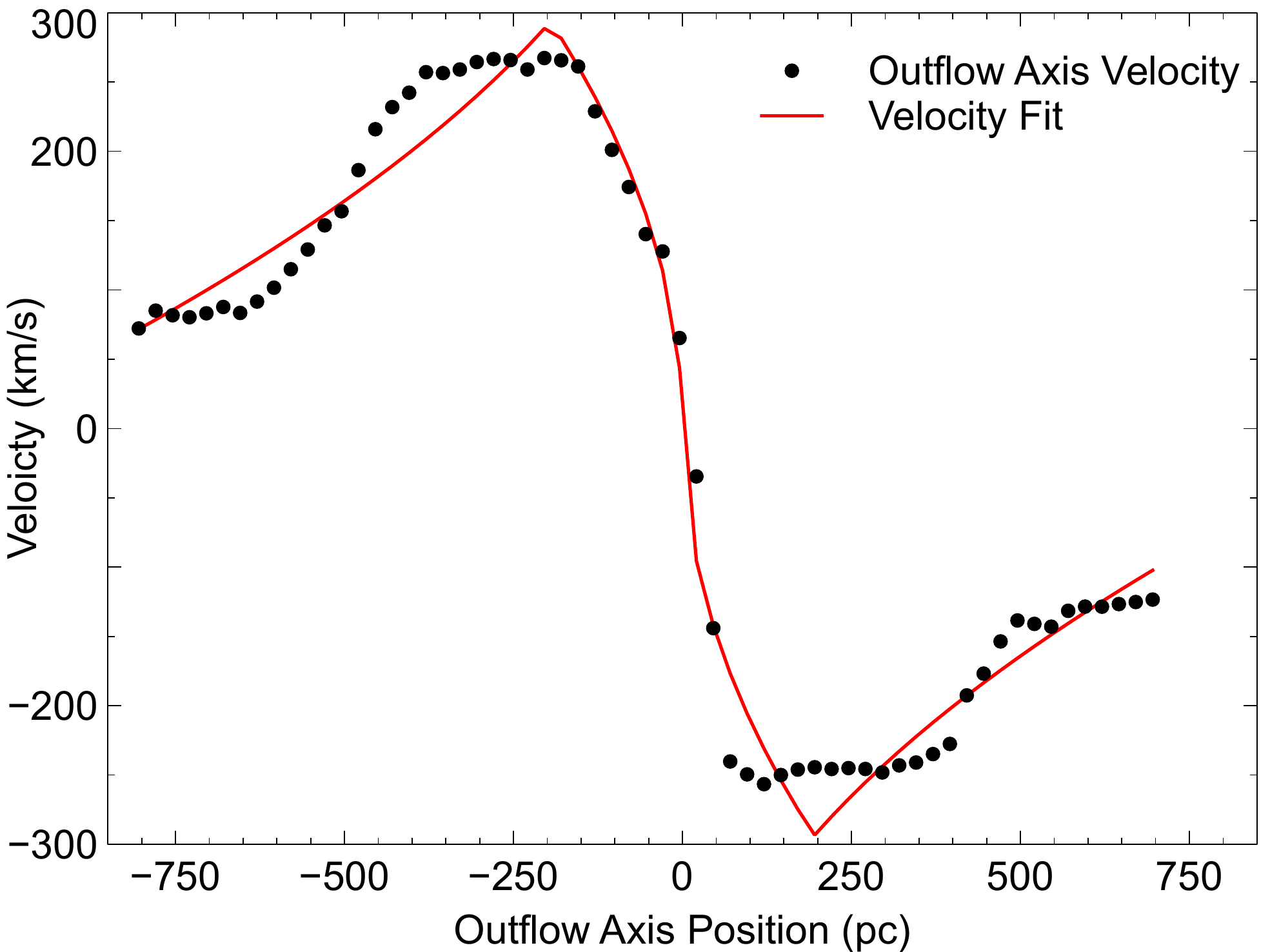}
	\caption{\pab{} velocity and fitted profile along the outflow axis. The fit is for $\sqrt{r}$ proportionality.}
	\label{fig:ngc5728velocityoutflows4}
\end{figure}

Following \cite{Das2005}, we fitted the outflow velocity map (derived from the \pab{} \textit{velmap} procedure) with all combinations of square root and linear accelerations and decelerations. The \pab{} map is used instead of \brg, as it encompasses a larger area and distance from the AGN, with the \brg{} map only extending to the turnover distance. The best fit was found with the $\sqrt{r}$ proportionality with a turnover velocity of 290 \kms{} at a distance of 200 pc from the nucleus. Fig. \ref{fig:ngc5728velocityoutflows4} shows the velocity values and fitted profile. As can be seen, the heuristic functional form is not a good fit around the turnover position; the competing forces on the gas of radiation pressure, magnetic fields, nuclear winds, gravity, ISM friction and gas pressure will produce a more complicated functional form than that given here. 

\cite{Muller-Sanchez2011} summarizes the state of knowledge of the dynamical models, noting that while drag forces can produce the required deceleration, the acceleration phase is more difficult to explain, with \cite{Everett2006} suggesting an accelerated wind close to the AGN (at pc scales) subsequently interacting with and accelerating the ISM over several 100 pc. This is similar to the two-stage outflow model of \cite{Hopkins2010a}.
\subsubsection{The Bicone Geometry}
\label{sec:ngc5728bicone}
We further examine the bicone geometry using the channel maps and \textit{PV} diagrams. The channel map plot for \Fe{} (Fig. \ref{fig:ngc5728channelmapshk1}) shows clear signs of limb-brightening of the outflow (at velocities -337 to +187 \kms), which would be expected for a hollow cone model. This is best shown in the velocity range -187 to +37 \kms, as the prominent `X'-shaped figure with an opening angle of $\sim70\degr$. The near wall of the approaching cone (-412 to -112 \kms) is clearly seen, as is the far wall of the receding cone (+187 to +562 \kms). A puzzling aspect in the absence of the far (near) wall in the approaching (receding) cone. It is possible that this has been terminated by the interaction with the plane of the nuclear structure (the inner disk or mini-spiral), and the resulting outflow is a half-cone or fan shape. 

By contrast, the \brg{} channel maps show little sign of the limb brightening, being more like a filled cone; supporting the concept of the \Fe{} jacketing the hydrogen recombination emission. The \textit{PV} diagrams for \brg{} (Figs. \ref{fig:ngc5728pvdiagram1} and \ref{fig:ngc5728pvdiagram3}) also show no or minimal signs of this structure.

\begin{figure}
	\centering
	\includegraphics[width=1\linewidth]{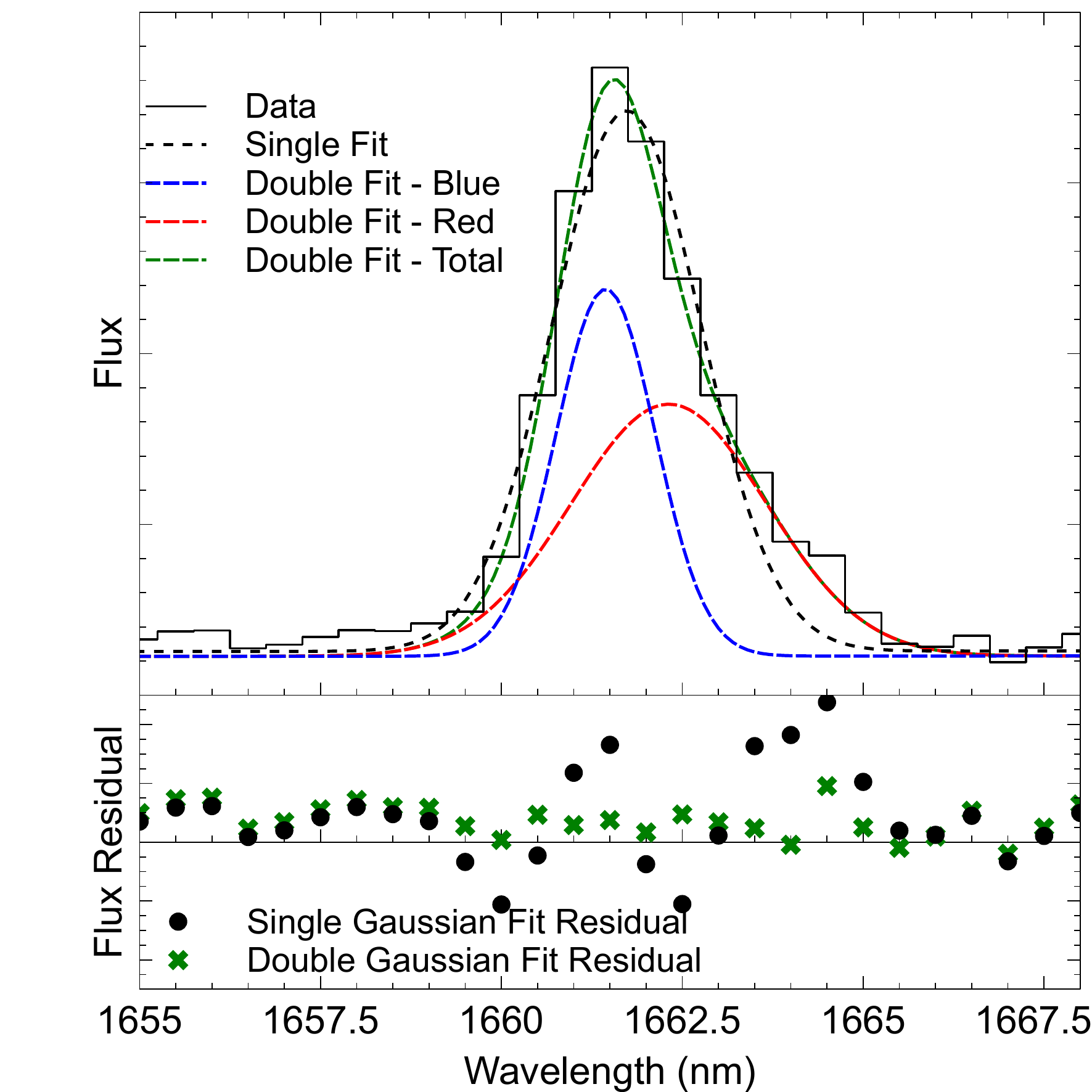}
	\caption{Outflow \Fe{} emission line profiles. Top panel: Single and double Gaussian fit to line profile at position 1\arcsec{} SE from the nucleus. Bottom panel: Fit residuals from single (black dots) and double (green crosses) Gaussian fits (at the same vertical scale).}
	\label{fig:ngc5728velocityoutflows1}
	\centering
	\includegraphics[width=1\linewidth]{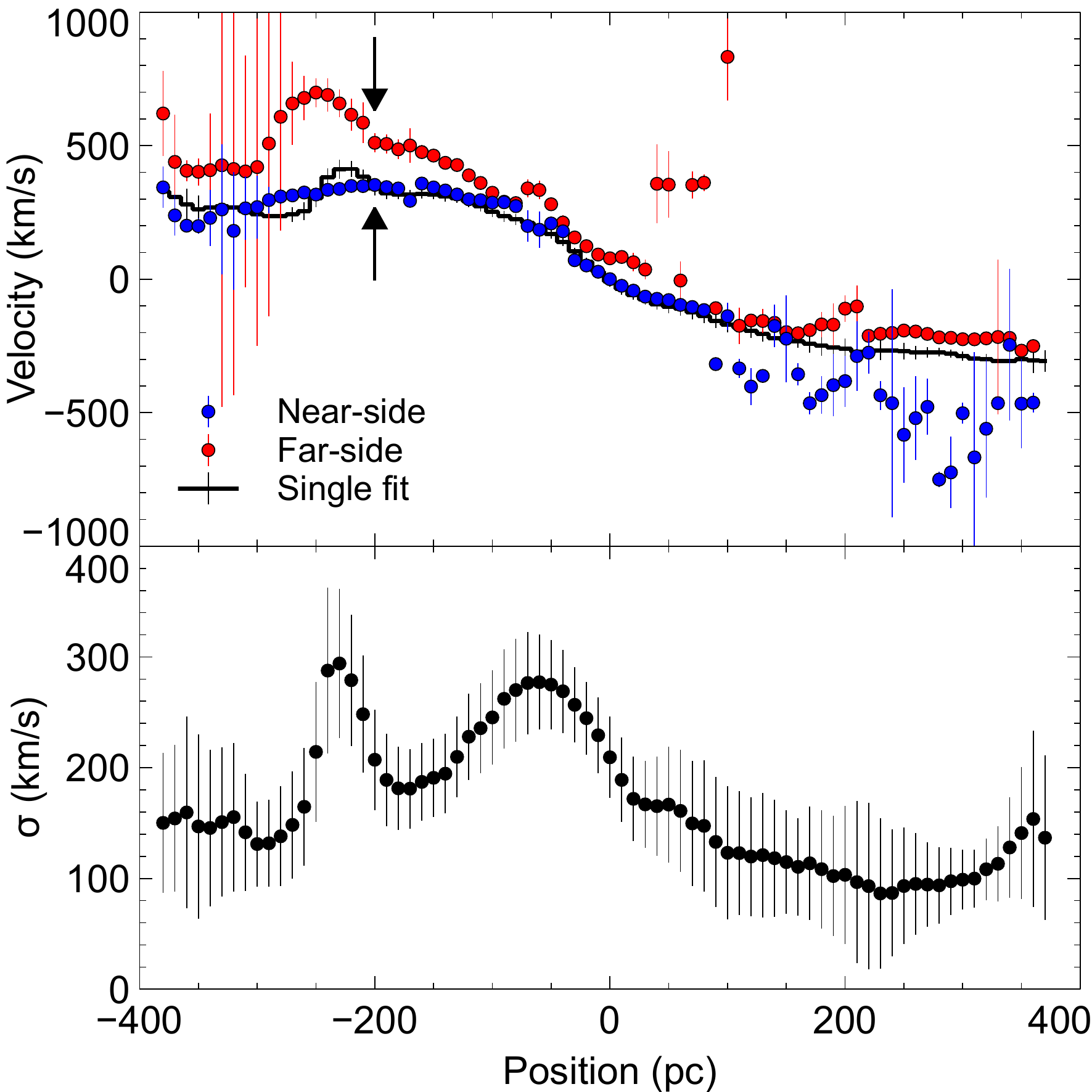}
	\caption{\Fe{} kinematics of the front and back wall of hollow bicone outflow. Top panel: Velocities of double Gaussian fit (red and blue points) along the center-line of the outflow, with uncertainties. The black line is the velocity derived from the \textit{velmap} procedure. The arrows indicate the position for the illustrative double Gaussian fit (Fig. \ref{fig:ngc5728velocityoutflows1}). Bottom panel: Velocity dispersion from \textit{velmap} procedure.}
	\label{fig:ngc5728velocityoutflows2}
\end{figure}

If the \Fe{} outflow structure is a hollow bicone, then we should be able to separate the LOS velocities of the front and back walls. Instead of a single Gaussian fit (as from the \textit{velmap} procedure), the asymmetric line profile of \Fe{} was fitted by a double Gaussian curve, as illustrated for the position 1\arcsec{} SE of the nucleus (Fig. \ref{fig:ngc5728velocityoutflows1}); this shows a double (green dashed line, with blue and red components)  vs. a single Gaussian fit (black dashed line). The fits for each pixel along the centerline of the outflow is shown in Fig. \ref{fig:ngc5728velocityoutflows2}, top panel, where the black line is the velocity from the \textit{velmap} procedure, and the blue and red fits are the shorter and longer wavelength of the double Gaussian fits. For the double Gaussian fit, one of the fits always dominates and produces corresponding uncertainties in the other fit, as shown by the error bars. This also corresponds with the low flux of either the far or near wall visibility in the channel maps. The bottom panel of Fig. \ref{fig:ngc5728velocityoutflows2} shows the \textit{velmap} measured velocity dispersion (i.e. that of the single Gaussian fit), with the uncertainties. Unlike the \textit{HST} STIS data for NGC~1068 \citep{Crenshaw2000}, the line profiles do not show bifurcation, just a red or blue skewness.

\begin{figure*}
	\centering
	\includegraphics[width=.7\linewidth]{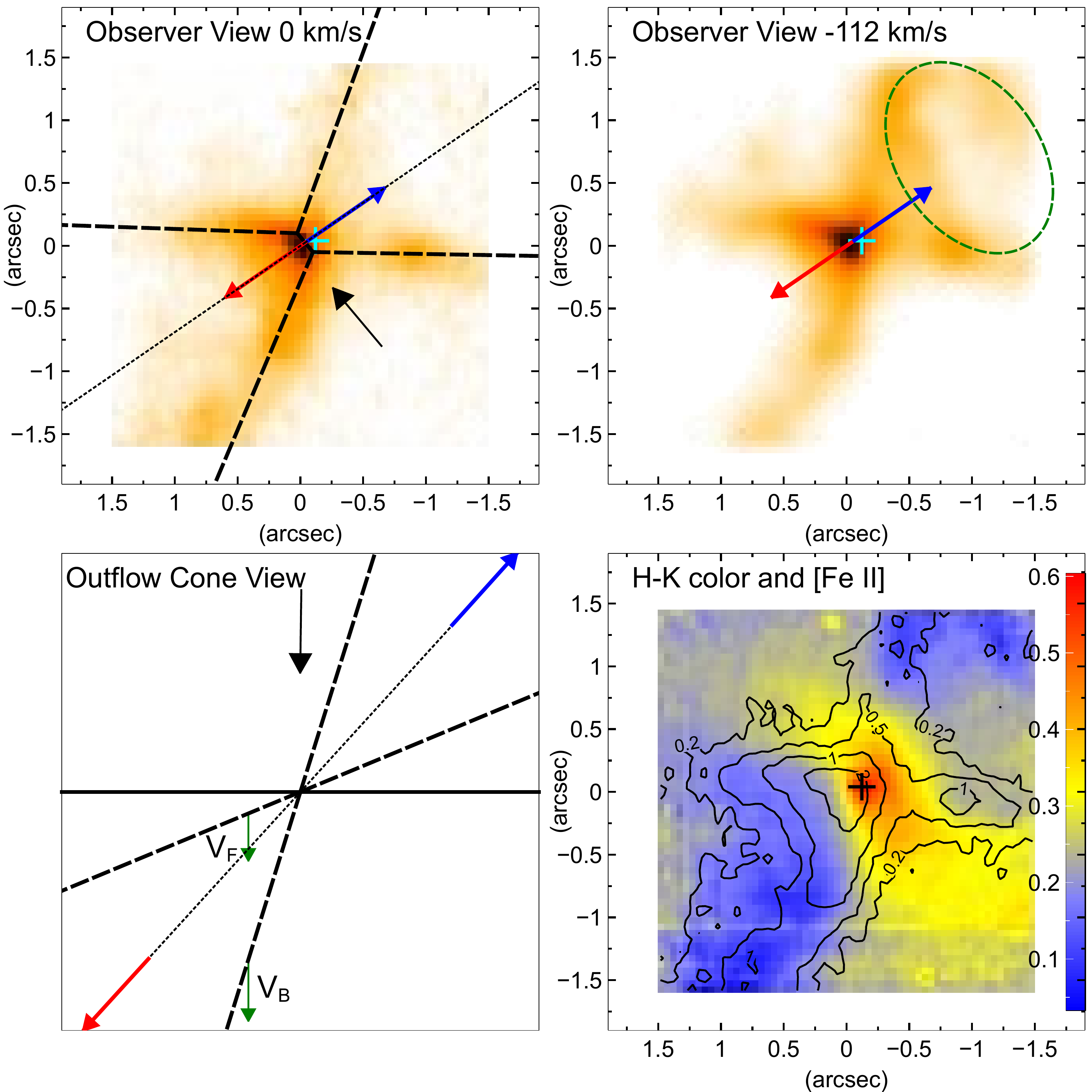}
	\caption{Hollow bicone outflow model. Top left: observer view of outflow boundaries (thick dashed lines) and center-line (thin dotted line), superimposed on the zero velocity channel map. Blue and red arrows indicate approaching and receding outflow velocity directions, the cyan cross indicates the AGN position and the black arrow shows the viewing angle along the plane of the sky for the right panel diagram. Top right: -112 \kms{} channel map, with overplotted outflow boundary of a cone slice projected to the derived inclination angle of 47\degr. Bottom left: Side-on (plane of the sky) view of the outflow cones with the lines and blue and red arrows as for the left panel, with the opening angle and inclination as derived. The black arrow is the observer LOS and the green arrows indicate the LOS front and back wall velocities. Bottom right: \Fe{} zero-velocity channel flux (contour) overlaid on the \hk{} color map. Contour values in 10\pwr{-18} \ecs, color-bar is \hk{} values. The association between the dust grain destruction and the \Fe{} emission is shown by the anti-correlation.}
	\label{fig:ngc5728velocityoutflows3}
\end{figure*}

From these velocity measurements, we can derive the outflow bicone parameters of internal opening angle and inclination to the LOS. The outflow is modeled as two cones truncated at the tip, and the sides aligned with the observed limb-brightening; this is shown in Fig. \ref{fig:ngc5728velocityoutflows3} - top left panel, where the colored map is the zero-velocity channel, the thick dashed lines are the sides of the outflow and the thin dotted line is the center-line of the outflow. The blue and red arrows illustrate the approaching and receding cones. The projected opening angle is measured from the zero velocity channel map at $\alpha = 71\degr$; this is larger than the value derived for the emission lines by \cite{Capetti1996} (55--65\degr) but somewhat narrower than the polarized continuum angle (94\degr); the \Fe{} jacket of the main flow is the source of the higher opening angle.

The cone internal angle, $\beta$, and the inclination, $i$, (where \textit{i}=0\degr{} means that the cone axis is in the plane of the sky and \textit{i}=90\degr{} means that the cone is face-on) can be derived from the following relationships:
\begin{align}
\tan(90-\beta/2)& = \tan(90-\alpha/2)/\cos(i) \\
V & = V_F / \sin(i-\beta/2) \\
& = V_B / \sin(i+\beta/2) 
\end{align}
where $V_F$ and $V_B$ are the measured front and back wall LOS velocities and $V$ is the true outflow velocity. The parameters were derived at several points along the outflow center-line, ensuring that the velocity difference $|V_B - V_F|$ was over 230 \kms, where the two Gaussian components were clearly separated. The equations were solved using the `GRG Nonlinear' algorithm, minimizing the difference between the outflow velocity derived from the front and back velocities, with the added check that the derived projected cone opening angle was close to the observed angle. 

The results at the several locations along the center-line agree very well, $\beta = 50\degr.2\pm2\degr.2$ and $i = 47\degr.6\pm3\degr.2$. It is noted that this inclination is close to that of the outer disk  (48\degr) as found in \citetalias{Durre2018a}, i.e. the bicone axis is nearly co-planar to the main galactic plane. This is also close to the value for the de-projection angle for the spiral-arm pitch fit (Section \ref{sec:ngc5728smbh}) of 43\degr. There is no requirement for AGN outflows to align to the galaxy geometry; \cite{Nevin2017}, \cite{Muller-Sanchez2011} and \cite{Fischer2013} all find no alignment between outflow and galactic photometric axes.

The biconal interpretation is further supported at the -112 \kms{} channel map slice, where a cone slice projected at the inclination can be overlaid on the elliptical feature (Fig. \ref{fig:ngc5728velocityoutflows3} - top right panel). In the bottom left panel of Fig. \ref{fig:ngc5728velocityoutflows3}, we show a diagram of the bicone seen side-on across the plane of the sky.

\cite{Muller-Sanchez2011} examined the biconal outflows of 7 Seyfert galaxies, and found an anti-correlation between the half-opening angle of the cone and the outflow maximum velocity. The values for NGC~5728 (50\degr{} and 260 \kms/$\cos{(i)}$ = 380 \kms) place it on the trend on the plot (their figure 27) from the objects they measured. They also found an anti-correlation between the \Htwo{} gas mass in the inner 30 pc and the half-opening angle; the \Htwo{} mass for NGC~5728 within that radius is $0.6\times10^8~\mmsun$, placing it $\sim0.5$ dex above the trend (their figure 28), but within uncertainties. The correlation between the opening angle and the gas mass (their figure 29) again places NGC~5728 $\sim0.5$ dex above the trend. They posit that a higher \Htwo{} mass provides increased confinement to the torus, which makes the outflow more highly collimated, which in turn increases the outflow velocity \citep[see also][]{Davies2006,Hicks2009,MullerSanchez2009}.

\begin{figure*}
	\centering
	\includegraphics[width=.9\linewidth]{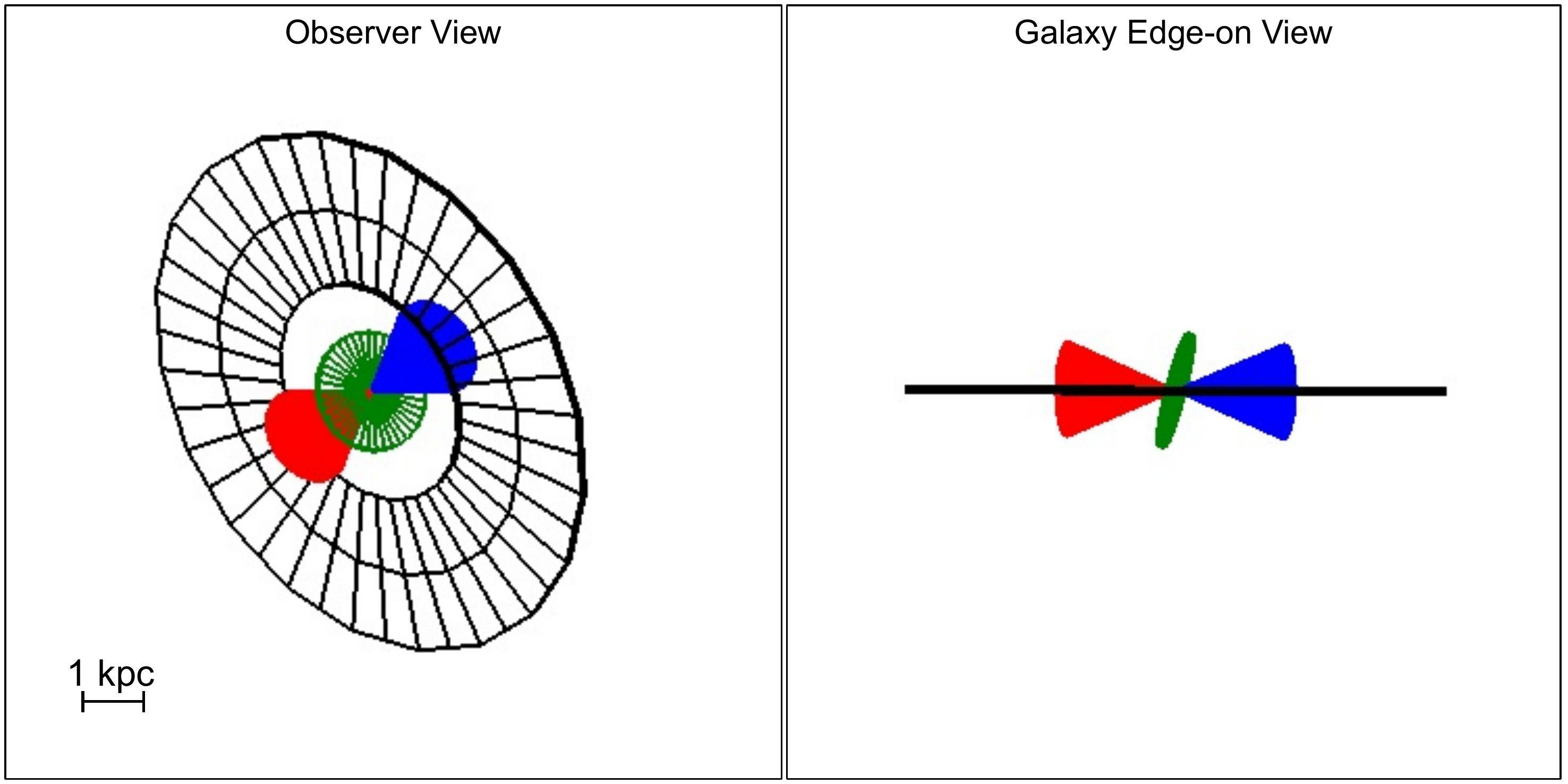}
	\caption{Model cartoon of galactic structure and outflows. Outflows: blue (approaching) and red (receding) cones. Main disk: black. Inner disk: green. Left panel: observer view. Right panel: edge-on to galactic plane view. The skewed morphology for the nuclear stellar structure vs. the main galactic plane is revealed.}
	\label{fig:ngc5728velocityoutflows6}
\end{figure*}

Fig. \ref{fig:ngc5728velocityoutflows6} shows cartoons of the model using the \texttt{Shape} 3D modeling software \citep{Steffen2010}, using the angular values and sizes derived here and in \citetalias{Durre2018a}. The left panel shows the structures as seen by the observer, the right panel shows the view edge-on to the main galaxy disk. The black disk represents the `Main Disk', with the assumption that the near side is to the NW, the blue (approaching) and red (receding) cones are the outflows and the `Inner Disk' is in green. The left panel shows the galaxy and its structures as seen by the observer, the right panel shows an edge-on view. The model cone length is 1.8 kpc, and the main disk has a 4.8 kpc radius. As can be seen, the bicone, inner disk and main disk axes are not aligned. The outflow axis is almost in the plane of the main disk, with the blue-shifted (approaching) cone just behind the disk from our perspective. This is supported by the dust lanes crossing the outflow on our LOS. The nuclear stellar structure is almost perpendicular to the main disk.

It is noted that the condition for visibility of the BLR and thus a Seyfert 1 classification, i.e. $|(i)| + \beta/2 > 90\degr{}$, is not met, which supports the unified model of AGN, where the nuclear dust structures outside of the BLR collimates the ionizing radiation \citep[see][]{Muller-Sanchez2011}.

As noted in \citetalias{Durre2018a}, the \Fe{} emission is well aligned with the \hk{} color; this is shown in Fig. \ref{fig:ngc5728velocityoutflows3} - bottom right panel. The edges and other features of the emission flux are anti-correlated with the \hk{} color (i.e. higher flux as aligned with lower color value). This shows that the dust is being removed at these locations. Alternative mechanisms are mechanical removal through entrainment in the outflows, or the grains being sublimated directly by high-energy photons or sputtered by shocks, releasing the iron to be excited by electron collisions \citep{Mouri2000}. The second process is probably the predominant one, given that the edges delineate the shock boundary of the outflows, as opposed to the lack of correlation between the \Fe{} velocity map and the \hk{} color map. This is a different mechanism for releasing iron from grains than the standard supernova shock scenario, and is a striking illustration of the process.
\subsubsection{Mass Outflow Rate}
\label{sec:NGC5728OutflowKinematics}
The maximum outflow velocity observed from \pab{} of 290 \kms{} gives a true outflow velocity of $\sim390$~\kms. We derive the mass outflow rate ($\dot{M}_{out}$) using the following procedure. To a first order, the cone can be modeled as a plane, where the LOS velocity and mass surface density represent the integrated chord through the cone. Each spaxel is considered as a cube (0\arcsec.05 = 10 pc a side, i.e. 100 pc\pwr{2}/$2\times10^{-4}$ arcsec\pwr{2}) moving at velocity $V = V_{out} / \sin{(i)}$, where $V_{out}$ is the observed LOS velocity. The mass in the cube is given by Equation \ref{eqn:odraHII} \citep{Riffel2013b,Riffel2015}
\begin{equation}
M_{HII} \approx 3 \times 10^{17} F_{Br\gamma}D^2~\mmsun \label{eqn:odraHII}
\end{equation}
where \textit{D} is the distance in Mpc and \textit{F} is the flux of the \brg{} emission, measured in erg cm\pwr{-2} s\pwr{-1}. This gives the density of hydrogen, which is assumed to be fully ionized in the outflow. (If there is significant mass in the form of neutral and atomic hydrogen in the outflow, this value will be underestimated, as will the corresponding kinetic power. This could be especially true away from the ionizing source, as the gas is coasting.) 

\begin{figure}
	\centering
	\includegraphics[width=1\linewidth]{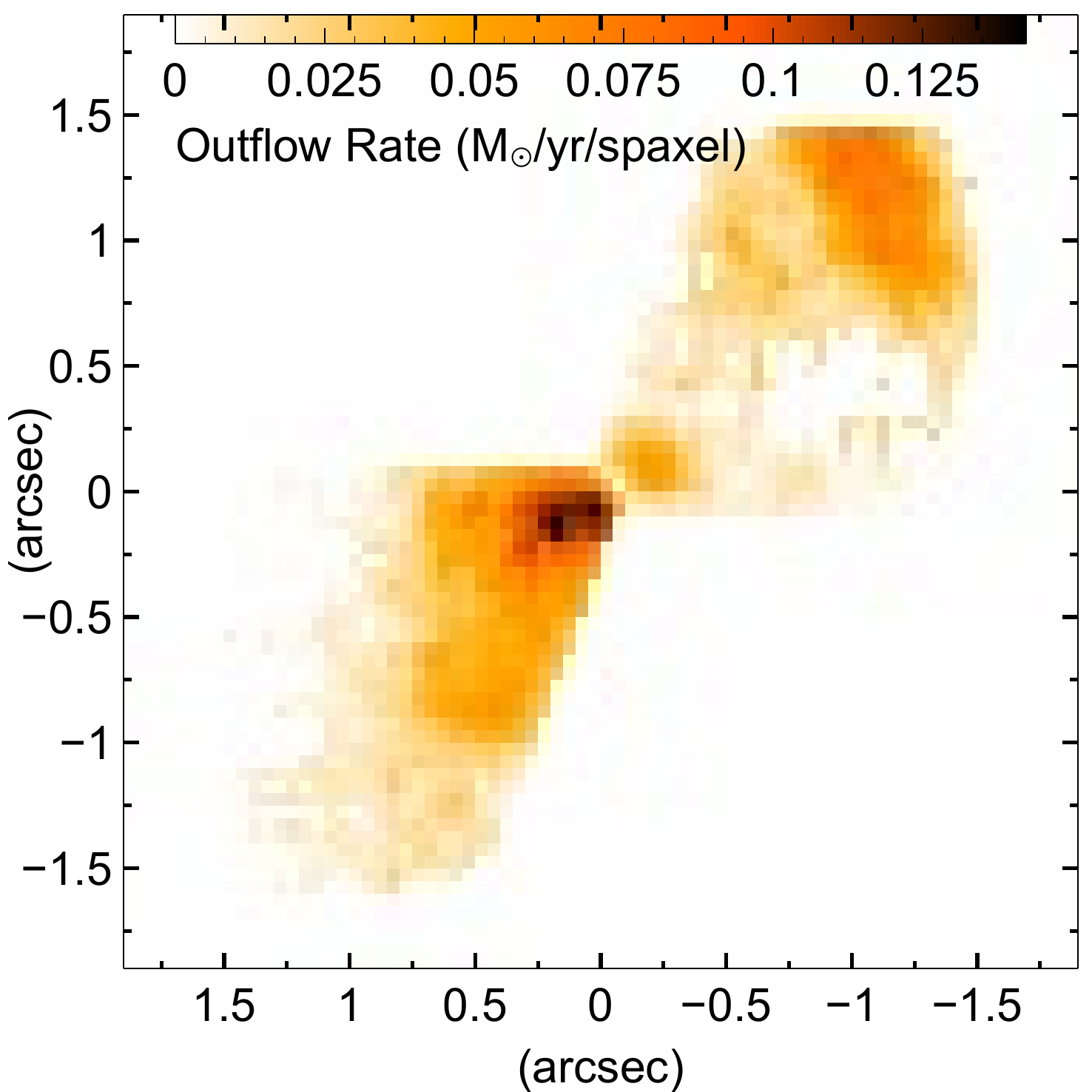}
	\caption{Outflow rate for bicone in \msy{} spaxel\pwr{-1} for hydrogen derived from the \brg{} flux and velocity fields, showing the SE peak close to the AGN and the NW feature at $\sim300$ pc where the radio jet impacts the ISM in the SF ring. (One spaxel equals 100 pc\pwr{2}, $2.5\times 10^{-4}$ arcsec\pwr{2}.)}
	\label{fig:ngc5728velocityoutflows5}
\end{figure}

Combined with the relationship (accurate to a few percent) of 1 \kms~=~1 pc Myr\pwr{-1}, and the distance of 41.1 Mpc, this gives:
\begin{equation}
\dot{M}_{out}~=~6.9~\times~10^{13}~V_{out}~F_{Br\gamma}~\textrm{[\msun~yr\pwr{-1}]} \label{eqn:ngc5728outflow}
\end{equation}
Fig. \ref{fig:ngc5728velocityoutflows5} shows the map of the outflow rates at each spaxel in the bicone in units of \msy.  The field has been masked to the cone outline. The total outflow rate over the whole field is 38 \msy; the total for just the SE cone is 20 \msy, showing that the NW cone is not heavily obscured (i.e. they have similar measured outflow rates). However, the outflow rate distribution in each cone is dissimilar; the SE cone shows that the maximum rate is at $\sim60$ pc from the AGN, whereas the NW cone has it at the supposed radio jet impact region at $\sim300$~pc. It is possible that the NW cone has some obscuration from the dust lane and thus has a higher mass outflow rate than that measured.

The ratio of the accretion and outflow rates can also be computed:
\begin{equation}
\dfrac{\dot{M}_{out}}{\dot{M}_{acc}} = 38/(2.7\times10^{-2}) = 1415
\end{equation}

The ratio of the flow rates ($\dot{M}_{out}/\dot{M}_{acc}$) can be compared with the results from \cite{Muller-Sanchez2011}, who find a range of $\sim100-8000$; our value (only exceeded by one of their objects, NGC~2992) supports their conclusion that Seyferts with strong and collimated radio emission are also hosts of powerful outflows of ionized gas. The ratio of outflow energy to bolometric luminosity is also consistent with the \cite{Hopkins2010a} models.
\subsubsection{Outflow Kinetic Power}
We compute the kinetic power or luminosity $L_{KE}$ in the outflows \citep[following e.g.][]{Storchi-Bergmann2010} from Equation \ref{eqn:ngc5728kinpwr1}, which after conversion to cgs units and using Equation \ref{eqn:ngc5728outflow} becomes Equation \ref{eqn:ngc5728kinpwr2}, where $\dot{M}_{out}$ is in units of \msun~yr\pwr{-1} and $V_{out}$ and $\sigma$ are in \kms. It is assumed that the dispersion measure is due to turbulence; if it is not included the derived power value is reduced by 15\%.
\begin{align}
L_{KE} &= \dot{M}_{out}~(V^2~+~\sigma^2) / 2 \label{eqn:ngc5728kinpwr1}\\
&= 3.15 \times 10^{35}~\dot{M}_{out}~(V_{out}^2/\sin^2(i)~+~\sigma^2)~\textrm{\es}\label{eqn:ngc5728kinpwr2}\\
\end{align}
This results in a total kinetic power of $\sim1.5 \times 10^{42}$ \es{}, which is $\sim0.01~L_{Bol}$; this is calculated for each pixel's mass and velocity.

An alternative method of calculating kinetic power is given in \cite{Greene2012} and references therein, where  
\begin{equation}
L_{KE} = 1.5~(\Omega/4\pi)~10^{46}~R^2_{10}~v^3_{1000}~n_0~\text{\es}\\
\end{equation}
where $R_{10}$ (the outflow extent) is in units of 10 kpc, $v_{1000}$ is the expansion velocity in units of 1000 \kms, $n_0$ is the ISM density in units of cm\pwr{-3} and $\Omega/4\pi$ is the covering factor of the outflow in steradians. Taking $R~=~2$ kpc, $v_{max}~=~390$\kms, the covering factor (for a bicone with opening angle of 50\degr) = 0.092 and a rough estimate of the ISM density $n_{0}~=~1$ cm\pwr{-3}, we derive $L_{KE} = 3.2\times10^{42}$ \es, very similar to the calculation above. Again, both calculations would underestimate the energetics if there is significant mass loading of neutral and molecular gas.

The gas mass that can be expelled from the galaxy can be estimated, following \cite{Storchi-Bergmann2010}. The projected outflow length and average velocity are 2 kpc and 175 \kms, respectively, which leads to the AGN active lifetime of at least 10\pwr{7} yr; this means a total kinetic energy of $\sim5 \times 10^{56}$~erg. We use  the estimated galaxy mass of $M_{Galaxy} = 7.2 \times 10^{10}~\mmsun$ (derived above) and the relationship:
\begin{equation}
M_{Gas} \approx \dfrac{E_{Bind}~\times~R_{Gas}}{G~\times~M_{Galaxy}} \label{eqn:GasExp}
\end{equation}
where $M_{Gas}$ is the expelled gas mass, $E_{Bind}$ is the total kinetic energy (equated to the binding energy) and $R_{Gas}$ is the observed outflow distance. This equates to $M_{Gas} = 1.6 \times 10^{8}~\mmsun$. 

This is of the order of the total gas outflow through the bicones over the activity cycle of $\sim4 \times 10^8~\mmsun$, and is also of the order of the estimated mass of molecular gas in the nuclear region ($7 \times 10^8~\mmsun$); this indicates that a large proportion of the gas will be blown away by the end of the AGN activity cycle, especially if the AGN lifetime is a few times that given here.

The galaxy mass value given above is likely to be underestimated, since we must include dark matter in the gravitational potential. From \cite{Rubin1980}, the rotational velocity is measured at $\sim175$ \kms{} at a radius of 11340 pc; given the disk inclination of 60\degr (a reasonable average of the last two values in Table 3 in \citetalias{Durre2018a}), the circular velocity is $\sim350$ \kms. From simple Keplerian rotation, this gives an enclosed value of $M_{Galaxy} = 3.2 \times 10^{11}~\mmsun$, i.e. a ratio of 4.5 over the purely baryonic mass. Substituting this value in Equation \ref{eqn:GasExp} gives an expelled gas mass $M_{Gas} = 3.6 \times 10^{7}~\mmsun$. 

The kinetic-to-bolometric luminosity ratio is larger than values found in \cite{Barbosa2009} for 6 nearby Seyfert galaxies, which were in the range  $10^{-5} \le L_{KE}/L_{Bol} \le 10^{-4}$; however, it is within the range of the values from \cite{Muller-Sanchez2011}, who find $2\times10^{-4} \le L_{KE}/L_{Bol} \le 5\times10^{-2}$ for 7 Seyfert galaxies with biconal outflows. In general, the former objects did not have collimated outflows and had significantly lower bolometric luminosities ($1\times10^{42} \le L_{Bol} \le 4.6\times10^{43}$ vs. $5\times10^{43} \le L_{Bol} \le 2.5\times10^{44}$~\es). The mass outflow to bolometric luminosity ratio is also compatible with the relationship found by \cite{Nevin2017} from summarized literature values.

The time-scale of the AGN activity implies a causal relationship, i.e. the evacuation of a large proportion of the available gas stops the feeding process and thus the AGN activity ceases. However, we note that the feeding to feedback mass ratios ($\sim0.1\%$) means that the AGN can be fed equatorially for a period after the polar gas has been expelled.

\subsubsection{The Source of The Outflow}
Could there be a contribution from nuclear star-formation to the outflow kinematics? Following the arguments from \cite{Muller-Sanchez2011} and \cite{Greene2012} and references therein, we can regard this as unlikely: 
\begin{itemize}
	\item The outflow geometry of starbursts and AGNs is different; star-burst galaxies show hollow truncated cone outflows \citep[see also][for IC~630]{Durre2017}, whereas this object shows a very small source area.
	\item The emission-line  diagnostics (\citetalias{Durre2018a}), show only AGN excitation at the nucleus (for both optical and infra-red diagnostics), with minimal or no SF mixing.
	\item Even if all the \HII{} emission was from star formation (which is, of course, inconsistent with the excitation mode deduced from the line ratios), the \brg{} flux from the central 1\arcsec{} would imply a SFR = 0.2 \msy{} (using the flux from Table 5 of \citetalias{Durre2018a}, the distance of 41.1 Mpc and the SFR relationship from \cite{Kennicutt2009} with $ T = 10^{4}~$K and $ n_{e}= 10^5$). \cite{Veilleux2005} gives a relationship between the SFR and the mechanical luminosity of $L_{KE}~=~7\times 10^{41}~(SFR/\text{\msy})$~\es, for solar metallicity and constant mass/energy injection after $\sim40$ Myr. This gives an energy rate of $1.4\times 10^{41}$ \es, an order of magnitude below the measured value ($\sim1.5 \times 10^{42}$ \es).
\end{itemize} 

\section{Conclusion}
We have examined the nuclear region of NGC~5728 using data at from optical and NIR IFU spectra, revealing a highly complex object showing gas with multiple morphologies and kinematic structures, driven by a powerful AGN. We have presented and analyzed gaseous and stellar kinematics by modeling emission and absorption spectral lines of several species, mapping velocity and dispersion, as well as using \textit{PV} diagrams and channel maps to resolve kinematic structures.

In summary, the nuclear gas and stellar kinematics present the following features:
\begin{itemize}
	\item The SMBH mass was determined to be $3.4^{+10.4}_{-2.7} \times 10^{7}$ \msun, derived from the \MSig{} relationship. The \MKs, $M-n${} and $M-\phi$ relationships produce values up to an order of magnitude higher, but these may have issues related to obscuration and the presence of a nuclear bar. 
	\item The bolometric luminosity of the AGN is $L_{Bol} = 1.46 \times 10^{44}$~\es, the Eddington ratio is $E_{Edd} = 3.3 \times 10^{-2}$ and the mass accretion rate necessary to power the system is $\dot{M}_{acc}  = 2.7~\times 10^{-2}$~\msy. The black hole sphere of influence is $\sim6$ pc.
	\item The model of an AGN-driven outflow (rather than rotation) is supported by the kinematics, from arguments of radio-jet alignment, non-alignment with the stellar rotation and compatibility of of the velocity and opening angle with other observations of outflows. As the outflow impacts on the main galactic disk far from the AGN, the kinematics show the gas being swept up by the ISM.
	\item The AGN central engine is obscured behind a broad dust lane; the position was determined by following the trajectory of the brightest pixel across the spectrum, which shows reduced obscuration at longer wavelengths. This deduced location was supported by the position of the base of the outflows as derived from the channel maps.
	\item The outflow velocity profile was traced using the \pab{} emission line kinematics, which showed a symmetric bicone pattern, accelerating away from the AGN to a LOS velocity maximum of about -250 \kms{} at $\sim220$ pc projected distance to the NW and 270 \kms{} at $\sim280$ pc SE of the nucleus, then decelerating to -150/110 \kms{} at double those distances. The velocity profile was modeled proportionately to the square-root of the distance from the AGN, but this did not fit the turnover well; the multiple competing astrophysical forces (radiation pressure, gravity, winds, gas pressure etc.) will affect the profile.
	\item The bicone geometry was examined using channel maps; the \brg{} emission seems to fill the cone, whereas the \Fe{} emission jackets it. Modeling the \Fe{} line profile along the bicone center with a double Gaussian fit, the internal opening angle (50\degr.2) and the inclination to the LOS (47\degr.6) were calculated; indicating that the bicone axis is nearly parallel to the plane of the galaxy. This supports the Unified Model of AGN, as these angles preclude seeing the accretion disk, which is obscured by the nuclear-scale dust structures. MUSE data also shows kinematic signatures of impact with the main galactic plane
	\item There is some evidence that the dust is being removed in the outflows, as seen with the \Fe{} emission being anti-correlated with the \textit{H-K} color. This could be by entrainment in the outflow or grain sublimation by ionizing radiation or shocks. This is also supported by the reduced extinction observed in the outflows.
	\item The mass outflow rate through the bicones is estimated at 38 \msy; the ratio of outflow rate to the accretion rate required to power the AGN is 1415. The total outflow kinetic power is estimated at $\sim1.5 \times 10^{42}$ \es{}, which is 1\% of the bolometric luminosity.
	\item Over the estimated AGN active lifetime of 10\pwr{7} years, the gas mass that can be completely unbound from the galaxies gravity is $1.6 \times 10^{8}~\mmsun$. This is of the order of the total gas outflow through the bicones over the activity cycle ($\sim4 \times 10^8~\mmsun$), and is also of the order of the estimated mass of molecular gas in the nuclear region ($7 \times 10^8~\mmsun$). A large proportion of the available gas for star formation can be expelled over the AGN life-cycle.
	%\item The Seyfert 1.9 classification has been revised to Seyfert 2, based on modeling of the optical nuclear spectrum from MUSE \Ha+\NII{} data. This shows that no additional broad-line component is required to fit the line profiles where the observations cross both bicones; this is an illustration how outflows combined with low spectral and spatial resolution can cause misclassification.
\end{itemize}

To understand the complex nature of AGNs, their associated nuclear gas and stars must be studied at the highest resolution possible using multi-messenger observations.
\acknowledgments
This work is based on observations collected at the European Organisation for Astronomical Research in the Southern Hemisphere under ESO programmes 093.B-0461(A), 093.B-0057(B) and 097.B-0640(A). We would like to thank Garrett Cotter for facilitating the access to the VLT observations, as well as our Triplespec colleagues at Caltech for their ongoing support of our survey. We acknowledge the continued support of the Australian Research Council (ARC) through Discovery project DP140100435. M.D. dedicates this paper (and \citetalias{Durre2018a}) to the memory of Dr. Ross McGregor Mitchell (1954-2018), atmospheric scientist  and astrophysicist. We would also like to thank the anonymous reviewer for a very detailed and constructive reports.

%References  
\facilities{VLT:Yepun (SINFONI, MUSE)}
\software{QFitsView \citep{Ott2012}, SPIRALITY \citep{Shields2015a,Shields2015}, SpArcFiRe \citep{Davis2014}, 2DFFT \citep{Davis2012}, Shape \citep{Steffen2010}}
\bibliographystyle{apj}
\bibliography{library.bib}

\begin{thebibliography}{66}
\expandafter\ifx\csname natexlab\endcsname\relax\def\natexlab#1{#1}\fi

\bibitem[{Bae \& Woo(2016)}]{Bae2016}
Bae, H.-J., \& Woo, J.-H. 2016, ApJ, 828, 97

\bibitem[{Barbosa {et~al.}(2009)Barbosa, Storchi-Bergmann, {Cid Fernandes},
  Winge, \& Schmitt}]{Barbosa2009}
Barbosa, F. K.~B., Storchi-Bergmann, T., {Cid Fernandes}, R., Winge, C., \&
  Schmitt, H. 2009, MNRAS, 396, 2

\bibitem[{Capetti {et~al.}(1996)Capetti, Axon, Macchetto, Sparks, \&
  Boksenberg}]{Capetti1996}
\href{http://adsabs.harvard.edu/doi/10.1086/177501}{{Capetti, A., Axon, D.~J.,
  Macchetto, F., Sparks, W.~B., \& Boksenberg, A.}} 1996, ApJ, 466, 169

\bibitem[{Cappellari \& Copin(2003)}]{Cappellari2003}
Cappellari, M., \& Copin, Y. 2003, MNRAS, 342, 345

\bibitem[{Cappellari \& Emsellem(2004)}]{Cappellari2004}
Cappellari, M., \& Emsellem, E. 2004, PASP, 116, 138

\bibitem[{Catinella {et~al.}(2005)Catinella, Haynes, \&
  Giovanelli}]{Catinella2005}
\href{http://adsabs.harvard.edu/cgi-bin/bib{\_}query?2005AJ....130.1037C
  http://vizier.u-strasbg.fr/viz-bin/VizieR-3?-source=J/AJ/130/1037{\&}-out.max=50{\&}-out.form=HTML
  Table{\&}-out.add={\_}r{\&}-out.add={\_}RAJ,{\_}DEJ{\&}-sort={\_}r{\&}-oc.form=sexa}{{Catinella,
  B., Haynes, M.~P., \& Giovanelli, R.}} 2005, AJ, 130, 1037

\bibitem[{Combes \& Leon(2002)}]{Combes2002}
\href{http://arxiv.org/abs/astro-ph/0209267}{{Combes, F., \& Leon, S.}} 2002,
  in SF2A-2002, ed. F.~Combes \& D.~Barret (Paris: EDP-Sciences), 1--2

\bibitem[{Crenshaw \& al.(2000)}]{Crenshaw2000a}
\href{http://arxiv.org/abs/astro-ph/0007017{\%}0Ahttp://dx.doi.org/10.1086/301574{\%}0Ahttp://stacks.iop.org/1538-3881/120/i=4/a=1731}{{Crenshaw,
  D.~M., \& al., E.}} 2000, AJ, 2, 1731

\bibitem[{Crenshaw \& Kraemer(2000)}]{Crenshaw2000}
\href{http://arxiv.org/abs/astro-ph/0002438}{{Crenshaw, D.~M., \& Kraemer,
  S.~B.}} 2000, ApJ, 532, L101

\bibitem[{Das {et~al.}(2007)Das, Crenshaw, \& Kraemer}]{Das2007}
\href{http://stacks.iop.org/0004-637X/656/i=2/a=699}{{Das, V., Crenshaw, D.~M.,
  \& Kraemer, S.~B.}} 2007, ApJ, 656, 699

\bibitem[{Das {et~al.}(2005)Das, Crenshaw, Hutchings, Deo, Kraemer, Gull,
  Kaiser, Nelson, \& Weistrop}]{Das2005}
\href{http://arxiv.org/abs/astro-ph/0505103}{{Das, V., Crenshaw, D.~M.,
  Hutchings, J.~B., {et~al.}}} 2005, AJ, 130, 945

\bibitem[{Davies {et~al.}(2006)Davies, Thomas, Genzel,
  M{\"{u}}ller-S{\'{a}}nchez, Tacconi, Sternberg, Eisenhauer, Abuter, Saglia,
  \& Bender}]{Davies2006}
\href{http://adsabs.harvard.edu/cgi-bin/nph-data{\_}query?bibcode=2006ApJ...646..754D{\&}link{\_}type=ABSTRACT{\%}5Cnpapers2://publication/doi/10.1086/504963}{{Davies,
  R.~I., Thomas, J., Genzel, R., {et~al.}}} 2006, ApJ, 646, 754

\bibitem[{Davies {et~al.}(2015)Davies, Burtscher, Rosario, Storchi-Bergmann,
  Contursi, Genzel, Carpio, Hicks, Janssen, Koss, Lin, Lutz, Maciejewski,
  S{\'{a}}nchez, de~Xivry, Ricci, Riffel, Riffel, Schartmann, M{\"{u}}ller,
  Sternberg, Sturm, Tacconi, \& Veilleux}]{Davies2015}
\href{http://stacks.iop.org/0004-637X/806/i=1/a=127?key=crossref.3005b9026bba182382c39c5b13098e2a}{{Davies,
  R.~I., Burtscher, L., Rosario, D., {et~al.}}} 2015, ApJ, 806, 127

\bibitem[{Davis {et~al.}(2012)Davis, Berrier, Shields, Kennefick, Kennefick,
  Seigar, Lacy, \& Puerari}]{Davis2012}
\href{http://stacks.iop.org/0067-0049/199/i=2/a=33?key=crossref.3b18c54a35b076b5e6ffd72756a56be7}{{Davis,
  B.~L., Berrier, J.~C., Shields, D.~W., {et~al.}}} 2012, ApJSS, 199, 33

\bibitem[{Davis {et~al.}(2017)Davis, Graham, \& Seigar}]{Davis2017a}
\href{http://arxiv.org/abs/1707.04001}{{Davis, B.~L., Graham, A.~W., \& Seigar,
  M.~S.}} 2017, MNRAS, 471, 2187

\bibitem[{Davis \& Hayes(2014)}]{Davis2014}
\href{http://iopscience.iop.org/article/10.1088/0004-637X/790/2/87/meta}{{Davis,
  D.~R., \& Hayes, W.~B.}} 2014, Am. Astron. Soc., 790, 2

\bibitem[{Diniz {et~al.}(2015)Diniz, Riffel, Storchi-Bergmann, \&
  Winge}]{Diniz2015}
Diniz, M.~R., Riffel, R.~A., Storchi-Bergmann, T., \& Winge, C. 2015, MNRAS,
  453, 1727

\bibitem[{Dopita {et~al.}(2015)Dopita, Shastri, Davies, Kewley, Hampton,
  Scharw‰chter, Sutherland, Kharb, Jose, Bhatt, Ramya, Jin, Banfield, Zaw,
  Juneau, James, \& Srivastava}]{Dopita2015a}
Dopita, M.~A., Shastri, P., Davies, R., {et~al.} 2015, ApJSS, 217, 12

\bibitem[{Durr{\'{e}} \& Mould(2018)}]{Durre2018a}
\href{http://arxiv.org/abs/1810.03258}{{Durr{\'{e}}, M., \& Mould, J.}} 2018,
  ApJ, In press

\bibitem[{Durr{\'{e}} {et~al.}(2017)Durr{\'{e}}, Mould, Schartmann, Uddin, \&
  Cotter}]{Durre2017}
\href{http://stacks.iop.org/0004-637X/838/i=2/a=102?key=crossref.4652be8fbbb86c6af289f4b20597b29d
  http://adsabs.harvard.edu/abs/2017ApJ...838..102D}{{Durr{\'{e}}, M., Mould,
  J., Schartmann, M., Uddin, S.~A., \& Cotter, G.}} 2017, ApJ, 838, 102

\bibitem[{Emsellem {et~al.}(2001)Emsellem, Greusard, Combes, Friedli, Leon,
  P�contal, \& Wozniak}]{Emsellem2001}
Emsellem, E., Greusard, D., Combes, F., {et~al.} 2001, A{\&}A, 368, 52

\bibitem[{Everett \& Murray(2006)}]{Everett2006}
\href{http://stacks.iop.org/0004-637X/656/i=1/a=93{\%}5Cnhttp://arxiv.org/abs/astro-ph/0610757}{{Everett,
  J.~E., \& Murray, N.}} 2006, ApJ, 656, 13

\bibitem[{Fischer {et~al.}(2013)Fischer, Crenshaw, Kraemer, \&
  Schmitt}]{Fischer2013}
Fischer, T.~C., Crenshaw, D.~M., Kraemer, S.~B., \& Schmitt, H.~R. 2013, ApJSS,
  209, 1

\bibitem[{Fischer {et~al.}(2015)Fischer, Crenshaw, Kraemer, Schmitt,
  Storchi-Bergmann, \& Riffel}]{Fischer2015}
Fischer, T.~C., Crenshaw, D.~M., Kraemer, S.~B., {et~al.} 2015, ApJ, 799, 234

\bibitem[{Gorkom(1982)}]{Gorkom1982}
Gorkom, J. H.~V. 1982, The Messenger, 33, 45

\bibitem[{Graham \& Scott(2013)}]{Graham2013}
Graham, A.~W., \& Scott, N. 2013, ApJ, 764, 151

\bibitem[{Greene {et~al.}(2012)Greene, Zakamska, \& Smith}]{Greene2012}
\href{http://stacks.iop.org/0004-637X/746/i=1/a=86?key=crossref.c507e1b74f8004d5bab3de34df83bb1e}{{Greene,
  J.~E., Zakamska, N.~L., \& Smith, P.~S.}} 2012, ApJ, 746, 86

\bibitem[{Hicks {et~al.}(2009)Hicks, Davies, Malkan, Genzel, Tacconi,
  {M{\"{u}}ller S{\'{a}}nchez}, \& Sternberg}]{Hicks2009}
Hicks, E. K.~S., Davies, R.~I., Malkan, M.~A., {et~al.} 2009, ApJ, 696, 448

\bibitem[{Ho(2008)}]{Ho2008}
Ho, L.~C. 2008, ARA{\&}A, 46, 475

\bibitem[{Hopkins \& Elvis(2010)}]{Hopkins2010a}
Hopkins, P.~F., \& Elvis, M. 2010, MNRAS, 401, 7

\bibitem[{Kennicutt {et~al.}(2009)Kennicutt, Hao, Calzetti, Moustakas, Dale,
  Bendo, Engelbracht, Johnson, \& Lee}]{Kennicutt2009}
Kennicutt, R.~C., Hao, C.-N., Calzetti, D., {et~al.} 2009, ApJ, 703, 1672

\bibitem[{Lin {et~al.}(2018)Lin, Davies, Hicks, Burtscher, Contursi, Genzel,
  Koss, Lutz, Maciejewski, M{\"{u}}ller-S{\'{a}}nchez, de~Xivry, Ricci, Riffel,
  Riffel, Rosario, Schartmann, Schnorr-M{\"{u}}ller, Shimizu, Sternberg, Sturm,
  Storchi-Bergmann, Tacconi, \& Veilleux}]{Lin2018}
\href{http://academic.oup.com/mnras/article/doi/10.1093/mnras/stx2618/4411840/LLAMA-Nuclear-stellar-properties-of-Swift-BAT-AGN}{{Lin,
  M.-Y., Davies, R., Hicks, E., {et~al.}}} 2018, MNRAS, 473, 4582

\bibitem[{Liu {et~al.}(2015)Liu, Arav, \& Rupke}]{Liu2015}
Liu, G., Arav, N., \& Rupke, D. S.~N. 2015, ApJSS, 221, 9

\bibitem[{McElroy(1995)}]{McElroy1995}
\href{http://adsabs.harvard.edu/doi/10.1086/192209}{{McElroy, D.~B.}} 1995,
  ApJSS, 100, 105

\bibitem[{Mediavilla \& Arribas(1995)}]{Mediavilla1995}
\href{https://academic.oup.com/mnras/article-lookup/doi/10.1093/mnras/276.2.579}{{Mediavilla,
  E., \& Arribas, S.}} 1995, MNRAS, 276, 579

\bibitem[{Mouri {et~al.}(2000)Mouri, Kawara, \& Taniguchi}]{Mouri2000}
Mouri, H., Kawara, K., \& Taniguchi, Y. 2000, ApJ, 528, 186

\bibitem[{{M{\"{u}}ller S{\'{a}}nchez} {et~al.}(2009){M{\"{u}}ller
  S{\'{a}}nchez}, Davies, Genzel, Tacconi, Eisenhauer, Hicks, Friedrich, \&
  Sternberg}]{MullerSanchez2009}
{M{\"{u}}ller S{\'{a}}nchez}, F., Davies, R.~I., Genzel, R., {et~al.} 2009,
  ApJ, 691, 749

\bibitem[{M{\"{u}}ller-S{\'{a}}nchez {et~al.}(2011)M{\"{u}}ller-S{\'{a}}nchez,
  Prieto, Hicks, Vives-Arias, Davies, Malkan, Tacconi, \&
  Genzel}]{Muller-Sanchez2011}
M{\"{u}}ller-S{\'{a}}nchez, F., Prieto, M.~A., Hicks, E. K.~S., {et~al.} 2011,
  ApJ, 739, 69

\bibitem[{Murayama \& Taniguchi(1998)}]{Murayama1998}
Murayama, T., \& Taniguchi, Y. 1998, ApJL, 497, L9

\bibitem[{Nevin {et~al.}(2017)Nevin, Comerford, M{\"{u}}ller-S{\'{a}}nchez,
  Barrows, \& Cooper}]{Nevin2017}
\href{http://academic.oup.com/mnras/article/doi/10.1093/mnras/stx2433/4209975/The-Origin-of-DoublePeaked-Narrow-Lines-in-Active}{{Nevin,
  R., Comerford, J.~M., M{\"{u}}ller-S{\'{a}}nchez, F., Barrows, R., \& Cooper,
  M.~C.}} 2017, MNRAS, 473, 2160

\bibitem[{Ott(2012)}]{Ott2012}
\href{https://ascl.net/1210.019}{{Ott, T.}} 2012, {QFitsView: FITS file
  viewer}, Astrophysics Source Code Library:1210.019

\bibitem[{Plummer(1911)}]{Plummer1911}
\href{https://academic.oup.com/mnras/article-lookup/doi/10.1093/mnras/71.5.460}{{Plummer,
  H.~C.}} 1911, MNRAS, 71, 460

\bibitem[{Prada \& Guti{\'{e}}rrez(1999)}]{Prada1999}
Prada, F., \& Guti{\'{e}}rrez, C.~M. 1999, ApJ, 20, 123

\bibitem[{Ricci {et~al.}(2014)Ricci, Steiner, \& Menezes}]{Ricci2014a}
Ricci, T.~V., Steiner, J.~E., \& Menezes, R.~B. 2014, MNRAS, 440, 2442

\bibitem[{Riffel \& Storchi-Bergmann(2011)}]{Riffel2011}
Riffel, R.~A., \& Storchi-Bergmann, T. 2011, MNRAS, 417, 2752

\bibitem[{Riffel {et~al.}(2015)Riffel, Storchi-Bergmann, \&
  Riffel}]{Riffel2015}
Riffel, R.~A., Storchi-Bergmann, T., \& Riffel, R. 2015, MNRAS, 451, 3587

\bibitem[{Riffel {et~al.}(2013)Riffel, Storchi-Bergmann, \&
  Winge}]{Riffel2013b}
Riffel, R.~A., Storchi-Bergmann, T., \& Winge, C. 2013, MNRAS, 430, 2249

\bibitem[{Rodr{\'{i}}guez-Ardila {et~al.}(2004)Rodr{\'{i}}guez-Ardila,
  Pastoriza, Viegas, Sigut, \& Pradhan}]{Rodriguez-Ardila2004}
Rodr{\'{i}}guez-Ardila, A., Pastoriza, M.~G., Viegas, S., Sigut, T. A.~A., \&
  Pradhan, A.~K. 2004, A{\&}A, 425, 457

\bibitem[{Rodr{\'{i}}guez-Ardila {et~al.}(2005)Rodr{\'{i}}guez-Ardila, Riffel,
  \& Pastoriza}]{Rodriguez-Ardila2005}
Rodr{\'{i}}guez-Ardila, A., Riffel, R., \& Pastoriza, M.~G. 2005, MNRAS, 364,
  1041

\bibitem[{Roth {et~al.}(1991)Roth, Mould, \& Davies}]{Roth1991}
\href{http://adsabs.harvard.edu/cgi-bin/bib{\_}query?1991AJ....102.1303R}{{Roth,
  J., Mould, J.~R., \& Davies, R.~D.}} 1991, AJ, 102, 1303

\bibitem[{Rubin(1980)}]{Rubin1980}
\href{http://adsabs.harvard.edu/doi/10.1086/158041}{{Rubin, V.~C.}} 1980, ApJ,
  238, 808

\bibitem[{Savorgnan {et~al.}(2013)Savorgnan, Graham, Marconi, Sani, Hunt, Vika,
  \& Driver}]{Savorgnan2013}
Savorgnan, G., Graham, A.~W., Marconi, A., {et~al.} 2013, MNRAS, 434, 387

\bibitem[{Schommer {et~al.}(1988)Schommer, Caldwell, Wilson, Baldwin, Phillips,
  Williams, \& Turtle}]{Schommer1988}
\href{http://adsabs.harvard.edu/doi/10.1086/165887}{{Schommer, R.~A., Caldwell,
  N., Wilson, A.~S., {et~al.}}} 1988, ApJ, 324, 154

\bibitem[{Shields {et~al.}(2015{\natexlab{a}})Shields, Boe, Pfountz, Davis,
  Hartley, {Pour Imani}, Slade, Kennefick, \& Kennefick}]{Shields2015a}
\href{http://adsabs.harvard.edu/abs/2015ascl.soft12015S}{{Shields, D.~W., Boe,
  B., Pfountz, C., {et~al.}}} 2015{\natexlab{a}}, {Astrophysics Source Code
  Library}

\bibitem[{Shields {et~al.}(2015{\natexlab{b}})Shields, Boe, Pfountz, Davis,
  Hartley, Imani, Slade, Kennefick, \& Kennefick}]{Shields2015}
\href{http://arxiv.org/abs/1511.06365}{{Shields, D.~W., Boe, B., Pfountz, C.,
  {et~al.}}} 2015{\natexlab{b}}, ArXiv, 1511.06365

\bibitem[{Shimizu {et~al.}(2018)Shimizu, Davies, Koss, Ricci, Lamperti, Oh,
  Schawinski, Trakhtenbrot, Burtscher, Genzel, Lin, Lutz, Rosario, Sturm, \&
  Tacconi}]{Shimizu2018}
\href{http://arxiv.org/abs/1710.09117{\%}0Ahttp://dx.doi.org/10.3847/1538-4357/aab09e}{{Shimizu,
  T.~T., Davies, R.~I., Koss, M., {et~al.}}} 2018, ApJ, 856, 154

\bibitem[{Steffen {et~al.}(2010)Steffen, Koning, Wenger, Morisset, \&
  Magnor}]{Steffen2010}
\href{http://arxiv.org/abs/1003.2012}{{Steffen, W., Koning, N., Wenger, S.,
  Morisset, C., \& Magnor, M.}} 2010, IEEE Trans. Vis. Comput. Graph., 17, 454

\bibitem[{Storchi-Bergmann {et~al.}(2010)Storchi-Bergmann, Lopes, McGregor,
  Riffel, Beck, \& Martini}]{Storchi-Bergmann2010}
Storchi-Bergmann, T., Lopes, R. D.~S., McGregor, P.~J., {et~al.} 2010, MNRAS,
  402, 819

\bibitem[{Tonry \& Davis(1979)}]{Tonry1979}
\href{http://adsabs.harvard.edu/cgi-bin/bib{\_}query?1979AJ.....84.1511T}{{Tonry,
  J., \& Davis, M.}} 1979, AJ, 84, 1511

\bibitem[{Urry \& Padovani(1995)}]{Urry1995}
Urry, C.~M., \& Padovani, P. 1995, PASP, 107, 803

\bibitem[{Vasudevan {et~al.}(2010)Vasudevan, Fabian, Gandhi, Winter, \&
  Mushotzky}]{Vasudevan2010}
Vasudevan, R.~V., Fabian, A.~C., Gandhi, P., Winter, L.~M., \& Mushotzky, R.~F.
  2010, MNRAS, 402, 1081

\bibitem[{Veilleux {et~al.}(2005)Veilleux, Cecil, \&
  Bland-Hawthorn}]{Veilleux2005}
\href{www.annualreviews.org}{{Veilleux, S., Cecil, G., \& Bland-Hawthorn, J.}}
  2005, Annu. Rev. Astron. Astrophys, 43, 769

\bibitem[{Wagner \& Appenzeller(1988)}]{Wagner1988}
\href{http://cdsads.u-strasbg.fr/cgi-bin/nph-bib{\_}query?1988A{\&}A...197...75W{\&}db{\_}key=AST}{{Wagner,
  S.~J., \& Appenzeller, I.}} 1988, A{\&}A, 197, 75

\bibitem[{Wilson {et~al.}(1993)Wilson, Braatz, Heckman, Krolik, \&
  Miley}]{Wilson1993}
\href{http://adsabs.harvard.edu/doi/10.1086/187137}{{Wilson, A.~S., Braatz,
  J.~A., Heckman, T.~M., Krolik, J.~H., \& Miley, G.~K.}} 1993, ApJ, 419, L61

\bibitem[{Winge {et~al.}(2009)Winge, Riffel, \& Storchi-Bergmann}]{Winge2009}
Winge, C., Riffel, R.~A., \& Storchi-Bergmann, T. 2009, ApJSS, 185, 186

\bibitem[{Winter {et~al.}(2012)Winter, Veilleux, McKernan, \&
  Kallman}]{Winter2012}
\href{http://stacks.iop.org/0004-637X/745/i=2/a=107?key=crossref.a502f77474a48e653b0adfd8eb9f3bf8}{{Winter,
  L.~M., Veilleux, S., McKernan, B., \& Kallman, T.~R.}} 2012, ApJ, 745, 107

\end{thebibliography}
\end{document}